\documentclass[a4paper,11pt]{article}
\pdfoutput=1
\usepackage[utf8x]{inputenc}
\usepackage{jheppub}
\usepackage{epsf}
\usepackage{graphicx}	
\usepackage{dcolumn}	
\usepackage{bm}			
\usepackage{amssymb, amsmath}
\usepackage{wasysym}
\usepackage{lipsum, color}

\usepackage{bbm}		
\usepackage{multirow}

\usepackage{slashed}
\usepackage{hyperref}
\hypersetup{colorlinks, citecolor=bluscuro, linkcolor=black, urlcolor=bluscuro}
\definecolor{rossos}{cmyk}{0,1,1,0.55}
\definecolor{bluscuro}{rgb}{0.15, 0.2, .85}
\definecolor{bluchiaro}{cmyk}{1,.3,0.,0.1}

\let\oldquote\quote
\renewcommand\quote{\scriptsize\oldquote}
\let\oldquotation\quotation
\renewcommand\quotation{\scriptsize\oldquotation}

\newcommand{\gsim}{\lower.7ex\hbox{$\;\stackrel{\textstyle>}{\sim}\;$}}
\newcommand{\lsim}{\lower.7ex\hbox{$\;\stackrel{\textstyle<}{\sim}\;$}}

\def\beq{\begin{equation}}
\def\eeq{\end{equation}}
\def\be{\begin{equation}}
\def\ee{\end{equation}}
\def\bea{\begin{eqnarray}}
\def\eea{\end{eqnarray}}
\def\bmat{\begin{pmatrix}}
	\def\emat{\end{pmatrix}}
\def\bei{\begin{itemize}}
	\def\eei{\end{itemize}}

\newcommand{\nn}{\nonumber}

\newcommand{\mpl}{M_{\rm{Pl} } }

\usepackage[usenames,dvipsnames,svgnames]{xcolor}
\usepackage[normalem]{ulem}

\makeatletter
\def\section{\@startsection {section}{1}{\z@}{-3.5ex plus -1ex minus
		-.2ex}{2.3ex plus .2ex}{\large\bf}}
\def\subsection{\@startsection{subsection}{2}{\z@}{-3.25ex plus -1ex
		minus -.2ex}{1.5ex plus .2ex}{\normalsize\bf}}
\makeatother
\makeatletter

\@addtoreset{equation}{section}

\makeatother

\usepackage{graphicx}  
\usepackage{dcolumn}   
\usepackage{bm,relsize}        
\usepackage{amssymb, amsmath}
\usepackage{wasysym}
\usepackage{slashed}
  \usepackage{bbold} 
\usepackage[usenames,dvipsnames,svgnames]{xcolor}
\usepackage{jheppub}

\usepackage{mathtools}
\usepackage{comment}
\usepackage{cancel}
\usepackage{ulem}

\usepackage{feyn}

\def\beq{\begin{equation}}
\def\eeq{\end{equation}}

\newcommand{\defeq}{\stackrel{\text{def}}{=}}

\begin{document}

\global\long\def\com#1#2{\underset{{\scriptstyle #2}}{\underbrace{#1}}}

\global\long\def\comtop#1#2{\overset{{\scriptstyle #2}}{\overbrace{#1}}}

\global\long\def\ket#1{\left|#1\right\rangle }

\global\long\def\bra#1{\left\langle #1\right|}

\global\long\def\braket#1#2{\left\langle #1|#2\right\rangle }

\global\long\def\op#1#2{\left|#1\right\rangle \left\langle #2\right|}

\global\long\def\opk#1#2#3{\left\langle #1|#2|#3\right\rangle }

\global\long\def\L{\mathcal{L}}

\title{Ripples in Spacetime from Broken Supersymmetry}
\abstract{We initiate the study of gravitational wave (GW) signals from first-order phase transitions in supersymmetry-breaking hidden sectors. Such phase transitions often occur along a pseudo-flat direction universally related to supersymmetry (SUSY) breaking in hidden sectors that spontaneously break $R$-symmetry. The potential along this pseudo-flat direction imbues the phase transition with a number of novel properties, including a nucleation temperature well below the scale of heavy states (such that the temperature dependence is captured by the low-temperature expansion) and significant friction induced by the same heavy states as they pass through bubble walls. In low-energy SUSY-breaking hidden sectors, the frequency of the GW signal arising from such a phase transition is guaranteed to lie within the reach of future interferometers given existing cosmological constraints on the gravitino abundance. 
Once a mediation scheme is specified, the frequency of the GW peak correlates with the superpartner spectrum. Current bounds on supersymmetry are compatible with GW signals at future interferometers, while the observation of a GW signal from a SUSY-breaking hidden sector would imply superpartners within reach of future colliders.}

\author[a]{Nathaniel Craig,}
\author[b]{Noam Levi,}
\author[c]{Alberto Mariotti,}
\author[d,e]{Diego Redigolo,}
\affiliation[a]{Department of Physics, University of California, Santa Barbara, CA 93106, USA}
\affiliation[b]{Raymond and Beverly Sackler School of Physics and Astronomy, Tel-Aviv University, Tel-Aviv 69978, Israel}
\affiliation[c]{Theoretische Natuurkunde and IIHE/ELEM, Vrije Universiteit Brussel,
and International Solvay Institutes, Pleinlaan 2, B-1050 Brussels, Belgium}
\affiliation[d]{CERN, Theoretical Physics Department, Geneva, Switzerland}
\affiliation[e]{INFN Sezione di Firenze, Via G. Sansone 1, I-50019 Sesto Fiorentino, Italy}
\date{\today}

\maketitle

\section{Introduction}

If supersymmetry is a property of our universe, how will it be discovered? Conventionally, searches for evidence of supersymmetry (SUSY) have focused on the Standard Model, looking for supersymmetric partners of Standard Model particles in direct production at colliders, scattering in dark matter experiments, and virtual effects in precision measurements. Thus far, no evidence has emerged of supersymmetry as it relates to the Standard Model, raising the prospect that it may lie outside the reach of the existing experimental program. Although this would pose a challenge to supersymmetry as a fully natural explanation for the scale of electroweak symmetry breaking, the abundance of remaining motivation (e.g. gauge coupling unification, dark matter, and straightforward string-theoretic embedding) favors continuing the search to shorter and shorter distances. While the LHC and proposed future colliders are promising tools in this search, the immense technical challenges of exploring energies far above the TeV scale in terrestrial experiments suggests casting a broader net. It invites identifying both new ways of accessing shorter distances and new sectors in which supersymmetry may be manifest.

A compelling avenue to shorter distances is to make use of the incredible energies of the Big Bang, searching for the imprint of supersymmetric phenomena on the early universe. In some sense, this is already the path taken by dark matter searches looking for the population of stable superpartners produced in the early universe, but it is not the only cosmological avenue for discovering SUSY. For example, spontaneous breaking of supersymmetry during inflation raises the prospect of observing signals in the three-point function of primordial curvature perturbations \cite{Baumann:2011nk, Craig:2014rta}, although the size of the signal depends on the strength of couplings between SUSY multiplets and the inflaton.

As for new sectors, at least one is {\it guaranteed} to exist in a supersymmetric universe: the sector responsible for breaking supersymmetry. Although there are many dynamical mechanisms for breaking supersymmetry, they typically possess a number of generic or universal features which can provide new ways of searching for supersymmetry even when superpartners of the Standard Model are decoupled. These include the goldstino, a goldstone fermion of spontaneous supersymmetry breaking (which becomes the longitudinal mode of the gravitino, the supersymmetric partner of the graviton, once gravity is accounted for), as well as a novel abelian global symmetry called the $R$-symmetry. The $R$-symmetry is generically spontaneously broken by the same dynamics that breaks supersymmetry, giving rise to a goldstone boson (the $R$-axion) and its scalar partner, a pseudo-modulus whose flat potential is protected by supersymmetry. In theories with low-energy supersymmetry breaking (LESB), in which the effects of SUSY breaking are communicated to the Standard Model by forces stronger than gravitation, these states may be accessible on their own. For example, the goldstino couples directly to Standard Model particles and may be produced at colliders, although the current reach of the LHC makes these searches less promising than continuing to look for Standard Model superpartners.

In this paper, we explore a new avenue for discovering supersymmetry in the physics of the early universe: using the stochastic gravitational wave background (SGWB) produced by a first-order phase transition to directly probe the sector responsible for breaking supersymmetry \cite{Craig:2009zx}. This makes use of the extraordinary opportunities afforded by the detection of gravitational waves (GW) by the LIGO-Virgo collaboration~\cite{Abbott:2016blz}, which has opened a new era in the exploration of the early universe. Sensitivity of current and proposed GW interferometers to stochastic gravitational wave backgrounds broadly motivates identifying beyond-the-Standard Model (BSM) scenarios whose first-order phase transitions may generate such a signal and exploring the complementarity of GW interferometry with other probes of new physics such as present and future colliders.\footnote{Note that the LIGO-Virgo collaboration 
already places direct constraints on SGWB~\cite{Abbott:2017mem,LIGOScientific:2019vic} beyond existing indirect limits.}

Among the most compelling scenarios for SGWB are those in which a first order phase transition (FOPT) is associated with the breaking of a global or gauge symmetry in the early universe~\cite{Witten:1984rs,Hogan:1984hx,Hogan:1986qda,Turner:1990rc,Caprini:2015zlo,Mazumdar:2018dfl}. As we will show, the supersymmetry-breaking sector is a natural candidate for such a phase transition because it generically possesses at least one pseudo-flat complex scalar direction, the pseudomodulus. In our constructions, the phase of this complex scalar direction is associated to the $R$-symmetry, which is in turn tied to the SUSY-breaking dynamics by many known theorems about SUSY quantum field theories~\cite{Nelson:1993nf,Intriligator:2007py,Komargodski:2009jf}. This complex scalar direction is lifted by quantum corrections and the resulting potential is likely to possess a metastable minimum at the origin (where $R$-symmetry is preserved), which will then decay to the true minimum through a FOPT in the early universe. At the true minimum the $R$-symmetry is broken, consistently with a realistic SUSY spectrum featuring Majorana masses for the fermionic partners of Standard Model gauge bosons. 

In our framework, the SUSY-breaking scale $\sqrt{F}$ correlates directly with the frequency range of the SGWB, such that theories of low-energy SUSY breaking feature a peak frequency accessible at LIGO-Virgo or proposed GW interferometers such as the Laser Interferometer Space Antenna (LISA), Einstein Telescope (ET), Cosmic Explorer (CE), DECi-hertz Interferometer Gravitational wave Observatory (DECIGO), and the Big Bang Observatory (BBO). In fact, a consistent cosmological history (in which the production of gravitinos in the early universe is consistent with the present dark matter abundance and small-scale structure constraints) guarantees that low-energy supersymmetry-breaking phase transitions produce a peak frequency in the range accessible to current and future interferometers~\cite{TheLIGOScientific:2014jea,Sathyaprakash:2012jk,Evans:2016mbw,Reitze:2019iox}. 

Once a mechanism is specified to mediate supersymmetry breaking to the Standard Model, the scale $\sqrt{F}$ is also correlated with the spectrum of Standard Model superpartners, allowing the possibility of cross-correlating GW and collider signals. As we will see, the non-observation of SUSY particles at the LHC leaves open the opportunity for seeing SGWB signatures from low-energy SUSY breaking, making this a leading avenue for the discovery of supersymmetry. In return, the observation of a SGWB signal from a low-energy supersymmetry breaking phase transition would imply the SUSY spectrum to be within the reach of future colliders such as FCC-hh, SPPC, or a high-energy muon collider, highlighting the strong complementarity between such SGWB signals and proposed colliders. 

Along the way, we identify a qualitatively new class of natural potentials capable of generating large GW signals from a first-order phase transition, corresponding to the scalar potential along the pseudo-flat direction associated with SUSY breaking. The size of the vacuum energy gap between the metastable and the true vacuum is set by the SUSY-breaking scale $\sqrt{F}$, which is necessarily smaller than the SUSY masses of other heavy fields in the sector in order to avoid tachyonic directions. As a consequence, the thermal corrections to the potential along the pseudomodulus direction are well described by a low-$T$ expansion.  Another distinctive feature of the pseudomodulus potential is the flatness at large field values, which is ensured by SUSY cancellations independently of the nature of SUSY-breaking deformations around the origin. These features combine to give strong first-order phase transitions, with the strongest transitions arising most naturally in models with two distinct SUSY-breaking scales. As we will discuss, a large amount of tuning would be necessary to realize a similar situation in non-SUSY scenarios, which explains why this possibility has not been explored so far in the literature~(see for instance \cite{Chung:2012vg} for a collection of potentials giving raise to FOPT for the SM Higgs).

The organization of our paper is intended to highlight the qualitative connections between low-energy supersymmetry breaking, first-order phase transitions, and stochastic gravitational wave signals before progressing into explicit examples, and does not presume deep familiarity with spontaneous supersymmetry breaking. We begin in Sec.~\ref{sec:frequencyvsspectrum} by giving a broad overview of the phenomenology of low-energy SUSY breaking, the parametrics of gravitational wave signals from first-order phase transitions, and the relationship between the frequency of the SGWB signal and spectrum of SUSY particles. In Sec.~\ref{sec:anatomy} we discuss the general features of the pseudomodulus potential and the properties of a first-order phase transition along this direction, highlighting their novelty compared to commonly-studied potentials for FOPT. We present a simple toy model that captures the main features of concrete SUSY potentials, showing how a promising GW signal from FOPT requires multiple SUSY-breaking scales.  

We then proceed to develop a series of increasingly realistic SUSY-breaking hidden sectors featuring FOPT in Sec.~\ref{sec:models}. In Sec.~\ref{sec:Orafe}, we derive the phase diagram of the simple O'Raifeartaigh model. In Sec.~\ref{sec:X3} we present the simplest single scale model featuring a FOPT, which is a simple deformation of the O'Raifeartaigh model with explicit $R$-symmetry breaking. Here, the FOPT takes place between the SUSY-breaking vacuum at the origin (which enjoys an unbroken $R$-symmetry) and the $R$-symmetry breaking minimum far away from the origin, where SUSY is restored unless coupled to an additional source of SUSY-breaking. In Sec.~\ref{sec:FD}, we develop a fully realistic model by introducing gauge interactions to the O'Raifeartaigh model, such that SUSY is broken in both the metastable vacuum at the origin and true vacuum. The presence of both $F$-term and $D$-term SUSY-breaking naturally gives rise to strong GW signals. 

In Sec.~\ref{sec:pheno} we further comment on the phenomenology of our setup and the complementarity between GW observatories and colliders, highlighting the sense in which the observation of a SGWB signal in our models would ensure further evidence for SUSY at future colliders.
We summarize our qualitative conclusions and future directions in Sec.~\ref{sec:conclusions}. Technical details are reserved for a series of appendices, including a review of the one-loop thermal effective potential in App.~\ref{app:potentials}, approaches to the calculation of the bounce action in App.~\ref{app:bounce} and inputs to our projections for GW interferometers in App.~\ref{sub:signal_detection}.

\begin{figure}[t!]
  \centering
 \includegraphics[width=1\textwidth]{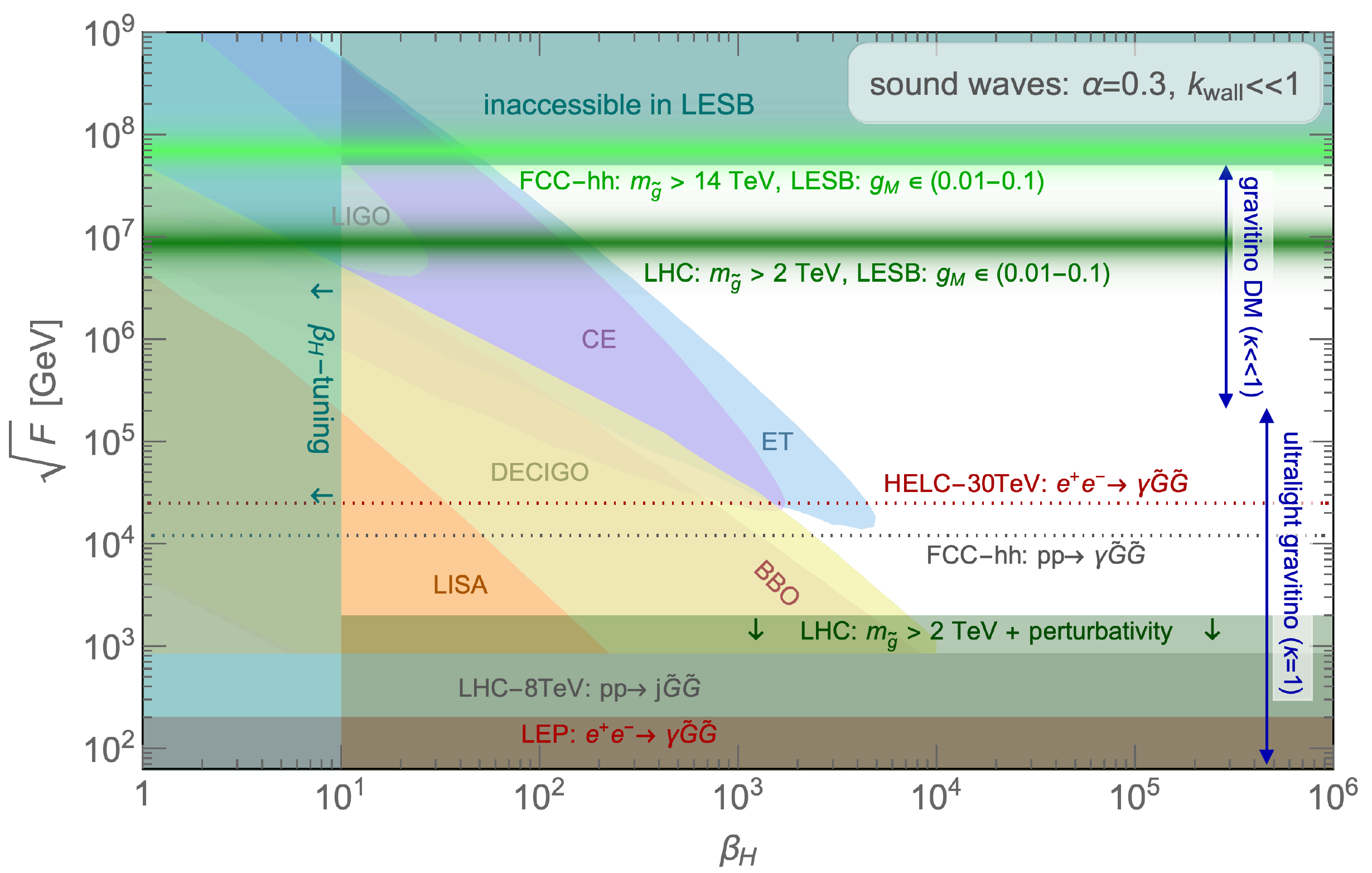}
\caption{Parameter space of low energy SUSY-breaking models in the $(\sqrt{F},\beta_H)$ plane, with $\alpha=0.3$ and $T_{\text{r.h.}}=\sqrt{F}$ (see Sec.~\eqref{sec:FOPTandGWs} for definitions). The GW reach is computed by requiring the strength of the SGBW signal at the peak frequency to intersect with the PLI curve of a given GW interferometer (see Appendix~\ref{ssub:pli_curves} for details). The {\bf colored regions} show the reach for a signal generated from plasma waves which is generically the dominant one in our scenarios (see Sec.~\ref{sec:FOPTandGWs}). The {\bf cyan region} with $\beta_H<10$ imples a large fine-tuning in our setups (see discussion around Eq.~\eqref{eq:betaFT}, the parametric discussion in Sec.~\ref{sec:anatomy} and the explicit evaluation in the models of Sec.~\ref{sec:models}). In the {\bf red shaded} region gravitino pair production is excluded by a $\gamma+\text{MET}$ search at LEP with $\mathcal{L}=0.24\text{ fb}^{-1}$~\cite{Brignole:1997sk}, the gray region is excluded by ATLAS bounds  $j+\text{MET}$ at $\sqrt{s}=8\text{ TeV}$ and $\mathcal{L}=10.5\text{ fb}^{-1}$~\cite{Brignole:1998me,Maltoni:2015twa}. The {\bf dotted gray} and {\bf dotted red} lines are the projection of the $\gamma+\text{MET}$ reach at FCC-hh with $\sqrt{s}=100\text{ TeV}$ and $\mathcal{L}=30\text{ ab}^{-1}$ and a future high energy lepton collider (HELC) with $\sqrt{s}=30\text{ TeV}$ and $\mathcal{L}=100\text{ ab}^{-1}$. The {\bf dark green shaded} region with {\bf dark green arrows} indicates the bound on the SUSY-breaking scale derived from the LHC bound on gluinos $m_{\tilde{g}}>2\text{ TeV}$, requiring the messenger sector to be perturbative. The two {\bf dark green} and {\bf light green} bands show the impact of the present LHC bounds~\cite{Aaboud:2018doq,Aaboud:2018mna,ATLAS:2019vcq,ATLAS-CONF-2020-047} and the future FCC-hh reach on gluinos~\cite{Arkani-Hamed:2015vfh} for perturbative messenger sectors with $g_M\in(0.01,0.1)$ (see Eq.~\eqref{eq:gluinomass} for a definition of $g_M$). The region between these two lines will be naturally populated by the model discussed in Sec.~\ref{sec:FD} and Sec.~\ref{sec:gaugemediation}. The {\bf dark blue arrows} on the right hand side shows the \emph{ultralight gravitino window} where $m_{3/2}\leq 16\text{ eV}$ and the gravitino does not poses any cosmological challenge with $\kappa=1$ (see Eq.~\eqref{eq:m32} for a definition) and the \emph{gravitino dark matter window} where  $\kappa\ll1$ and the full gravitino mass is heavier than the gravitino mass contribution set by $\sqrt{F}$. The {\bf dark cyan} region marked as \emph{inaccessible in LESB} is always excluded by a combination of gravitino overabundance~\cite{Hall:2013uga} and BBN constraints~\cite{Jedamzik:2006xz} (see Sec.~\ref{sec:pheno} for details).}\label{fig:moneyplot}
\end{figure}

\section{Detectable GW signals in low energy SUSY-breaking}\label{sec:frequencyvsspectrum}

In this section we illustrate the correlation between possible signals at present and future GW interferometers and the phenomenology of low energy SUSY-breaking (LESB).   The underlying assumption is that the SGWB is produced through a first-order phase transition controlled by the SUSY-breaking hidden sector. As we will show, this connection is relatively insensitive to the details of the dynamics in the hidden sector. In Sec.~\ref{sec:LESB} we summarize the structure and the parametric predictions of LESB theories, while in Sec.~\ref{sec:FOPTandGWs} we go through the field theory inputs that are necessary to compute the spectrum of GWs from a FOPT.  In Sec.~\ref{sec:summaryparspace} we combine the results of the preceding sections to delineate the parameter space of possible gravitational wave signals from low-energy supersymmetry breaking, illustrated in Fig.~\ref{fig:moneyplot}.

\subsection{Low-energy SUSY-breaking}\label{sec:LESB}
Here we briefly review the structure of LESB and its broad parametric predictions, remaining agnostic as to the particular model realization. This general discussion is buttressed by Sec.~\ref{sec:pheno}, where we will present the predictions of a simple, explicit model which gives rise to GW signals. For the purposes of this paper, we adopt a phenomenological definition of low-energy SUSY breaking: LESB models are those in which \emph{the gravitino is the lightest supersymmetric particle (LSP)}. This requirement has deep implications for collider searches, precision observables, and cosmology, which we summarize in turn. 

The first ingredient in low energy SUSY-breaking scenarios is a hidden sector at a high scale $m_*$, which breaks supersymmetry (and $R$-symmetry) spontaneously. At energies much below $m_*$, the spontaneous breaking of both supersymmetry and the $R$-symmetry can be encoded in a model-independent manner through the $F$- and scalar components of a single chiral superfield

\begin{equation}
\langle X\rangle= \frac{f_a}{\sqrt{2}} e^{2i a/f_a}+\sqrt{2}\theta\tilde{G}+\theta^2 F\ ,\label{eq:spurion}
\end{equation}
where $f_a$ is the order parameter for the $R$-symmetry breaking while $\sqrt{F}$ is the SUSY breaking scale, corresponding to an $R$-charge $R_X=2$ for the superfield $X$. The parameter $\theta$ is a constant, complex anti-commuting two-component spinor enabling component fields of different spin to be united into a single superfield. The Majorana fermion in the multiplet is the Goldstino $\tilde{G}$, the goldstone fermion associated with spontaneous SUSY-breaking, while the compact scalar field $a$ is the $R$-axion, the goldstone boson associated with spontaneous $R$-symmetry breaking. 

Switching on gravity, the Goldstino becomes the longitudinal component of the gravitino via the super-Higgs mechanism~\cite{Deser:1977uq} while the $R$-axion is lifted by an unavoidable explicit symmetry-breaking contribution arising from the fine-tuning of the cosmological constant~\cite{Bagger:1994hh,Bellazzini:2017neg}. The gravitino and $R$-axion masses can be written as
\begin{align}
&m_{3/2}=\frac{F_0}{\sqrt{3}M_{\text{Pl}}}\simeq24\text{ keV}\left(\frac{1}{\kappa}\right)\left(\frac{\sqrt{F}}{10^7\text{ GeV}}\right)^2\ ,\label{eq:m32}\\
&m_a^{\text{grav.}}=m_{3/2}\left[\frac{6^{3/2}M_{\text{Pl}}}{f_a}\right]^{1/2} \simeq5.2\text{ GeV}\left(\frac{m_{3/2}}{24\text{ keV}}\right)\left(\frac{10^7\text{ GeV}}{\sqrt{F}}\right)\left(\frac{1}{\sqrt{\epsilon_R}}\right)\ ,\label{eq:mRax}
\end{align}
where $M_{\text{Pl}}=2.4\cdot10^{18}\text{ GeV}$ is the reduced Planck scale and we have defined 
\begin{equation}
\kappa\defeq F/F_0\qquad,\qquad \epsilon_R\defeq2F/f_a^2\ .\label{eq:defs}
\end{equation}
This reflects the fact that the gravitino mass is set by the sum of supersymmetry-breaking contributions from all sectors, corresponding to a total SUSY-breaking scale $\sqrt{F_0}$ that may be larger than the scale $\sqrt{F}$ in the LESB sector under consideration (i.e. $\kappa\lesssim1$). Similarly, the $R$-symmetry breaking scale $f_a$ may exceed the scale of supersymmetry breaking $\sqrt{F}$ (i.e. $\epsilon_R\lesssim1$), as is often the case in calculable hidden sectors. In writing the gravity contribution to the $R$-axion mass in Eq.~\eqref{eq:mRax} we saturated the upper bound on the superpotential vacuum expectation value (VEV)~\cite{Dine:2009sw}. In the presence of a possible explicit $R$-symmetry breaking term in the hidden sector $\epsilon_{\slashed{R}}$, the $R$-axion mass will receive an extra contribution 
\begin{equation}
m_a^{\slashed{R}}= \sqrt{\epsilon_{\slashed{R}}F}\simeq 10^3\text{ TeV}\left(\frac{\epsilon_{\slashed{R}}}{0.01}\right)^{1/2}\left(\frac{\sqrt{F}}{10^7\text{ GeV}}\right)\ ,
\end{equation}
making the $R$-axion heavier than the superpartners of Standard Model fields and hence phenomenologically irrelevant. 

Most of the universal phenomenological predictions of low-energy SUSY breaking follow from the gravitino's role as the LSP~\cite{Giudice:1998bp}.  First, the gravitino is the endpoint of every superpartner decay. In particular, the lifetime of the next-to-lightest supersymmetric particle (NLSP) is determined by its decay into the gravitino plus a Standard Model state,
\begin{equation}
\tau_{\text{NLSP}}=\frac{48\pi}{c_{\text{NLSP}}} \frac{M_{\text{Pl}}^2 m_{3/2}^2}{m_{\text{NLSP}}^5}\simeq10^2\text{ sec}
\left(\frac{1}{c_{\text{NLSP}}}\right)\left(\frac{m_{3/2}}{24\text{ keV}}\right)^2\left(\frac{500\text{ GeV}}{m_{\text{NLSP}}}\right)^5\ , \label{eq:NLSPdecay}
\end{equation} 
where $c_{\text{NLSP}}$ is an $\mathcal{O}(1)$ coefficient which depends on the particulars of the NLSP. Second, the gravitino may be directly produced in pairs with a rate controlled by dimension-eight contact operators suppressed by $1/F^2$ in the $m_{\text{soft}}\gg m_{3/2}$ limit~\cite{Brignole:1997pe,Brignole:1997sk}. These operators lead to a total cross section at lepton colliders for pair production in association with a photon of the form
\begin{equation}
\sigma(e^+e^-\to \tilde{G}\tilde{G}\gamma)\simeq \frac{\alpha_{\text{em}}s^3}{160\pi^2 F^4}\left[\frac{247}{60}+\log\left(\frac{4E_{\text{min}}^2}{s}\right)\right]\log\left(\frac{1-\cos\theta_{\text{min}}}{1+\cos\theta_{\text{min}}}\right)\ ,\label{eq:directpairxsec}
\end{equation}
where $\sqrt{s}$ is the beam energy, $E_{\text{min}}$ is the minimal photon energy, and $\theta_\text{min}$ is the minimal photon angle with respect to the beam direction. Here we have expanded in $E_{\text{min}}\ll \sqrt{s}$ (see Ref.~\cite{Brignole:1997sk} for the full formula). A similar formula can be derived for $\sigma(pp\to \tilde{G}\tilde{G}j)$ as shown in Ref.~\cite{Brignole:1998me}. Using these formulas and rescaling the Standard Model backgrounds to higher energies and luminosities, we may determine the sensitivity of missing energy searches at future colliders gravitino pair production; see Sec.~\ref{sec:pheno} for details. These searches lead to projected direct constraints on the SUSY-breaking scale $\sqrt{F}$ as shown in Fig.~\ref{fig:moneyplot}.  
  
Finally, the stability of the gravitino LSP typically results in a cosmological hazard. This is the well known ``gravitino problem'' of LESB theories~\cite{Moroi:1993mb,Kawasaki:1994af,Moroi:1995fs}. For sufficiently high reheating temperature (i.e. $T_{\text{r.h.}}>45 m_{3/2}^2 M_{\text{Pl}}/M_3^2$, where $M_3$ is the soft mass of the gluino), the gravitino is in thermal equilibrium with the Standard Model bath. At freeze-out, the gravitino is still relativistic and its abundance is bounded from above by small-scale cosmological observables~\cite{Pierpaoli:1997im,Viel:2005qj}. The latter imply $m_{3/2}\lesssim 16\text{ eV}$, which corresponds to $\sqrt{F}<260\text{ TeV}$. Alternately, if the reheating temperature is low enough, the gravitino is never in equilibrium with the Standard Model bath but is typically overproduced by a combination of UV scattering contributions~\cite{Bolz:2000fu,Pradler:2006qh,Pradler:2006hh,Rychkov:2007uq}, freeze-in from the decays of superpartners~\cite{Cheung:2011nn}, and decay of the NLSP relic abundance after freeze-out~\cite{Feng:2003xh,Feng:2003uy}. 

In order for SUSY-breaking sectors to generate sizable GW signals, the hidden sector needs to be reheated after inflation. Fixing the reheating temperature $T_{\text{r.h.}}=\sqrt{F}$ and requiring the gravitino to not overclose the universe implies  
\begin{equation}
C_{\text{UV}} \frac{M_{3}^2 T_{\text{r.h.}}}{m_{3/2}}+C_{\text{F.O.}} \frac{m_{3/2} m_{\text{NLSP}}}{\alpha_{\text{eff}}^2}\lesssim 0.27\, T_{\text{eq}}M_{\text{Pl}}\ ,\label{eq:m32abundance}
\end{equation}
where $C_{\text{UV}}=45\sqrt{5}f_3/(8\pi^{13/2} g_*^{3/2})\simeq 4\cdot 10^{-5}$, $g_*\simeq230$, and $f_3\simeq 18$ encodes the thermal corrections as computed in Ref.~\cite{Rychkov:2007uq}; $C_{\text{F.O.}}=x_{F.O.}/(4\pi \sqrt{g_*})=0.12$ for $x_{F.O.}=23$, and $\alpha_{\text{eff}}\simeq0.01$ is chosen to match the correct dark matter relic abundance in the pure Higgsino case~\cite{ArkaniHamed:2006mb}.  Eq.~\eqref{eq:m32abundance} reflects a similar expression in Ref.~\cite{Hall:2013uga}, although here we have dropped the freeze-in contribution from superpartner decays because it is always subdominant compared to UV scattering contributions. 

For $\kappa=1$, the only region where the gravitino is not overabundant for $T_{\text{r.h.}}=\sqrt{F}$ corresponds to $\sqrt{F}<260\text{ TeV}$, while for $\kappa\ll1$ one can decouple the gravitino mass and push the SUSY-breaking scale to be as high as $\sqrt{F}\simeq 5\cdot 10^7\text{ GeV}$. In this case, the upper bound is obtained by combining the overclosure bound, LHC bounds on Standard Model superpartner masses, and the BBN bounds on NLSP decays into the gravitino through the universal two-body decay in Eq.~\eqref{eq:NLSPdecay} as derived in Ref.~\cite{Jedamzik:2006xz}. This bound could slightly vary depending on the NLSP type and the detailed features of the spectrum, but this does not alter the primary message: requiring a reheating temperature $T_{\text{r.h.}}=\sqrt{F}$ to obtain sufficiently strong gravitational wave signals implies a quite stringent upper bound on $\sqrt{F}$ as long as the gravitino is required to be the LSP. 

Thus far, our discussion has not correlated the scale $\sqrt{F}$ of supersymmetry breaking with the mass spectrum of Standard Model superpartners. Supersymmetry breaking in the hidden sector is transmitted to the visible sector (which we will take to be the minimal supersymmetric Standard Model, or MSSM, in this paper) through a mediation mechanism. The simplest possibility is to assume that a certain number of messengers $N_{\text{mess}}$ in a given representation of the SM gauge group are coupled to the SUSY-breaking field $X$ via the superpotential $W_{\text{mess}}=y_{\text{mess}} X\Phi \tilde{\Phi}$. Given this coupling, the $R$-symmetry breaking scale $f_a$ controls the masses of the fermionic messengers, while the SUSY-breaking scale $\sqrt{F}$ gives an off-diagonal mass to the scalar messengers. The non-supersymmetric splitting between scalar and fermionic messengers is then transmitted to MSSM superfields via Standard Model gauge interactions. The resulting gaugino and squark masses are those of standard gauge mediation~\cite{Giudice:1998bp},  
\begin{align}
M_{I}=\frac{\alpha_I N_{\text{mess}} s_M}{4\pi}\left(\frac{\sqrt{2}F}{f_a}\right) \qquad ,\qquad m_{\tilde{f}}^2=\sum_IC_{\tilde{f}}(I)\left(\frac{\alpha_I N_{\text{mess}}}{4\pi}\right)^2 \left(\frac{\sqrt{2}F}{f_a}\right)^2\ ,\label{eq:softmasses}
\end{align}
where $C_{\tilde{f}}(I)$ is the quadratic Casimir of the representation of the MSSM sfermion $\tilde{f}$ under the $I$th Standard Model gauge group and for simplicity we have considered messengers in the $5+\bar{5}$ representation of $SU(5)$. The additional coefficient $s_M\lesssim 1$ appearing in the gaugino masses accounts for the phenomenon of ``gaugino screening''~\cite{ArkaniHamed:1998kj,Komargodski:2009jf,Cohen:2011aa}. In the simple scenarios discussed here, the ratio between the gluino and squark soft masses $M_3/m_{\tilde{q}}\simeq \sqrt{N_{\text{mess}}} s_M\lesssim 1$, so that the most relevant collider bounds at current and future colliders can be framed purely in terms of the gluino mass, assuming the squarks to be decoupled and the lightest gaugino to be the next-to-lightest SUSY particle (NLSP). 

Writing the $R$-symmetry breaking VEV as in Eq.~\eqref{eq:defs}, the final gluino mass can be simply written in terms of underlying parameters as
\begin{equation}
\!\!\!\!\!m_{\tilde{g}}\simeq 7\text{ TeV}\left(\frac{g_M}{0.1}\right)\frac{\sqrt{F}}{10^{7}\text{ GeV}}\ ,  \quad g_M\defeq N_{\text{mess}} \sqrt{\epsilon_R} s_M\left[1 + \frac{\alpha_3}{4\pi}(9 + 6 \log\frac{Q}{M_3})\right]\,,\label{eq:gluinomass}
\end{equation}  
where we have collected various coefficients into a model-dependent prefactor $g_M$ which encodes i) the suppression of the gaugino masses due to $f_a\gg\sqrt{F}$ (i.e. $\sqrt{\epsilon_R}\ll1$), ii) the enhancement for $N_{\text{mess}}\gg1$, iii) the gaugino screening controlled by $s_M\lesssim 1$, and iv) the relation of the gluino soft mass to its pole mass, correctly accounting for the one loop running of the gluino soft mass at low energies in the limit of heavy squarks~\cite{Martin:1993yx,Martin:1997ns}. 

Broadly speaking, Eq.~\eqref{eq:gluinomass} establishes an interesting relation between the SUSY-breaking scale and the present and future collider bounds on the gluino. Given the current LHC bound on gluino masses, which ranges between $m_{\tilde{g}}\gtrsim 2-2.5\text{ TeV}$~\cite{Aaboud:2018doq,Aaboud:2018mna,ATLAS:2019vcq,ATLAS-CONF-2020-047}, Eq.~\eqref{eq:gluinomass} indicates the lowest values of the SUSY-breaking scale $\sqrt{F}$ consistent with data. 

Depending on the model, $g_M$ can span many orders of magnitude, but there are three parametric regimes of interest:
\begin{itemize} 
\item $1\ll g_{M}\lesssim 160$, which is realized in strongly-coupled messenger sectors that are at the boundary of perturbativity. The upper bound on $g_{M}$ is indeed obtained by requiring the SM gauge couplings and $y_{\text{mess}}$ to be perturbative at the scale of the hidden sector.
\item $g_{M}\simeq 1$, which is realized in weakly-coupled messenger sectors if $M_{\text{mess}}\simeq \sqrt{F}$ and the gaugino masses are not screened. The latter requirement requires non-trivial dynamics in the hidden sector, as shown in Ref.~\cite{Komargodski:2009jf}.
\item $g_{M}\ll 1$, which is typical of models where the soft masses are suppressed compared to the SUSY-breaking scale because $f_a\gg \sqrt{F}$ and the gaugino masses may be further screened compared to the squark masses. As we will show in Sec.~\ref{sec:pheno}, this is the typical situation in simple, explicit setups featuring SGWB signals.    
\end{itemize}

One of the most appealing features of LESB mediated via gauge interactions is the flavor-preserving nature of the MSSM superpartner spectrum. This is because the flavor-blind contributions to superpartner masses transmitted by gauge interactions vastly exceeds the omnipresent, flavor-violating ``gravity-mediated'' contributions. However, these gravity-mediated contributions reflect {\it all} contributions to SUSY breaking, while the gauge-mediated contributions reflect only the SUSY-breaking in the sector of interest. Thus when the gravitino mass is enhanced by $\kappa\ll1$, the contribution from gravity mediation increases relative to the contribution from gauge mediation, and may eventually run afoul of bounds on flavor violation. In particular, this implies that $\kappa$ is bounded from below by bounds from flavor-changing neutral currents (FCNCs). For instance, considering the slepton contributions to $\mu\to e\gamma$~\cite{Barbieri:1995tw,Hisano:1995nq} and the squark contributions to $\Delta m_K$~\cite{Gabbiani:1996hi,Ciuchini:1998ix} leads to the bounds
\begin{equation}
\kappa\vert_{\mu\to e\gamma}\gtrsim 10^{-9}\cdot\left(\frac{1}{\epsilon_R}\right)\quad , \quad \kappa\vert_{\Delta m_K}\gtrsim10^{-9}\cdot\left(\frac{\sqrt{F}}{10^7\text{ GeV}}\right)\cdot\left(\frac{1}{\epsilon_R}\right)^{3/4}\ ,\label{eq:FCNCvskappa}
\end{equation}
where $\kappa$ and $\epsilon_R$ are defined in Eq.~\eqref{eq:defs} and we set $\text{BR}(\mu\to e\gamma)<4.2\cdot 10^{-13}$ and $\Delta m_K=(3.479\pm0.001)\cdot 10^{-12}\text{ MeV}$, asking for the squark contribution to be less then present experimental uncertainty. These constraints give a robust upper bound on the gravitino mass in our framework. Finally, even in the absence of flavor violating effects, the electric dipole moments arising from the relative phase between gaugino and higgsino masses can be probed in precision experiments such as ACME~\cite{Baron:2013eja,Andreev:2018ayy}. The current limit are already challenging a CP-violating phase of order $10^{-2}$ with gauginos below the TeV scale and the future experimental program will sensibly improve this reach making it one of the most interesting indirect probes of LESB~\cite{Nakai:2016atk,Cesarotti:2018huy}.

\subsection{First order phase transitions and SGWB}\label{sec:FOPTandGWs}
Phase transitions in field theory are triggered by the nucleation of vacuum bubbles and their subsequent percolation in the space-time volume. The vacuum bubbles can be found in Euclidean signature as the stationary minimum-energy bounce solutions interpolating between the false and true vacuum~\cite{Coleman:1977py,Callan:1977pt}. In all cases we consider here, the thermal fluctuations will dominate so that the total decay rate per unit volume can be approximated as 
\begin{equation}\label{eq:decay}
\Gamma(T) \simeq T^{4}\left(\frac{S_{3}}{2 \pi T}\right)^{\frac{3}{2}} \exp \left(-S_{3} / T\right)\ ,
\end{equation}
where $S_3$ is the 3 dimensional Euclidean action for the $O(3)$-symmetric bounce~\cite{Linde:1980tt,Linde:1981zj}. The decay rate encodes the probability of true vacuum bubbles to be nucleated in a spacetime region where the false vacuum dominates. 

The time evolution of the phase transition can be described in terms of different temperatures. First of all, a necessary condition for nucleation is that the universe reaches temperatures below the critical temperature $T_c$, where the false and true vacuum are degenerate. At the nucleation temperature $T_n<T_c$, one bubble will nucleate per Hubble volume, corresponding to\footnote{The nucleation temperature is formally defined by the integral $1=\int_{T_{n}}^{T_{c}} \frac{d T}{T} \frac{\Gamma(T)}{H(T)^{4}}$,  which is well approximated by Eq.~\eqref{eq:Tn} since $\Gamma(T)$ depends exponentially on the temperature.}
\begin{equation}
\frac{\Gamma(T_n)}{H(T_n)^4}=1\quad \Rightarrow\quad \frac{S_3(T_n)}{T_n}\simeq-9.2\log \frac{g_*}{230}+4\log\frac{M_{\text{Pl}}}{T_n}+\frac{3}{2}\log \frac{S_3(T_n)}{T_n}\ ,\label{eq:Tn}
\end{equation}
where we assumed that the phase transition happens during radiation domination so that $H^2(T)=\frac{\pi^2 g_* T^4}{90 M_{\text{Pl}}^2}$ and normalized $g_*$ to the number of the degrees of freedom in the MSSM. In all the cases we  will consider, the last term in Eq.~\eqref{eq:Tn} can be neglected together with the constant term, so that solving 
\begin{equation}
\frac{S_3(T_n)}{T_n}\simeq \mathcal{C}(T_n)\qquad ,\qquad \mathcal{C}(T_n)\defeq104\log\left(\frac{10^7\text{ GeV}}{T_n}\right)\label{eq:approxnuc}
\end{equation}
 is always a good approximation. Since $\mathcal{C}(T)$ is a slowly-varying function of $T$, we can further simplify this equation assuming $\mathcal{C}(T_n)\simeq\mathcal{C}(T_c)$; this number is always going to be $\mathcal{O}(10^2)$ in the temperature range of interest. 

After one bubble per volume has been nucleated at $T_n$, the bubbles expand to fill the space-time volume. The phase transition is considered to be completed at the percolation temperature $T_p$, when a small fraction of the total volume remains in the false vacuum. For fast phase transitions like the ones discussed here, one can show that $T_p\simeq T_n$ so that we can neglect this difference and take $T_n$ as the temperature at which the phase transition completes. This sets the relevant dimensionful scale controlling the frequency range of the SGWB spectrum. 

The shape and amplitude of the SGWB spectrum strongly depends on the amount of energy released into GWs during the FOPT, the duration of the phase transition, and the behavior of the bubbles in the cosmic fluid. The two first ingredients can be easily quantified in terms of field theory data via the quantities
\begin{align}
&\alpha(T_n)= \frac{30}{\pi^2 g_*  T_n^4} \left ( \Delta V(T_n)-T_n \left.\frac{d \Delta V(T)}{dT}\right\vert_{T=T_n} \right)\ , \label{eq:alpha}\\
&\beta_H(T_n)\defeq\frac{\beta(T_n)}{H(T_n)}=T_n \frac{d}{dT} \left( \left.\frac{S_3}{T} \right) \right\vert_{T=T_n}\ ,\label{eq:beta}
\end{align}
where $\Delta V(T_n)$ is the potential energy difference between the true and the false vacuum at $T_n$. The amount of energy released into GWs is quantified by $\alpha$, the latent heat relative to the radiation energy density $\rho_R=\frac{\pi^2 g_* T^4}{30}$~\cite{Kamionkowski:1993fg}. The duration of the phase transition is quantified by $\beta_H$, the inverse of the typical timescale of the transition normalized with respect to Hubble; it is defined under the assumption that the nucleation rate rises exponentially~\cite{Linde:1980tt,Linde:1981zj} as $S(t)=e^{\beta_H H(t)(t-t_n)}$. Using the approximate nucleation condition in Eq.~\eqref{eq:approxnuc} we can write 
\begin{equation}
\beta_H(T_n)\simeq S'(T_n)-\mathcal{C} \ ,\label{eq:naturalbeta}
\end{equation}
 where $S'(T_n)\gtrsim \mathcal{C}$ in order for the nucleation rate to rise as a function of time, and $\beta_H\gtrsim \mathcal{C}\sim 100$ unless there is some measure of fine-tuning between the first and the second terms of the above expression. To evaluate the fine-tuning associated with $\beta_H$ in explicit models, we define
\begin{equation}
\Delta_{\beta_H}\defeq\text{Max}_{\{p_i\}}\Delta_{\beta_H}^{p_i}=\text{Max}_{\{p_i\}}\left\vert\frac{d \log \beta_H}{d \log p_i}\right\vert\ ,\label{eq:betaFT}
\end{equation}
where $\Delta_{\beta_H}^{p_i}$ are the individual tunings with respect to the underlying parameters of the theory $p_i$. As we will discuss in Sec.~\ref{sec:anatomy}, the parametric dependence of the fine-tuning can be derived for our general class of models and then computed explicitly in the models of Sec.~\ref{sec:models}. As a result, obtaining $\beta_H<10$ would imply a large amount of fine tuning (in the sense of being a non-generic prediction of a given model). This is illustrated in Fig.~\ref{fig:moneyplot}, where it provides a meaningful bound on the parameter space of GW signals in LESB . 

The dominant production mechanism of gravity waves during the first-order phase transition depends on the dynamics of the bubbles in the cosmic fluid. If the mean free path of the particles is much longer than the width of the bubble wall, the velocity of the wall $v_{\text{w}}$ can be determined by equilibrating the pressure on the bubble wall induced by the difference in potential energy $\Delta V$ with the friction forces exerted by the surrounding plasma~\cite{Arnold:1993wc,Mancha:2020fzw}. The latter are induced by states whose mass changes in passing from the false to the true vacuum. For  $v_{\text{w}}\to1$, the total pressure can be derived in a quasi-classical approximation~\cite{Bodeker:2009qy,Bodeker:2017cim,Mancha:2020fzw} and reads   
\begin{equation}
p= \Delta V- \Delta P_{\text{LO}}-\gamma \Delta P_{\text{NLO}}\ ,\quad \Delta P_{\text{LO}}=\frac{\Delta m^2T^{2}}{24}\ ,\quad \Delta  P_{\text{NLO}}\simeq\frac{1}{16 \pi ^2}
\gamma g^{2} \Delta m_V T^{3}\ ,\label{eq:pressure}
\end{equation}
where the Lorentz gamma factor is $\gamma=1/\sqrt{1-v_{\text{w}}^2}$ and the leading-order plasma friction $P_{\text{LO}}$ depends on the change in the masses-squared $\Delta m^2$ of all the states in the thermal bath~\cite{Bodeker:2009qy}. Since $\Delta m^2=m_{\text{true}}^2-m_{\text{false}}^2$, the approximate expression in Eq.~\eqref{eq:pressure} is only valid when both $\gamma T\gtrsim m_{\text{true}}$ and $ T\gtrsim m_{\text{false}}$. The first condition ensures that particles in the false vacuum have enough energy to pass through the wall, while the second forestalls Boltzmann suppression of the pressure~\cite{Mancha:2020fzw}. The next-to-leading order radiation pressure $P_{\text{NLO}}$ is instead induced by the change in mass of the vector bosons and it is $\gamma$-enhanced for $v_{\text{w}}\to1$, as first derived in~\cite{Bodeker:2017cim}. 

The pressure in Eq.~\eqref{eq:pressure} determines both how much the bubble wall accelerates as a function of the bubble radius~\cite{Darme:2017wvu,Ellis:2019oqb}, and the fraction of the FOPT energy which is in the bubble wall at the time of collision $T_{*}$ (traditionally called $k_{\text{coll}}$). Since we will be dealing with fast phase transitions, we take  $T_{*}\simeq T_p\simeq T_n$. 

In the absence of friction, the acceleration of the bubble wall grows linearly with the bubble radius until the gamma factor reaches a terminal value  
\begin{equation}
\gamma_{*}\simeq\frac{2}{3}\frac{R_{*}}{R_0}\simeq2.6\cdot10^8\sqrt{\frac{230}{g_*(T_n)}}\left(\frac{100}{\beta_H}\right)\left(\frac{10^7\text{ GeV}}{T_n}\right)\left(\frac{\Delta V(T_n)}{T_n^4}\right)^{1/3}\ ,\label{eq:gammastar}
\end{equation}
where we took the initial radius to be $R_0\simeq R_c=\left(\frac{3}{2\pi}\frac{S_3(T_n)}{\Delta V(T_n)}\right)^{1/3}$ estimated in the thin wall approximation~\cite{Coleman:1977py}, estimated $R_*$ as in Ref.~\cite{Enqvist:1991xw}, and assumed radiation domination. The last term is $\mathcal{O}(1)$ in phase transitions which do not have a supercooling phase since $\Delta V(T_n)/T_n^4< 75.6 (g_*/230)$. 

In the FOPTs discussed here, the bubble growth is generically stopped by plasma effects from heavy states. This is due to a novel effect in which the relevant energy scale for particles interacting with the wall reaches values $\sim \gamma T_n$ much larger than the intrinsic scales associated with the bubble. This is particularly relevant for SUSY-breaking hidden sectors, where there is a large separation of scales that can be spanned by these ultra-relativistic effects.  In particular, the bubbles expand and accelerate linearly with the radius until the boost factor is large enough to allow heavy states of mass $m_{\text{true}}\gg T_n$ to cross the bubble wall. The significant mass change of these states induces a new  source of LO friction,
\begin{equation}
\Delta P_{\text{LO}}^{\text{heavy}}\simeq\frac{1}{24}(m_{\text{true}}^2-m_{\text{false}}^2)^2 T^2_n e^{-m_{\text{false}}/T_n}\ .\label{eq:heavypressure}
\end{equation}
If $\Delta V-\Delta P_{\text{NLO}}^{\text{heavy}}\leq 0$, the gamma factor of the bubble wall and the bubble radius at equilibrium are approximately
\begin{equation}
\gamma_{\text{eq}}^{\text{heavy}}\simeq \frac{m_{\text{true}}}{T_n}\quad ,\quad R_{\text{eq}}^{\text{heavy}}\simeq\frac{3}{2}\gamma_{\text{eq}}^{\text{heavy}}R_c\ .
\end{equation}
This effect is very similar to the pressure term from mixing discussed in Ref.~\cite{Vanvlasselaer:2020niz}, but here we typically pay the Boltzmann suppression of $m_{\text{false}}$. 

The resulting fraction of the energy in the bubble wall at the time of collisions is generically very suppressed, 
\begin{equation}
k_{\text{coll}}\simeq\frac{R_{\text{eq}}}{R_*}\left(1-\frac{\Delta P_{\text{LO}}}{\Delta V}\right)\simeq 4\cdot10^{-9}\gamma_{\text{eq}}^{\text{heavy}}\left(\frac{2.6\times10^8}{\gamma_{*}}\right)\left(1-\frac{\Delta P_{\text{LO}}}{\Delta V}\right)\ .
\end{equation}
As such, most of the energy released in the FOPT goes into the plasma, giving rise to sound waves propagating through the cosmic fluid. These sound waves source gravitational waves from the motion of the plasma with an efficiency determined by
\begin{equation}\label{eq:ksw}
k_\mathrm{sw}\simeq\frac{\alpha}{0.73+0.083 \sqrt{\alpha}+\alpha}\ ,
\end{equation}
where we have expanded the general formula of Ref.~\cite{Hindmarsh:2013xza,Hindmarsh:2015qta} for $k_{\text{coll}}\ll1$. The resulting GW spectral density is 
\begin{equation}
\Omega_{\text{sw}}^{*}=
3.8\left(1\over\beta_H^2\right)\left(\frac{\kappa_{\mathrm{sw}} \alpha}{1+\alpha}\right)^{3/2}
\left(\frac{f}{f_{\mathrm{sw}}^*}\right)^{3}\left[1+\frac{3}{4}\left(\frac{f}{f_{\mathrm{sw}}^*}\right)^{2}\right]^{-\frac{7}{2}}\quad ,\quad  
f_\text{sw}^*=1.2 \beta_H^* H_*\ ,\label{eq:swsignal}
\end{equation}
where $H_*^2=\frac{\pi^2 g_*(T_*)}{90}\frac{T_*^4}{\mpl^2}(1+\alpha)$ to account for the reheating of the plasma and $\beta_H^*$ is normalized accordingly following Eq.~\eqref{eq:beta}. The sound wave spectrum is a broken power law which drops like $\Omega_{\text{sw}}^{*}\sim f^3$ for $f\ll f_{\mathrm{sw}}^*$, as expected from causality in a radiation dominated universe, and as $\Omega_{\text{sw}}^{*}\sim f^{-4}$ for $f\gg f_{\mathrm{sw}}^*$. The high frequency behavior of the spectrum is likely to be affected by the turbulence contribution, whose size is still subject to large theoretical uncertainties~\cite{Caprini:2015zlo,Caprini:2019egz}. Here, we include for simplicity only the sound waves contribution to the GW spectrum in Eq.~\eqref{eq:swsignal}, which will mainly determine the detectability of a given GW signal. 
After redshift is taken into account, assuming that the entropy per comoving volume remains constant~\cite{Kamionkowski:1993fg}, the GW spectrum today reads 
\begin{equation}
\Omega_{\text{sw}}^0 h^2= 
\left( \frac{a_*}{a_0}\right)^4 \left(\frac{H_*}{H_0} \right)^2\Omega_{\text{sw}}^*
=2.8 \cdot 10^{-5} \left( \frac{230}{g_*}\right)^{1/3}\Omega_{\text{sw}}^*\ ,\label{eq:GWspectrumtoday}
\end{equation}
where the peak frequency and the power at the peak frequency scale as  
\begin{align}
&f_\mathrm{sw}^0
=f_\mathrm{sw}^* 
\left( \frac{a_*}{a_0}\right)
=1.1 \times 10^{2} \text{ Hz} \left( \frac{g_*}{230}\right)^{1/6}
\left(\frac{\beta_H}{50}\right)
\left( \frac{T_n}{10^7 \mathrm{GeV}}\right)\left(\frac{1.3}{1+\alpha}\right)^{1/4}\ ,\label{eq:ftoday}\\
&\Omega_{\text{GW}}^{\mathrm{sw},0} h^2\simeq
10^{-10}
\left( \frac{230}{g_*}\right)^{1/3}\left(50\over\beta_H\right)^2
\left(\frac{\kappa_{\mathrm{sw}} \alpha}{0.08}\right)^{3/2}\left(\frac{1.3}{1+\alpha}\right)^{3/2}\ .\label{eq:poweratpeak}
\end{align}
Here we have taken $T_*\simeq T_n$ and normalized the scalings for $\alpha=0.3$, $\beta_H=100$ and $T_n=10^7\text{ GeV}$, which will be the typical values for  FOPTs related to fully calculable SUSY-breaking hidden sectors explored in the following sections.
\begin{figure}[t!]
  \centering
   \includegraphics[width=0.48\textwidth]{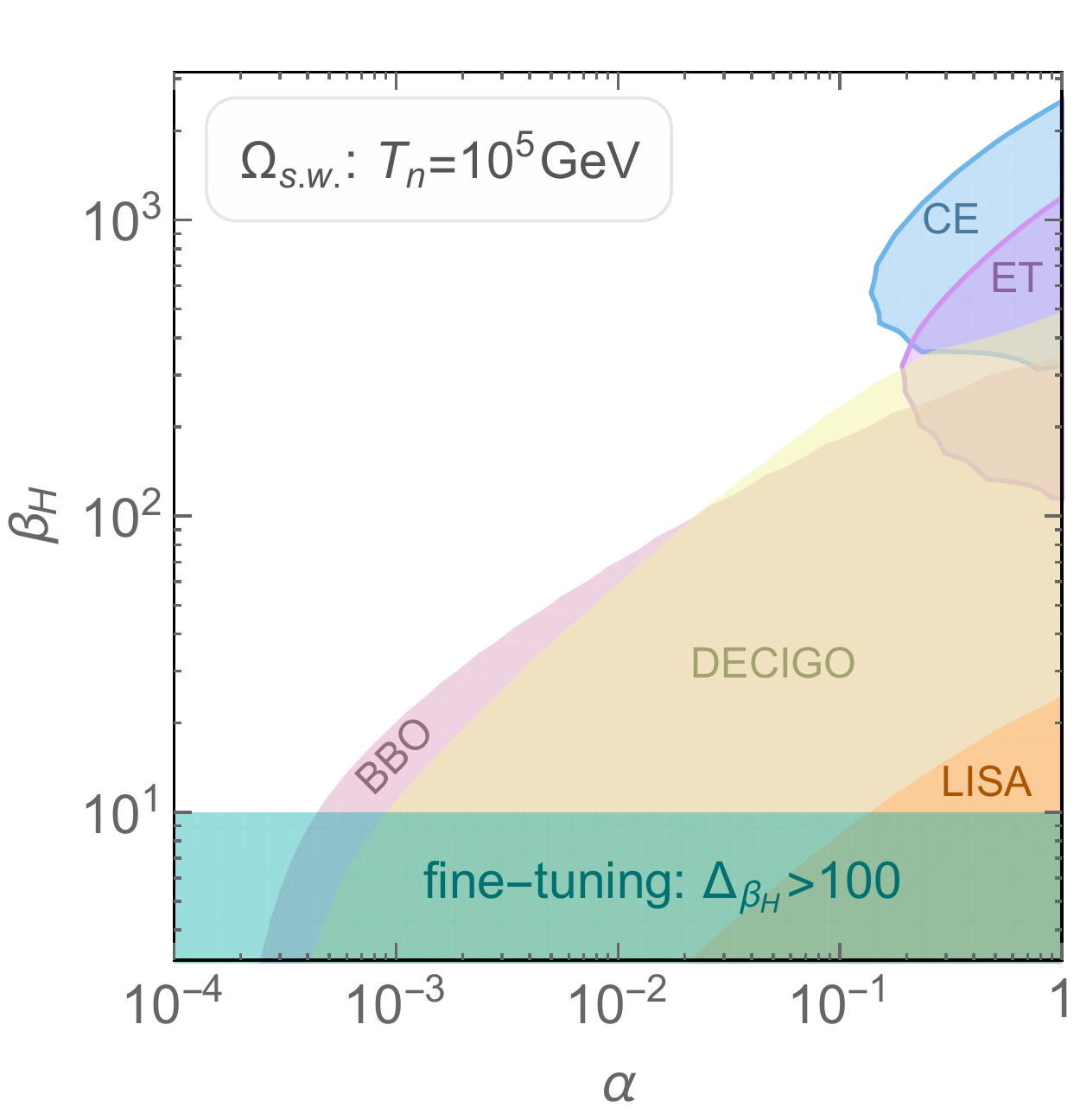}\hfill
 \includegraphics[width=0.48\textwidth]{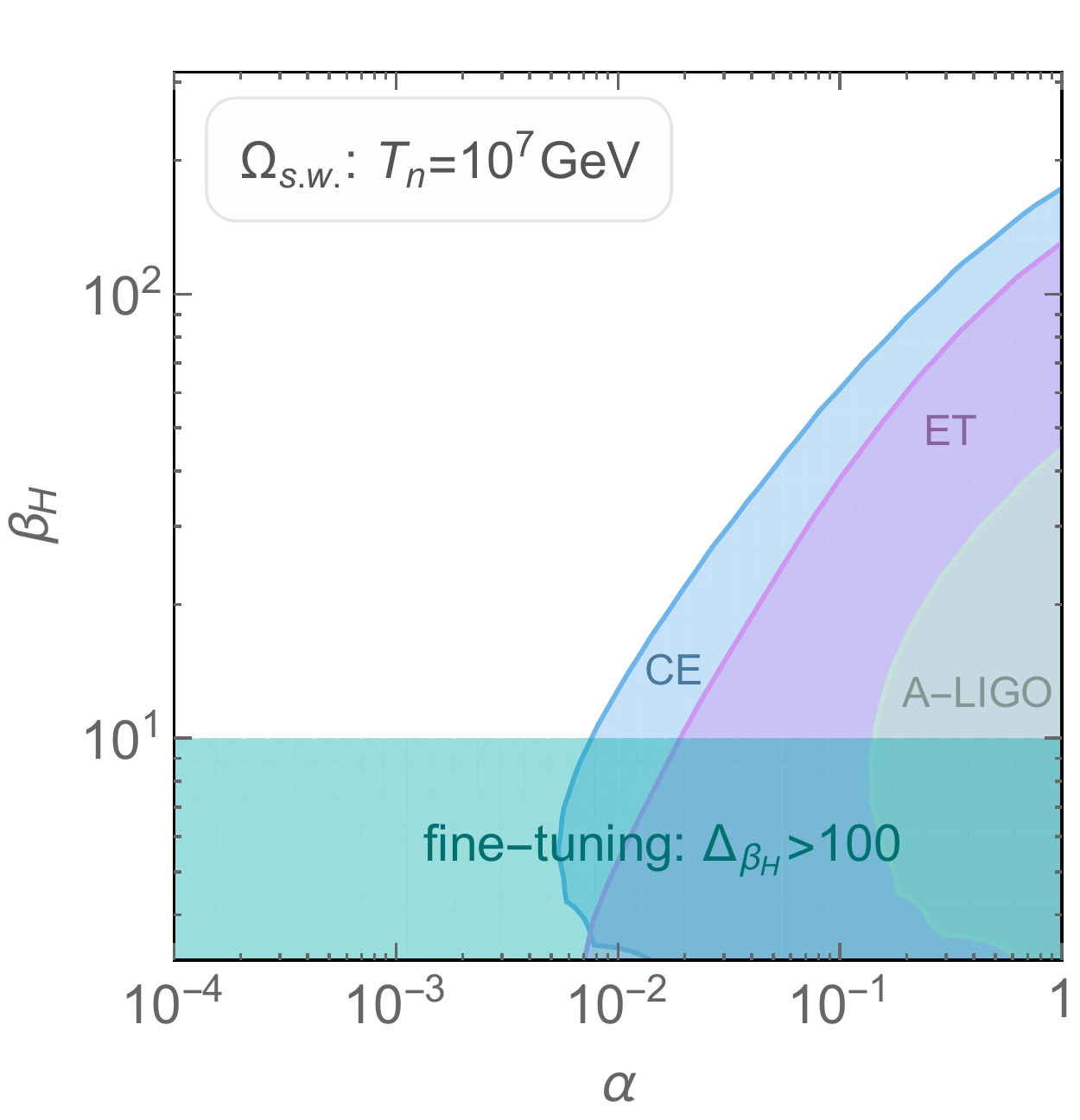}
\caption{The reach of future GW interferometers in the $(\alpha,\beta_H)$ plane for two different scales of FOPTs, assuming the signal is dominated by sound waves given by Eq.~\ref{eq:swsignal}. The shaded regions are obtain by requiring the signal at the peak in Eq.~\eqref{eq:poweratpeak} to be inside the PLI curve of a given experiment. {\bf Left:} $T_n=10^5\text{ GeV}$ corresponds to LESB scenarios with an ultralight gravitino LSP and $\kappa=1$.  {\bf Right:} $T_n=10^7\text{ GeV}$ corresponds to LESB scenarios with gravitino DM and $\kappa\ll1$. We will exhibit calculable scenarios of this type in Sec.~\ref{sec:models}.}\label{fig:alphabeta}
\end{figure}

Having derived the expected GW spectrum, we can determine the region in the $(\alpha,\beta_H)$ plane where we expect the SBGW to be detectable at future interferometers. Given the fraction of energy density in GWs today in Eq.~\eqref{eq:GWspectrumtoday}, the sensitivity of a given interferometer is controlled by the time integrated signal-to-noise ratio
\begin{equation}
\rho^2= t_{\text{obs}}\int_{f_{\text{min}}}^{f_{\text{max}}}\left[\frac{\Omega_{\text{GW}}(f,\alpha,\beta,v_{\text{w}})}{\Omega_{\text{noise}}(f)}\right]^2\ ,\label{eq:sens}
\end{equation}
where  $\Omega_{\text{noise}}(f)$ is the effective noise of the interferometer within a given frequency band $(f_{\text{min}},f_{\text{max}})$ and $t_{\text{obs}}$ is the observation time. A detectable stochastic GW background is defined to have $\rho>10$.  The Power Law Integrated (PLI) curves are generated by considering a power law function of the frequency $f$ for the GW signal shape in Eq.~(\ref{eq:sens}). The PLI curves for each GW interferometer considered here are given in Appendix~\ref{sub:signal_detection} for completeness. 

In Fig.~\ref{fig:alphabeta} we show the regions in the $(\alpha, \beta_H)$ plane where the power at the peak frequency in Eq.~\eqref{eq:poweratpeak} lies within the reach of future interferometers for two different nucleation temperatures.  Low nucleation temperatures such as $T_n=10^5\text{ GeV}$ can be probed over a wide frequency range depending on $\beta_H$ (i.e. the duration of the FOPT) while high nucleation temperatures such as $T_n=10^7\text{ GeV}$ will be accessible only at future high frequency interferometers such as Advanced LIGO (A-LIGO)~\cite{TheLIGOScientific:2014jea}, the Einstein Telescope (ET)~\cite{Sathyaprakash:2012jk} and the Cosmic Explorer (CE)~\cite{Evans:2016mbw,Reitze:2019iox}. In the next section we show that $T_n\sim \sqrt{F}$ in our LESB scenarios, so that high nucleation temperatures in fully calculable SUSY-breaking scenarios correspond to superpartners lying out of the reach of the LHC.

\subsection{LESB in the future: GW interferometers vs.~colliders}\label{sec:summaryparspace}

We are now ready to establish a connection between the SGWB signals and SUSY-breaking phenomenology described in the previous two sections. The first step is to relate the nucleation temperature relevant for the SGWB signal to the scales in a SUSY-breaking hidden sector. As we will see, the nucleation temperature is essentially set by the SUSY-breaking scale $\sqrt{F}$ in our scenarios.

Focusing on FOPT where the barrier between the false and the true vacuum is present at $T=0$, $S_3$ is bounded from below by a constant and we can define $T_{\text{min}}$ as the temperature where 
\begin{equation}
\beta_H\vert_{T=T_{\text{min}}}=0\qquad \Rightarrow\qquad T_{\text{min}}<T_{n}<T_{\text{c}}\ .\label{eq:Tmin}
\end{equation}
If $S_3(T)$ is monotonic for $T>T_{\text{min}}$, the solution of the equation above is unique. The nucleation temperature is then bounded from above by $T_c$, where $\beta_H\to \infty$ and $\alpha$ in Eq.~\eqref{eq:alpha} is suppressed and dominated by $\frac{d \Delta V(T_n)}{dT}$. It is further bounded from below by $T_{\text{min}}$ where $\beta_H\to 0$ and $\alpha$ is dominated by $\Delta V(T_n)$. 

More importantly, $T_n$ can be directly related to the SUSY-breaking scale $\sqrt{F}$ which sets the size of the  $O(3)$-symmetric bounce action. A simple way of seeing this is to note that the bounce action at $T_n$ is itself set by the scale of relevant features in the potential,
\begin{equation}
S_3(T_n)\simeq c_3\sqrt{F}\quad \Rightarrow\quad T_n= \frac{c_3}{\mathcal{C}} \sqrt{F} \ ,\label{eq:general}
\end{equation}
where $c_3$ is a model-dependent function of the parameters controlling the shape of the potential which we assume to be temperature independent for simplicity (an approximation that is certainly justified if $T_n$ is close enough to $T_{\text{min}}$). Here we assume that $c_3/\mathcal{C}\sim \mathcal{O}(1)$, an assumption that will turn out to be justified analytically in Sec.~\ref{sec:anatomy} and numerically in the explicit models of Sec.~\ref{sec:models}.

Having established a relation between $T_n$ and the SUSY-breaking scale $\sqrt{F}$, we identify two different viable regions of the LESB parameter space satisfying the following simple requirements: 
\begin{itemize}
\item the gravitino is the lightest supersymmetric particle (LSP) as required by LESB, and
\item the reheating temperature $T_{\text{r.h.}}$ is as high as $\sqrt{F}$ to generate GW signals from the hidden sector. We take $T_{\text{r.h.}}=\sqrt{F}$ in Fig.~\ref{eq:FCNCvskappa} to maximize the allowed parameter space.
\end{itemize}

The two viable regions satisfying the above requirements are 

\paragraph{Gravitino Dark Matter window:} where $260\text{ TeV}<\sqrt{F}\lesssim 50 \text{ PeV}$ and $\kappa\ll1$ so that the gravitino mass is larger than the nominal value set by the $F$-term in Eq~\eqref{eq:spurion}. The upper bound on the SUSY-breaking scale is obtained by combining the constraints on gravitino overabundance in Eq.~\eqref{eq:m32abundance}, BBN constraints on NLSP decays, and the LHC bound on the gluino mass $m_{\tilde{g}}>2\text{ TeV}$. The precise upper bound is potentially dependent on further model-building epicycles; the value here is meant to be indicative. In this window, the gravitino abundance can match the observed dark matter relic abundance today, while the soft masses are still dominated by the gauge mediation contributions in Eq.~\eqref{eq:softmasses} so that flavor constraints are under control when Eq.~\eqref{eq:FCNCvskappa} is satisfied. 
Perturbative gauge mediation models with $f_a\gtrsim \sqrt{F}$ and gaugino screening will naturally live in the upper end of this window for $\sqrt{F}\simeq 1-50\text{ PeV}$. As shown in Fig.~\ref{fig:moneyplot}, the future reach on gluinos at FCC-hh~\cite{Arkani-Hamed:2015vfh} could provide a further direct test of these models. Future interferometers in the LIGO frequency band such as A-LIGO~\cite{TheLIGOScientific:2014jea}, ET~\cite{Sathyaprakash:2012jk} and CE~\cite{Evans:2016mbw,Reitze:2019iox} have the unique opportunity to probe these scenarios as long as the thermal transition to the SUSY-breaking vacuum is associated with a sufficiently strong FOPT~(see Fig.~\ref{fig:alphabeta}). In the rest of the paper, we discuss explicit scenarios of this type.

\paragraph{Ultralight gravitino window:} where $m_{3/2}<16\text{ eV}$ and $\sqrt{F}<260\text{ TeV}$. This region has no cosmological issues for $\kappa=1$, but it requires $g_M\gtrsim1$ to satisfy the LHC bound on the gluino mass given the low SUSY-breaking scale (see Eq.~\eqref{eq:gluinomass} for definition and comments). A lower bound on the gravitino mass can be derived from direct searches for gravitino pair production at LEP in $\gamma+\text{MET}$ and at the LHC in $j+\text{MET}$. As shown in Fig.~\ref{fig:moneyplot}, present direct bounds on the gravitino are not competitive with the bound on $\sqrt{F}$ obtained by requiring $m_{\tilde{g}}>2\text{ TeV}$ and perturbativity in the messenger sector. The HL-LHC will not improve much on that. Future colliders -- in particular, high energy lepton colliders (HELCs) -- can drastically improve the reach on gravitino pair production and meaningfully probe this window even if MSSM superpartners remain inaccessible. As shown in Fig.~\ref{fig:alphabeta}, these scenarios can be probed across a wide frequency range by future GW interferometers depending on the strength and the duration of the FOPT. Building explicit calculable models in this window presents challenges~\cite{Hook:2015tra,Hook:2018sai}, and we leave a study of possible GW signals for a future work.

\section{Anatomy of the SUSY-breaking phase transition}\label{sec:anatomy}

In this section we describe the generic features of FOPT occurring in calculable SUSY-breaking hidden sectors. First, we discuss how a large class of perturbative hidden sectors can be encoded in the effective field theory of the \emph{universal pseudomodulus}, which is the scalar component $x$ of the chiral superfield $X$ in Eq.~\eqref{eq:spurion}, universally related to the spontaneous breaking of supersymmetry~\cite{Nelson:1993nf,Intriligator:2007py,Komargodski:2009jf}. 

Second, we show how the flatness of the pseudomodulus potential gives rise to a new class of FOPTs with a very distinctive feature: the nucleation temperature is generically small  compared to the SUSY mass scale, $T_n\leq m_{*}$, so that the thermal potential is well approximated in the low-$T$ expansion. As we will discuss, non-supersymmetric realizations of this class of FOPT typically entail a large amount of fine-tuning. 

Finally, we derive parametric estimates for $T_n$, $\alpha$ and $\beta_H$ for this new class of FOPTs using the triangular barrier approximation~\cite{Duncan:1992ai,Amariti:2009kb} and comment on a universal feature of bubble dynamics in our FOPTs.  The observations of this section will find a concrete realization in the working examples of Section \ref{sec:models}.

\subsection{The SUSY-breaking pseudomodulus}\label{sec:pseudopot}
The existence of flat directions is a trademark of hidden sectors with spontaneous SUSY breaking. Here we focus on a large class of SUSY breaking sectors where the dynamics of both SUSY and $R$-symmetry breaking can be embedded in
a single chiral superfield $X$ parametrized as in Eq.~\eqref{eq:spurion}
\be
 X=\frac{x}{\sqrt{2}} e^{2i a/f_a}+\sqrt{2}\theta\tilde{G}+\theta^2 F\ ,\label{eq:pseudo}
\ee
where the R-charges of the components are respectively $R[x]=2, R[\tilde G]=1, R[F]=0$. The scalar component $x$ (the \emph{universal pseudomodulus}) tracks the breaking of the $R$-symmetry, while $\langle F \rangle$ sets the SUSY breaking scale.\footnote{In more general scenarios there could be multiple different field directions associated to SUSY-breaking and $R$-symmetry breaking~\cite{Komargodski:2009jf} or even multiple pseudo-flat directions from multiple sources of $F$-term SUSY breaking \cite{Curtin:2012yu}.} The phase transition occurs along $x$ from a local minimum at the origin $x=0$ (where $R$-symmetry is preserved) to the $T=0$ vacuum of the theory where $\langle x \rangle = f_a$ and $R$-symmetry is broken. Hence $\langle x \rangle= f_a$ is the order parameter of the phase transitions of interest here, parameterizing the spontaneous breaking of the $R$-symmetry.

\begin{figure}[t!]
  \centering
 \includegraphics[width=0.49\textwidth]{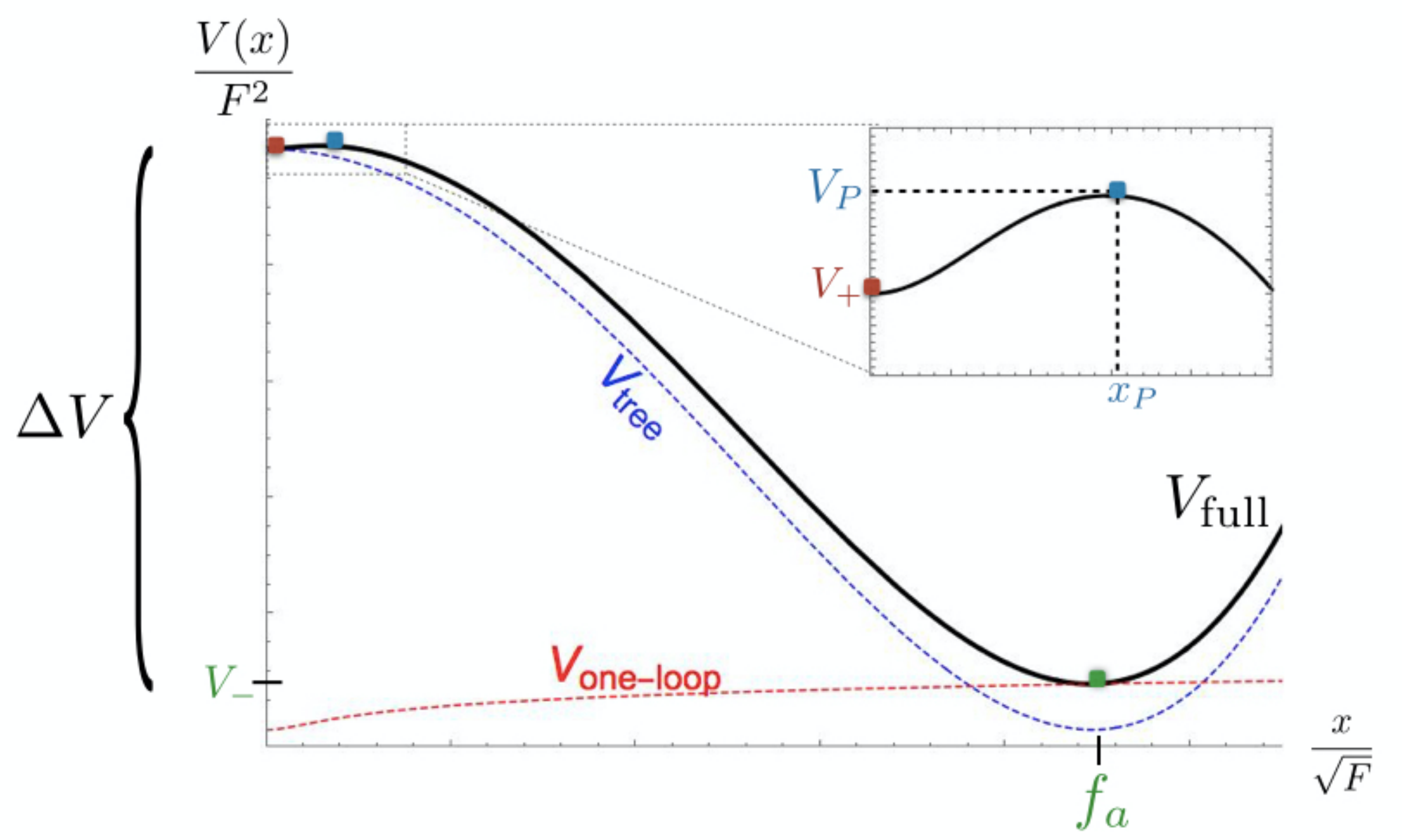}\hfill
  \includegraphics[width=0.49\textwidth]{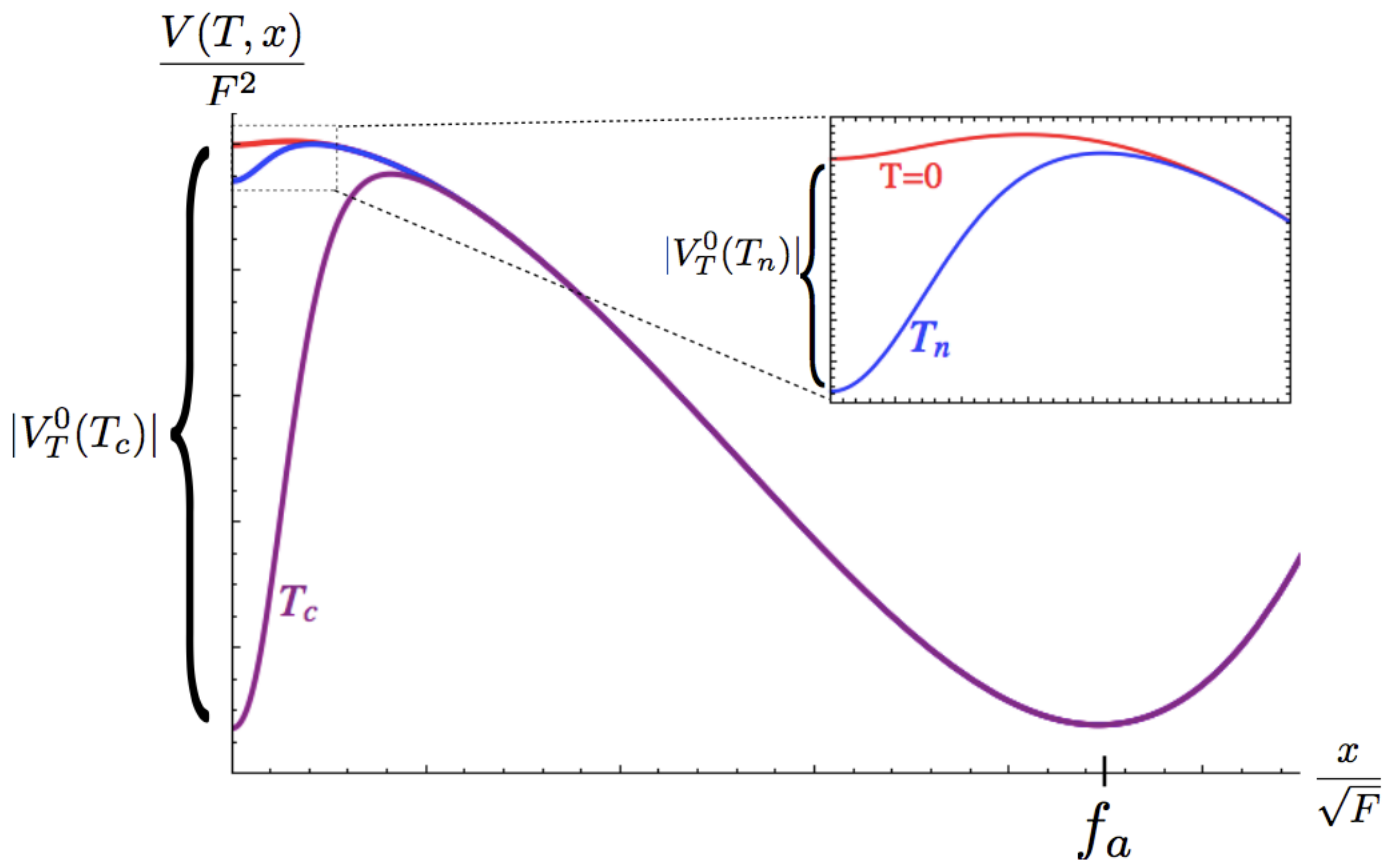}
\caption{Qualitative features of the pseudomodulus potential relevant to the FOPT in SUSY-breaking hidden sectors.  {\bf Left:} Sketch of the zero-temperature potential as described in Sec.~\ref{sec:pseudopot}, exhibiting the following features: i) the distance between the two minima is larger than their potential difference, $f_a^4\gtrsim\Delta V$, and ii) the height of the peak between the two minima is loop-suppressed compared to the potential difference, $V_P\ll \Delta V$. An explicit realization of this potential is presented in Sec.~\ref{sec:toy}. The tree level potential ({\bf dashed blue}) generated by explicit $R$-symmetry breaking destabilizes the origin, giving rise to a minimum at $\langle x\rangle=f_a$ where the $R$-symmetry is further spontaneously broken by the VEV of $x$. Quantum corrections ({\bf dashed red}) generate a local minimum at origin.  {\bf Right:} Behavior of the temperature corrections described in Eq.~\eqref{VT_toy} at $T=0$, $T=T_c$, and $T= T_n$. The thermal corrections give a contribution to the potential at the origin which at $T_n$ is typically much smaller than $F^2$. The barrier and the true vacuum are essentially unchanged. The approximations in Sec.~\ref{sec:approxFOPT} are then justified.
\label{fig:albe_cart}
}
\end{figure}

In hidden sectors which admit a weakly coupled description, the phase transition can be fully described by studying the effective potential of the pseudomodulus $x$, whose mass is typically well below the mass  $m_*$ of the heavy SUSY states in the hidden sector. 
As we will see, the unique features of the pseudomodulus potential leave a strong imprint on the properties of the phase transition. The full effective potential for the pseudomodulus can be written as 
\begin{equation}
V_{\text{eff}}(x)=V_0(x)+V_T(x)\ ,
\end{equation}
where $V_0(x)$ encodes the zero-temperature quantum corrections and $V_T(x)$ the thermal ones. 

The zero-temperature part of the effective potential $V_0(x)$ is flat at tree level, up to explicit $R$-symmetry breaking effects. Along this so-called $F$-flat direction, the size of the potential energy is set by supersymmetry breaking, $V \sim F^2$. Interactions that explicitly violate the $R$-symmetry typically destabilize the origin and give a slope to the pseudomodulus potential at tree level, but these features are usually small compared to the scale $\sqrt{F}$. At one loop, quantum corrections lift the pseudomodulus potential; these corrections are present even in the absence of explicit $R$-symmetry breaking. The combination of tree-level explicit $R$-symmetry breaking and one-loop quantum corrections give rise to the schematic zero-temperature potential shown in Fig.~\ref{fig:albe_cart}. Assuming the quantum corrections exceed the $R$-symmetry breaking effects, at zero temperature this creates a metastable vacuum at the origin that is separated by a barrier from the true vacuum at $\langle x \rangle_{\text{true}}=f_a$. The energy difference between the two vacua $\Delta V$ is proportional to the SUSY-breaking scale. The barrier is located at a distance $x_P$ from the origin; at this point, the barrier height is $V_P$. The essential features characterizing the zero temperature potential are:
\begin{itemize}
\item \emph{The potential is flat.} This means that the distance $f_a$ in field space between the false vacuum and the true vacuum is larger than the size of the potential energy difference $\Delta V$: 
\begin{equation}
f_a^4 > \Delta V\ ,\label{eq:flatness}
\end{equation}
where this hierarchy assumes that $R$-symmetry breaking effects are parametrically smaller than the loop corrections. This will be manifest in the toy model of Sec.~\ref{sec:toy}. Under this assumption, the flatness of the potential is a direct consequence of the fact that SUSY loop corrections asymptote to a logarithm at large field values (see for instance~\cite{Witten:1981kv}). Obtaining similar quantum corrections in non-supersymmetric theories with a field-independent mass gap is notoriously difficult without fine-tuning.\footnote{A well studied example of a flat potential is that of the dilaton of spontaneously broken conformal symmetry. Here, however, the mass gap is field-dependent and as a consequence the theory becomes strongly coupled at the origin~\cite{Rattazzi:2000hs}. The features of the dilaton phase transition are consequently very different from the one described here.}   
\item \emph{The barrier between the two vacua is small.}  Given that the potential is generated by loop effects (and subleading $R$-symmetry breaking effects), the size of the barrier $V_P$ is one-loop suppressed with respect to the energy difference between the true and the false vacuum $\Delta V$: 
\begin{equation}
\frac{V_P}{\Delta V}= \frac{\lambda_{\text{eff}}^2}{16\pi^2}\ ,\label{eq:smallbar}
\end{equation}
where $\lambda_{\text{eff}}\sim\mathcal{O}(1)$ should be thought of as the effective coupling determining the height of the barrier. The position of the barrier $x_P$ is model-dependent, but will not play a critical role in the determination of the bounce action as long as Eq.~\eqref{eq:flatness} is satisfied. 
\end{itemize}

We now turn to the finite temperature corrections. First, as it is well known, finite temperature effects break SUSY and thus significantly
modify the pseudo-modulus potential. The thermal effects are dominated by the loops of heavy fields in the SUSY hidden sector coupled to $x$, 
whose mass is of order $m_{*}$. Since $m_{*}$ is by construction larger than the SUSY-breaking scale $\sqrt{F}$ setting the zero-temperature potential, the relevant temperatures for the phase transition are smaller than the mass scale $m_{*}$ of the particles running in thermal loops. This implies that the correct approximation of the thermal potential is the \emph{low temperature expansion} (see Appendix~\ref{app:potentials} for explicit formulas).
This makes the finite-temperature potential of the pseudomodulus qualitatively different from ordinary non-SUSY models, where the high temperature approximation applies since the typical scalar potential curvature is of the same order of the highest mass scale in the theory.

The effect of thermal corrections on the pseudomodulus potential takes the schematic form
\be
\label{VT_toy}
V_T(x) \simeq -N \, T^4  \left( \frac{\lambda^2 x^2 +m_{*}^2}{(2 \pi T)^2} \right)^{3/4} e^{-\sqrt{\frac{\lambda^2 x^2 +m_{*}^2}{T^2}}}\ ,
\ee
where we assumed the presence of $N$ degrees of freedom with masses-squared $\sim \lambda^2 x^2 +m_{*}^2$. Notice that $N$ counts all the heavy degrees of freedom, both bosonic and fermionic, which contribute with the same sign to the thermal potential. This enhances the importance of thermal effects compared to zero-temperature loops, where cancellations occur between states of different statistics. 

The thermal correction constitutes a negative contribution to the potential which is maximal (in absolute value) at the origin of the pseudomodulus, when 
$x \sim 0$ and the Boltzmann suppression factor is minimized. As a consequence, thermal corrections in our scenarios have an exponentially larger impact at the origin relative to the true vacuum or the barrier. This behavior is explicitly shown in the right panel of Fig.~\ref{fig:albe_cart}. As we will show in Sec.~\ref{sec:models}  small deviations from this generic feature can be induced by heavy states becoming lighter at large field values of the pseudomodulus.

\subsection{First order phase transitions in the low-$T$ expansion}\label{sec:approxFOPT}

Given the shape of our potential as shown in Fig.~\ref{fig:albe_cart}, we can approximate the bounce action in the triangular barrier approximation \cite{Duncan:1992ai,Amariti:2009kb}. Within this approximation, we will be able to capture the parametric behavior of the FOPTs analytically and in Sec.~\ref{sec:models} we will show how the full analytical solution in explicit models reflects the general features explored here. A more in-depth discussion about the computation of the bounce action in the various cases can be found in Appendix~\ref{app:bounce}.  

Taking the false vacuum to be at the origin of field space, we can write the bounce action in the triangular barrier approximaton as
\be
\label{S3overTfull}
\frac{S_3}{T} = 
\frac{144 \sqrt{2}\pi}{5}   \frac{(V_P-V_+)^{5/2}}{(V_P-V_-)^3}\frac{f_a^3}{T} f(r_\lambda)\quad\ ,\quad \text{for}\quad \frac{f_a}{x_P}>g(r_\lambda)\ ,
\ee
where we have defined the variables 
\be
r_\lambda \defeq\frac{\lambda_-}{\lambda_+}\quad ,\quad \lambda_- \defeq \frac{V_{P}-V_-}{f_{a}-x_{P}}\quad ,\quad  
 \lambda_+\defeq \frac{V_{P}-V_{+}}{x_{P}}\ ,
\ee
and the functions
\begin{align}
&f(r_\lambda)=\frac{r_\lambda^3(1+r_\lambda)}{3\sqrt{3}(3+2r_\lambda-3(1+r_\lambda)^2/3)^{3/2}}\underset{r_\lambda\to0}{\simeq}1+\frac{5}{3}r_\lambda\ ,\\
&g(r_\lambda)= 1+\frac{r_\lambda}{3+2r_\lambda-3(1+r_\lambda)^{2/3}}\underset{r_\lambda\to0}{\simeq}\frac{3}{r_\lambda}+\frac{7}{3}\ .
\end{align}
The expansion for $r_{\lambda}\to0$ is justified as long as both Eq.~\eqref{eq:flatness} and Eq.~\eqref{eq:smallbar} are satisfied and the true vacuum VEV $f_a$ sets the largest scale in the pseudomodulus potential. The triangular barrier approximation can be extended beyond the region set by $f_a/x_P>g(r_\lambda)$, but the range of validity of Eq.~\eqref{S3overTfull} is sufficient to capture the parametrics of the phase transitions of interest. We give the full expression of the triangular barrier approximation in Appendix~\ref{app:bounce}. 

The triangular barrier approximation depends in general on only five parameters characterizing the potential : the three values of the potential at the critical points, $V_{\pm}, V_P$, and the position of the two critical points, $f_a, x_P$. For a given theory we can compute these temperature-dependent quantities explicitly, and find that the bounce action in Eq.~\eqref{S3overTfull} is an excellent match to the full numerical result.\footnote{Throughout this paper we make use of the Mathematica package FindBounce~\cite{Guada:2020xnz} for our numerical analysis, which we further validate using CosmoTransitions~\cite{Wainwright:2011kj}.} 

At the leading order in the $r_{\lambda}\to0$ expansion, the bounce action is independent of $x_P$; our analytical estimates will assume that this holds. To further simplify our analytical treatment, we approximate the thermal potential in Eq.~\eqref{VT_toy} by only including thermal corrections at $x=0$, where the exponential suppression is minimized, and neglecting the temperature dependence of $V_{-}$ and $V_P$. Within this approximation we obtain 
\begin{align}
&V_+= V_T^0\quad ,\quad V_T^0\defeq V_T(x=0)\quad ,\quad V_{-}= -\Delta V\  , 
\end{align}
where we set $V_{+}$ to be exactly zero at zero temperature so that its (strictly negative) value is purely controlled by the thermal corrections at the origin. The value of the potential at the true vacuum is $-\Delta V$, and independent of temperature in this approximation. 

With these approximations, the bounce action becomes simply 
\be
\label{S3oTsimp}
\frac{S_3}{T} \simeq \frac{144 \sqrt{2} \pi }{5T} \frac{ (V_{P}-V_T^0)^{5/2} f_a^3}{ (\Delta V )^3}\quad \ ,\quad \frac{3V_T^0}{\Delta V}+1>0, 
\ee
and we are now ready to describe the shape of $S_3/T$ as a function of $T$. First we define the critical temperature $T_c$, where the thermal corrections at the origin balance the zero-temperature potential difference between the two minima:
\be
\vert V_{T_c}^0\vert\simeq\Delta V \quad \Rightarrow \quad
T_c \simeq 
\frac{2}{5}
\frac{m _{*}}{\mathcal{W}\left(0.13 \left(N\frac{m_{*}^{4}}{F^2} \right)^{2/5}\right)}\ , 
\label{Tcrit}
\ee
where $\mathcal{W}(x)$ is the Lambert function, defined as the solution to the equation $\mathcal{W}(x) e^{\mathcal{W}(x)} = x$. At large $x$ the function $\mathcal{W}(x)$ behaves approximately like $3/4\log (1+ x)$, and this simple approximation can be used for all practical purposes here (see Appendix~\ref{app:potentials} for a short summary of the properties of the Lambert function). Using this, the low-$T$ expansion will apply in regions of parameter space where 
\begin{equation}
T_c\lesssim m_*\quad \Rightarrow \quad \sqrt{F}\lesssim 0.8 \left(\frac{N}{10}\right)^{5/8}m_*\ ,\label{eq:lowTvalidity}
\end{equation}
where we have normalized the number of degrees of freedom in the thermal loops to the typical order of magnitude we will find in the explicit examples of Sec.~\ref{sec:models}. The low-$T$ approximation is then valid whenever Eq.~\eqref{eq:lowTvalidity} is satisfied, making it a generic feature of the pseudomodulus potential where the vacuum energy is protected from quantum corrections induced by heavy SUSY states. 

From the definition of $T_c$ in Eq.~\eqref{Tcrit}, we can immediately see that the triangular approximation in Eq.~\eqref{S3oTsimp} breaks down in this regime and should be extended (see Appendix \ref{app:bounce}). However, the nucleation temperature in our setup is generically very far from $T_c$, so that Eq.~\eqref{S3oTsimp} is always a good approximation at the temperatures relevant for the FOPT. As the temperature decreases below $T < T_c$, $S_3/T$ decreases as long as $\vert V_T^0\vert>V_P$, since $\vert V_T^0\vert$ decreases exponentially with the temperature. When the temperature approaches $T_{\text{min}}$ defined in Eq.~\eqref{eq:Tmin}, then $\vert V_T^0\vert \simeq V_P$ and $S_3/T$ attains a minimum value. As the temperature decreases further below $T_{\text{min}}$, $S_3/T$ grows as $1/T$.

Plugging the simplified bounce action in Eq.~\eqref{S3oTsimp} into the $T_{\text{min}}$ definition in Eq.~\eqref{eq:Tmin}, we can easily obtain an analytic expression for $T_{\text{min}}$ which reads 
\be\label{Tmin_triangular}
\quad T_{\text{min}} = \frac{2 m_{*}}{3}\frac{1}{ \mathcal{W} \left( \frac{5^{2/3}}{3 \pi} \left( N\frac{m_{*}^4}{V_{P}} \right)^{2/3}\right)}\quad \Rightarrow \quad \frac{T_{\text{min}}}{T_c}\lesssim 0.2\ ,
\ee
where to obtain the first expression we assumed $T_{\text{min}}\lesssim 0.48 m_{*}$ and the second inequality follows from approximating the Lambert function $W(x)\simeq 3/4 \log(x+1)$, assuming $N\sim \mathcal{O}(10)$ and using the Eq.~\eqref{eq:smallbar} for the scaling of $V_P$ with $\lambda_{\text{eff}}\sim\mathcal{O}(1)$ and $\Delta V\sim F^2$. Higher values of $\lambda_{\text{eff}}$ or a suppressed value of $\Delta V$ will lead to a reduction of the hierarchy between $T_{\text{min}}$ and $T_c$. The latter cases are less interesting from the point of view of the expected GW signal.

We are now ready to verify that there exists a nucleation temperature $T_n$ where $S_3/T$ satisfies the nucleation condition Eq.~\eqref{eq:approxnuc}. As discussed in Eq.~\eqref{eq:Tmin}, the nucleation temperature is always within the interval $(T_{\text{min}},T_c)$. Scenarios where $T_{n}$ is closer to $T_{\text{min}}$ have a larger $\alpha$ (see Eq.~\eqref{eq:alpha}) and a smaller $\beta_H$ (see Eq.~\eqref{eq:beta}), favorable for generating an observable GW signal. Understanding the scaling of $T_n$ with respect to $T_{\text{min}}$ and $T_c$ thus provides valuable information about the strength of the FOPT.

Even approximating the nucleation condition in Eq.~\eqref{eq:approxnuc} with a constant $\mathcal{C}$, solving the equation analytically with respect to $T$ using $S_3/T$ given by Eq.~\eqref{S3oTsimp} is not possible. We may, however, expand in $V_P/\vert V^0_T\vert\ll 1$ and solve for $T_n$ order by order in this expansion. This is always a good approximation as long as $T_n$ does not approach $T_{\text{min}}$ too closely. At first order, writing $T_n=T_n^0(1+\delta T^1_n)$ we find 
\be
\label{Tnucl_1}
T_{n} \simeq
T_{n}^{0}\left( 
1-\frac{7}{\mathcal{C}^{2/5}} \frac{V_P}{m_{*}^4} \left(\frac{T_{n}^0}{m_{*}}\right)^{3/5} 
\left(\frac{f_a m_{*}^3}{\Delta V}\right)^{6/5}\right)\ ,
\ee
where 
\be
\label{Tnucl_0_bis}
T_{n}^{0} =  0.48 m_{*}\frac{1}{\mathcal{W}\left(
0.32\left(\frac{N^{5}}{\mathcal{C}^{2}}\right)^{2/21} \left(\frac{f_a m_*^3}{\Delta V}\right)^{4/7}
\right)}\ ,
\ee
and we have again assumed $T_{n}^0 < 0.48 m_{*}$. Given that the argument of the Lambert function is much larger than one, $T_{n}^{0}$ depends only logarithmically on the parameters $N, \mathcal{C}, f_a,\Delta V$, and can be taken proportional to $m_{*}$ for simplicity.

The leading scaling of $T_n$ with respect to the parameters shaping the potential is captured by the leading corrections proportional to $V_P$ in \eqref{Tnucl_1}.
Indeed, we observe that by increasing $V_P$ (i.e. the height of the barrier), or by increasing $f_a$, the nucleation temperature decreases, approaching
 the region of parameter space where nucleation does not occur.
The border between the nucleation and the non-nucleation areas is the portion of parameter space which is optimal for gravitational waves, since it is where $\beta_H$ is minimal. This behavior is in good agreement with the numerical results of Sec.~\ref{sec:models}, and one can verify that Eq.~\eqref{Tnucl_1} reproduces the behavior of the full numerical result when properly matched to the models in Sec.~\ref{sec:models} up to an overall scaling of the bounce action.

\subsection{$\alpha$, $\beta_H$ and fine-tuning}
Now we can use our prediction for $T_n$ to compute the parameters characterizing the FOPT:
\begin{itemize}
\item Within our analytical approximation, the temperature corrections only affect the 
potential at the origin of field space and are exponentially suppressed for $T<m_{*}$. Therefore, we approximate $\alpha$ as
\be
\alpha \simeq \frac{30}{g_{*}(T_n)\pi^2 }\frac{ \Delta V}{T_{n}^4}\ ,\label{eq:alphagen}
\ee
where the scaling of $T_n$ can obtained by using \eqref{Tnucl_1}. Within this approximation, $\Delta V$ is temperature-independent and the largest values of $\alpha$ correspond to $T_n$ closer to $T_{\text{min}}$. 
\item 
The inverse time scale of the phase transition can be computed explicitly from \eqref{S3oTsimp}, giving
\begin{equation}
\beta_H\simeq \mathcal{C} \left(\frac{1.1 N}{\mathcal{C}^{2/5}} 
e^{-\frac{m_{*}} {T_{n}}}
\left(\frac{T_{n}}{m_{*}} \right)^{11/10}
\left(\frac{f_a m_{*}^3}{\Delta V} \right)^{6/5}
 -1\right)\ .\label{eq:betaapprox}
\end{equation}
One can easily verify that if $T_n=T_{\text{min}}$, then $\beta_H\simeq 1$ within the small $V_P$ expansion. Moreover, the exponential dependence on $T_n$ makes $\beta_H$ very sensitive to the underlying parameters. 
\end{itemize}
We now use the approximate $\beta_H$ formula in Eq.~\eqref{eq:betaapprox} to estimate the $\beta_H$-tuning defined in Eq.~\eqref{eq:betaFT}. 
We compute first the tuning with respect to $V_P$, which is encoded in Eq.~\eqref{eq:betaapprox} through the dependence of $T_n$ on $V_P$. At leading order in $V_P/m_*^4\ll1$ we obtain
\be
\label{beta_tune2}
\left|  \frac{d \log \beta_H }{d \log V_P} \right|= \left| \left(1-\frac{\beta_0}{\beta_H}\right) \right| \gtrsim
\left| 4 \frac{\mathcal{C}}{\beta_H} \right|
\ee
where in the last step we used the fact that
\be
\beta_0 \simeq \mathcal{C} \left(-1 +\frac{5}{2}  \frac{m_{*}}{T_{n}^0} \right)
\gtrsim 4 \mathcal{C} 
\ee
since $T_{n}^0 < \frac{10}{21} m_{*}$ and $\beta_H \lesssim \mathcal{C}$ in the interesting region of parameter space. The tuning associated with the barrier height is the dominant one, given that the tuning with respect to vacuum distance $f_a$ is suppressed by an extra factor of  $T_{n}^0/m_{*}$. As in Eq.~\eqref{eq:naturalbeta}, we see that the natural value of $\beta_H$ is $\beta_H\simeq\mathcal{C}(T_n) \simeq 100$ for the scales of interest in this study. Smaller values of $\beta_H$ can be obtained at the price of fine-tuning the barrier height at the percent level. This might imply an even larger tuning with respect to the fundamental parameters of a given model, as we will show in a concrete example in Section \ref{sec:X3}.

\subsection{A toy example: fine-tuning vs.~single SUSY-breaking scale}\label{sec:toy}

We now present a simple toy model which captures most of the features of the pseudomodulus potential in the explicit SUSY-breaking hidden sectors we will encounter in Sec.~\ref{sec:models}. We take the zero-temperature potential to be
\be
\label{eq:simplified_pot}
V_0(x) = \kappa_D^2\left(F-\epsilon_{\slashed{R}} x^2\right)^2  + \frac{\lambda ^2}{32 \pi^2} |F|^2 \log \left( \frac{\lambda^2 x^2 +m_{*}^2}{m_{*}^2}\right)\ ,
\ee
which reproduces the shape of the potential sketched in Fig.~\ref{fig:albe_cart}. 
The first term captures tree-level effects, while the second term captures one-loop quantum corrections. The $x$ potential is flat at tree-level up to $R$-symmetry breaking operators parametrized by $\epsilon_{\slashed{R}}$.\footnote{As shown in Sec.~\ref{sec:X3}, the potential controlled by $\epsilon_{\slashed{R}}$ can be obtained from a marginal operator breaking $R$-symmetry in the superpotential. Similarly, one could study explicit $R$-breaking operators of arbitrary dimension in the superpotential $\mathcal{W}_{\slashed{R}}=\frac{\epsilon_{\slashed{R}} X^n}{n\Lambda^{n-3}}$ which correspond to tree level potentials of the form $V(x) = \left(F-\frac{\epsilon_{\slashed{R}}x^{n-1}}{\Lambda^{n-3}} \right)^2$. These types of operators would naturally be generated by UV dynamics as in Ref.~\cite{Intriligator:2006dd}.} SUSY-breaking corrections induced by heavy fields lift the $x$ potential around the origin, giving a mass to the pseudomodulus, but ultimately become subdominant for $x\gg\sqrt{F}$ where SUSY is restored in the direction associated to the $F$-term. This large-field behavior is a unique characteristic of SUSY models. 

As long as the explicit $R$-symmetry breaking is parametrically small, the position of the true vacuum and the zero temperature difference energy between the true vacuum and false vacuum are
\begin{equation}
\left\langle x\right\rangle_{\text{true}}=f_a=\sqrt{\frac{F}{\epsilon_{\slashed{R}}}}\quad ,\quad \Delta V=(\kappa_D F)^2\ ,
\end{equation}
where we have introduced the parameter $\kappa_D$ to allow the scale controlling the difference in vacuum energy to vary relative to the scale controlling the loop corrections along the pseudomodulus potential. We will exhibit a concrete realization of such a model in Sec.~\ref{sec:FD}. Requiring the potential to be flat as in Eq.~\eqref{eq:flatness} requires $\epsilon_{\slashed{R}}<1/\sqrt{\kappa_D}$.

Following the triangular barrier prescription, we need to find the position of the barrier and the value of the potential at the barrier; for the toy model these take the form
\begin{align}
& x_P\simeq\frac{\lambda}{8\pi\kappa_D}f_a \ , \label{eq:pospeak}\\
& V_P\simeq \frac{\lambda^2F^2}{32\pi^2}\left(2\log\left(\frac{\lambda^2 f_a}{8\pi \kappa_D m_* }\right)-1\right)\ .\label{eq:Vpeak}
\end{align}
From the last equation we see that the loop suppression of the zero-temperature barrier $V_P$, as assumed in Eq~\eqref{eq:smallbar}, is here an automatic consequence of  the fact that the pseudomodulus direction is lifted by quantum corrections. For a single-scale model (i.e. $\kappa_D=1$) the position of the peak $x_P$ is fixed in terms of the one of the true vacuum $f_a$, while for a two-scale model, $\kappa_D\gg1$ can enhance the hierarchy between $x_P$ and $f_a$. 

We are now ready to use the triangular barrier approximation in Eq.~\eqref{S3overTfull} to compute the bounce action and the features of the FOPT between the origin and the true vacuum. For $\epsilon_{\slashed{R}}<1/\sqrt{\kappa_D}$, $f_a$ is the largest scale in the problem and the approximation in Eq.~\eqref{S3oTsimp} is justified. If the general features of the bounce action characterize the FOPT in the low-$T$ expansion discussed above, this simple toy model allows us to say something more precise about the scaling of the energy released during the FOPT. From Eq.~\eqref{eq:alphagen} we have
\begin{equation}
\alpha=\frac{30}{g_{*}(T_n)\pi^2 }\left(\frac{\kappa_D F}{T_n^2}\right)^2\sim 10^{-2} \kappa_D^2\left(\frac{F}{m_*^2}\right)^2\left(\frac{230}{g_*(T_n)}\right) ,\label{eq:alphatoy}
\end{equation} 
where we normalized the number of relativistic degrees of freedom at $T_n$ to be close to the MSSM value and we substituted $T_n\sim T_n^0\sim0.5 m_* $, which is the natural value of the nucleation temperature unless either $V_P$ or $f_a$ are tuned to suppress it (see Eq.~\eqref{Tnucl_1}). In a single-scale model where $\kappa_D=1$, tuning $T_n\ll \sqrt{F}$ is the only way to enhance the strength of the FOPT. The same tuning will allow $\beta_H$ to be small. Conversely, in a two-scale model of SUSY-breaking, having $\kappa_D\gg1$ can compensate the suppression in Eq.~\eqref{eq:alphatoy} without any tuning. We will show an explicit example of this class of hidden sectors in Sec.~\ref{sec:FD}. These are clearly the best candidates to be probed by future GW interferometers.

\section{Explicit Models}\label{sec:models}
In this section we provide two working examples of the general idea described in the previous sections. Both models are straightforward deformations of the minimal O'Raifeartaigh model, which is the simplest theory of chiral superfields that breaks SUSY spontaneously~\cite{ORaifeartaigh:1975nky}. The O'Raifeartaigh model involves three chiral superfields, namely the SUSY-breaking field $X$ containing the pseudomodulus and two messenger fields $\Phi_{1,2}$. The dynamics are determined by three parameters: the SUSY-breaking scale $\sqrt{F}$, the SUSY-preserving mass $m$ of the messengers, and the coupling $\lambda$ between the three fields. To set the stage for our analysis, we begin in Sec.~\ref{sec:Orafe} by determining the phase diagram of the minimal O'Raifeartaigh model which can be described as a function of the dimensionless parameter 
\begin{equation}
y_F\defeq\frac{\lambda F}{m^2}\ .
\end{equation}
The model exhibits a rich phase structure as a function of temperature and the underlying parameters; for $y_F \sim 1$ the origin of the pseudomodulus is the global minimum at all $T$ and no interesting phase transitions occur, while for $y_F \ll 1$ a second minimum develops away from the origin that may become the global minimum at intermediate temperatures, leading to a variety of phase transitions. Unfortunately, as we will see, none of these phase transitions are sufficiently strongly first-order to generate an observable GW signal. However, this minimal O'Raifeartaigh model serves as the foundation for SUSY-breaking hidden sectors that {\it do} generate observable GW signals.

In Sec.~\ref{sec:X3}, we present the simplest SUSY-breaking hidden sector featuring a strong FOPT like the ones describe in Sec.~\ref{sec:anatomy}. This hidden sector involves a marginal deformation in the superpotential of the minimal O'Raifeartaigh model, breaking the $R$-symmetry explicitly and obtaining a pseudomodulus potential very similar to the one described in the toy model in Sec.~\ref{sec:toy}. We show that in such a simple single-scale model, $\alpha$ will be generically suppressed as predicted in Eq.~\eqref{eq:alphatoy}, and discuss quantitatively the fine-tuning of $\beta_H$ defined in Eq.~\eqref{eq:betaFT}. Phenomenologically, this model is unsatisfactory since the global minimum restores SUSY, although this may be remedied by the introduction of external SUSY-breaking effects.

In Sec.~\ref{sec:FD}, we show how both the shortcomings of the simple model of Sec.~\ref{sec:X3} are resolved in hidden sectors with two SUSY-breaking scales, in keeping with our expectations from Sec.~\ref{sec:toy}. We make this concrete by gauging a $U(1)$ flavor symmetry of the messengers in the minimal O'Raifeartaigh model, which admits an additional source of SUSY breaking via the Fayet-Iliopoulos term. This additional ``$D$-term'' supersymmetry breaking provides a second SUSY-breaking scale, which both ensures that supersymmetry is broken everywhere on the pseudomoduli space and increases $\alpha$, leading to observable GW signals.

\subsection{Warm up: The O'Raifeartaigh model at finite temperature}
\label{sec:Orafe}
In the minimal O'Raifeartaigh model, the pseudo-modulus is stabilized at the origin by quantum corrections. Since the $R$-symmetry is unbroken in the global minimum at $T=0$, one would expect that including finite temperature corrections will not induce any phase transitions. Instead, the dynamics of the O'Raifeartaigh model at finite temperature presents rich features that we discuss here in detail (see Refs.~\cite{Craig:2006kx,Katz:2009gh} for
earlier works on related issues).  

\begin{figure}[t!]
  \centering
\includegraphics[width=1\textwidth]{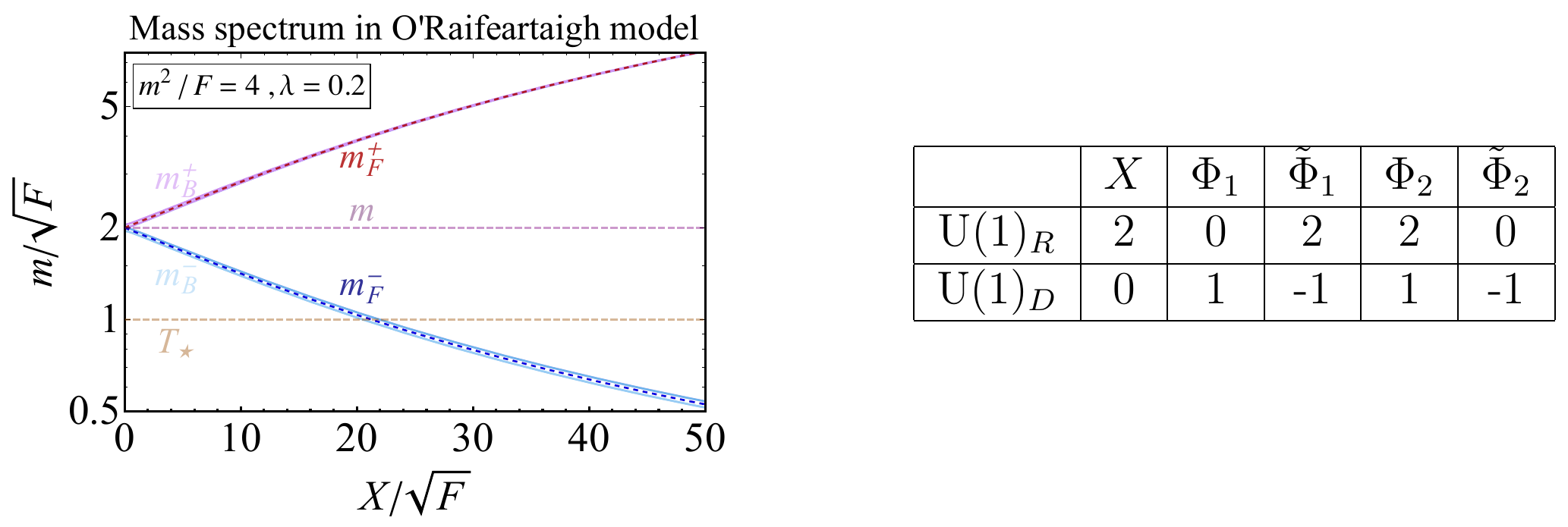}
\caption{{\bf Left:} Behavior of the hidden sector spectrum in the simple O'Raifeartaigh model as a function of the pseudomodulus direction $x$. The {\bf dashed dark red/blue} line indicates the fermionic eigenvalues growing/going to zero like $x^{\pm2}$ (see Eq.~\eqref{eq:massesOR}). The two {\bf pink} and {\bf light blue} solid lines indicate the scalar mass states splitted in pairs around the fermionic ones. The {\bf dashed light magenta} line indicates the states that remain independent on $x$. The {\bf dashed peach} line shows $T_\star$ for this particular benchmark, where the new vacuum induced by thermal corrections becomes degenerate with the origin (see Eq.~\eqref{eq:crazyvacuum}).    {\bf Right:} Unbroken symmetries of the chiral superfields in the O'Raifeartaigh model superpotential in Eq.~\eqref{Orafe_W}. The model enjoys a $U(1)_R$ symmetry and an extra $U(1)_D$ flavor symmetry. The first will be explicitly broken in the model in Sec.~\ref{sec:X3} while the second one will be gauged in the model in Sec.~\ref{sec:FD}. }\label{fig:OR_spectrum}
\end{figure}

Having in mind applications to gauge mediated SUSY breaking, we consider the vector-like version of the minimal O'Raifeartaigh model, which is described by the superpotential
\be
\label{Orafe_W}
W = - F X + \lambda X \Phi_1 \tilde \Phi_2 +m (\Phi_1 \tilde \Phi_1+\Phi_2 \tilde \Phi_2)\ ,
\ee
encoding the interactions of the SUSY-breaking chiral superfield $X$ and two vector-like sets of messenger superfields $\Phi_{i}, \tilde \Phi_{i} \, (i = 1,2)$. The first term is a tadpole ensuring that supersymmetry is broken at the scale $\sqrt{F}$, while the second term encodes interactions among the fields with strength $\lambda$. We take the masses of the two pairs of messengers to be equal for simplicity. The superpotential above enjoys an unbroken $R$-symmetry under which $X$ carries $R[X] = +2$, as well as a $U(1)_D$ flavor symmetry under which the messengers $\Phi$ and $\tilde \Phi$ have opposite charges (see Fig.~\ref{fig:OR_spectrum} right for a summary table with the full charge assignment).  

The potential for the scalar components of the chiral superfields is
\be
V = |F - \lambda \phi_1 \tilde \phi_2|^2 + |\lambda X \tilde \phi_2 + m \tilde \phi_1|^2+|\lambda X \phi_1 + m  \phi_2|^2+|m \phi_1|^2+|m \tilde \phi_2|^2\ ,
\ee
where $X=\frac{x}{\sqrt{2}}$ denotes the scalar component of the pseudomodulus in the notation of Eq.~\eqref{eq:pseudo}.
For $\lambda F \leq m^2$, the tree level vacuum of the theory is at $\phi_i = \tilde \phi_i =0$ with $x$ undetermined, and SUSY is broken at a scale $\sqrt{F}$.
Radiative corrections from loops of the messenger fields $\phi$ and $\tilde \phi$ generate a potential for $x$ that stabilizes it at the origin,
and thus
the global vacuum at zero temperature lies at  $\phi_i=\tilde \phi_i =0$ and $\langle x \rangle = 0$.
Note that the one-loop corrections have the shape described in Sec.~\ref{sec:anatomy}, being polynomial close to the origin of the pseudomodulus potential and
logarithmic for large field values. 
Expanding for $y_F\equiv \frac{\lambda F}{m^2}\sim 1$ we obtain
\bea
\label{V_1loop_approx}
&&
V^{\text{1-loop}}_{x \to 0} \simeq 
\frac{\lambda^3 F}{16 \pi ^2} \left( \log 4-1 \right)  x^2
-\frac{\lambda ^4}{384 \pi ^2}  (12 \log 2-7)x^4 + O(x^6)\ ,
\\
&&
\label{V_1loop_large}
V^{\text{1-loop}}_{x \to \infty} \simeq 
\frac{\lambda ^2 F^2 }{16 \pi ^2} \log \left(\frac{ x^2}{m^2}\right)\ ,
\eea
where we have fixed the renormalization scale to the messenger mass $m$. The thermal corrections to the $x$ potential can be added with standard formulas that we review in the Appendix \ref{app:potentials}.  

The shape of the thermal corrections is set by the $x$ dependence of the mass eigenvalues for the scalar and fermionic components of 
the messengers. From \eqref{Orafe_W} we can distinguish two classes of mass-squared eigenvalues: i) the ones growing quadratically with $x$, and ii) the ones decreasing 
as $1/x^2$ and asymptotically going to zero in the large-$x$ region. Specifically, the fermionic eigenvalues scale as 
\be
m^2_{\pm} = m^2 +\frac{\lambda^2 x^2}{4} \left(  1 \pm \sqrt{1+ \frac{8 m^2}{\lambda^2 x^2}} \right)
 = \left \{ 
 \begin{array}{c}
 m \quad   \text{for} \quad x \to 0  \\
 \sim x^{\pm2} \quad \text{for} \quad x \to \infty
 \end{array}
\right. \ ,\label{eq:massesOR}
\ee
and the bosonic eigenvalues are split in pairs  around the fermionic ones, e.g.~at the origin the bosonic eigenvalues are $\{ m^2, m^2,m^2 + \lambda F, m^2 -\lambda F \}$.
The behavior of the full spectrum as a function of $x$ is shown in Figure \ref{fig:OR_spectrum} (right). We also observe that at large $x$, the spectrum asymptotes to a supersymmetric one.

\begin{figure}[t!]
  \centering
 \includegraphics[width=0.9\textwidth]{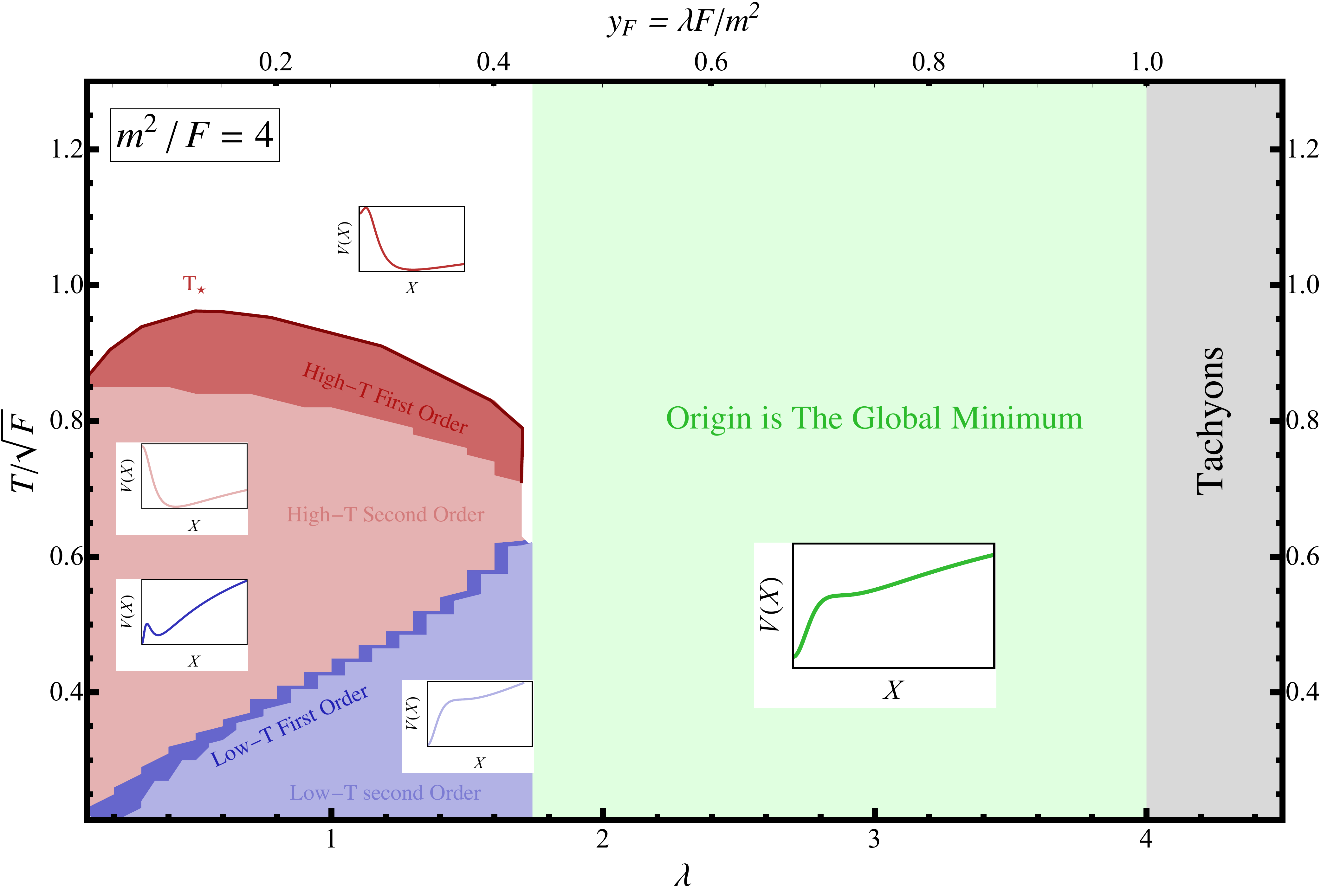}
\caption{Phase diagram of the O'Raifeartaigh model at fixed $F/m^2=4$. For {\bf large $\lambda$} the quantum corrections dominate and the origin is the global minimum at all temperatures. For {\bf small $\lambda$}, at $T=T_\star$ a new vacuum develops as a consequence of the interplay between the thermal and the loop corrections as shown in Eq.~\eqref{eq:crazyvacuum}. In {\bf dark red} we show the range of temperatures where a barrier is present between the origin and the true minimum, in {\bf light red} we show the range of temperatures where the barrier disappears. At lower temperatures, the origin again becomes the global minimum, and the second minimum decays back into the origin. In {\bf blue} we show the range of temperatures where a barrier separates the two minima and in {\bf light blue} the region when the barrier disappears.}\label{fig:NoamOR}
\end{figure}

For low temperatures (i.e. $ T < m$), the induced thermal corrections are a decreasing function of $x$, since they are mainly controlled by the lightest 
eigenstates. These corrections are mildly Boltzmann suppressed at large $x$ and modify
the pseudo-modulus potential as soon as $T^4 \sim \frac{\lambda^2 F^2}{16 \pi^2}$. 
For larger temperatures, the contribution from the other mass eigenstates and in particular from the ones  
growing with $x$ become relevant, and the thermal potential is a growing function of $x$. Hence at temperatures $T \sim m$ we expect the global minimum to be at the origin of the field space. However, for intermediate temperatures the thermal corrections can make the origin of the field space unstable, leading to a very rich evolution of the potential with temperature.

The thermal corrections compete with the loop corrections in the large $x$ region (see Eq.~\eqref{V_1loop_large}), 
eventually leading to a minimum of the potential at 
\begin{equation}
x_{\star} \simeq \frac{2 \sqrt{2} \pi  T}{\lambda y_F}\quad ,\quad T_{\star} \sim 0.23 \sqrt{ y_F} m\ ,\label{eq:crazyvacuum}
\end{equation}
where $x_\star$ is obtained using the high-$T$ expansion for the thermal potential up to $T^2$, assuming 2 bosons and 2 fermions with masses-squared 
$\simeq \frac{2 m^4}{\lambda ^2 x^2} $, and $T_\star$ is an estimate of the temperature where the new minimum can be the global one. The latter is estimated by requiring the temperature corrections at $x_{\star}$ to be comparable to the height of the one loop potential. If $T_\star$ is close to $m$, then the neglected contributions from the states whose masses grow with $x^2$ lifts again the minimum at $x_{\star}$, which will then never be the global minimum at any temperature. In conclusion, we expect that depending on the hierarchy between $\lambda F$ and $m^2$, the minimum at 
$x_{\star}$ could become the global minimum in a certain temperature range around $T_\star$. This complicated phase diagram is well summarized in Fig.~\ref{fig:NoamOR}, where we have fixed the ratio $ \frac{ F}{m^2}$ to a representative value and explore the dynamics 
of the model as a function of the temperature and coupling $\lambda$.

For large $\lambda$, corresponding to $y_F\sim1$, the minimum at $X_{\star}$ is never the global minimum of the scalar potential (green region in the plot).
For small $\lambda$, i.e. $y_F\ll1$ two phase transitions occur while lowering the temperature. 
Specifically, at very high temperature the global minimum is at the origin, as explained above. At intermediate temperatures the global minimum is at $X_{\star}$, and finally
at zero temperature the global vacuum is again at the origin. The corresponding two phase transitions can be first or second order.
We have explored the parameter space of the model for different values of $y_F$ and $\lambda$, and found that these phase transitions are never strongly first order (i.e. small $\beta_H$ and large $\alpha$) in the regime of perturbative $\lambda$.

Although the minimal O'Raifeartaigh model is itself not a good candidate for a strong FOPT, it nonetheless provides the foundation for simple variations that are. We explore these variations in the following subsections, restricting our attention to the region of parameter space in which the minimal O'Raifeartaigh model exhibits a simple thermal history corresponding to the
green region in Figure \ref{fig:NoamOR}. Deformations of the minimal O'Raifeartaigh model will endow this region with phase transitions as a function of temperature, while avoiding the complications of new minima arising from the interplay of thermal and loop corrections shown in the red and blue regions.

\subsection{O'Raifeartaigh model with explicit $R$-symmetry breaking}
\label{sec:X3}

Now we turn to a simple, concrete realization of a SUSY-breaking hidden sector whose pseudomodulus potential exhibits the properties exlpored in Sec.~\ref{sec:anatomy}. This model simply amounts to deforming the minimal O'Raifeartaigh model studied in the previous section with the following marginal, $R$-symmetry-breaking term in the superpotential:
\be
W_{\cancel{R}}(X) =\frac{1}{3} \epsilon X^3\ .\label{eq:X3def}
\ee
The complete tree-level scalar potential of the model is  
\be
V = |-F + \epsilon X^2+\lambda \phi_1 \tilde \phi_2|^2 + |\lambda X \tilde \phi_2 + m \tilde \phi_1|^2+|\lambda X \phi_1 + m  \phi_2|^2+|m \phi_1|^2+|m \tilde \phi_2|^2
\ee
and assuming $y_F \leq 1$, the global minimum sits at $\langle x \rangle_{\text{true}} = \sqrt{\frac{2 F}{\epsilon}}$ and $\phi_i = \tilde \phi_i =0$. In contrast to the minimal O'Raifeartaigh model, the $R$-symmetry-breaking deformation destabilizes the origin at tree level and restores supersymmetry in the true vacuum.

The radiative corrections are identical to the ones in the O'Raifeartaigh at zeroth order in $\epsilon$, and they
tend to stabilize the pseudo-modulus at $x=0$, competing with the tree-level contributions induced by the $\epsilon$
deformation.
Close to the origin, the effective potential for the pseudomodulus obtained by integrating out the $\phi_i$ and $\tilde \phi_i$ fields reads (up to quartic order)
\be
V_0(x) \underset{x\to0}{\simeq} F^2+ \frac{m_{\text{eff}}^2}{2} x^2 -
\frac{\lambda_{\text{eff}}}{4} x^4\quad ,\quad 
\begin{cases} &m_{\text{eff}}^2 = \left( \frac{\lambda^3}{8 \pi ^2} \left( \log 4-1 \right) - 2\epsilon \right)F\\
& \lambda_{\text{eff}}= \frac{\lambda ^4}{96 \pi ^2}  (12 \log 2-7) -\epsilon^2
\end{cases} \ ,\label{eq:X3potential}
\ee
where again we have approximated the loop corrections in the leading order in $y_F=\frac{\lambda F}{m^2} \sim 1$. 
For $\epsilon <  \frac{\lambda^3}{16 \pi ^2} \left( \log 4-1 \right)$, the radiative corrections are sufficient to create a metastable vacuum at the origin of $x$. In this regime, the $\epsilon$ contribution to the quartic is always negligible. 
Along the pseudomodulus direction, there is now a true vacuum created by the $R$-symmetry-breaking deformation and a false vacuum created by radiative corrections. The height and location of the barrier between these two vacua may be approximated as
\be
V_P -V_{+} \simeq \frac{m_{\text{eff}}^4}{4\lambda_{\text{eff}}} \sim 
\frac{24 \pi^2}{\lambda^4}
\left( \frac{\lambda^3}{8 \pi ^2} \left( \log 4-1 \right) - 2\epsilon\right)^2 F^2
\qquad \qquad
x_P \simeq \frac{m_{\text{eff}}}{\sqrt{\lambda_{\text{eff}}}}\ ,
\label{eq:VpeakX3}
\ee
This approximation is valid if $x_P \lesssim \frac{m}{\lambda}$, that is if there is a cancellation between the two terms in $m_{\text{eff}}^2$
such that $m^2_{\text{eff}} \lesssim \frac{\lambda^3}{96 \pi^2}F$. The global minimum far away from the origin is not modified by the quantum corrections since SUSY is effectively restored there (we will come back to this point in Sec.~\ref{sec:X3GWpheno}) and stays at $\langle x \rangle_{\text{true}} = \sqrt{\frac{2F}{\epsilon}}$,
so that $\Delta V = F^2$.

In summary, this hidden sector provides a concrete realization of the toy model discussed in Sec.~\ref{sec:toy}. A direct consequence of having a single SUSY-breaking scale $\sqrt{F}$ is that the potential difference $\Delta V$ and  the quantum corrections determining the barrier are both controlled by the same scale. This corresponds to $\kappa_D = 1$ in the toy model of Sec.~\ref{sec:toy}, and typically leads to suppressed $\alpha$ as we will show below.  The formulae above allow straightforward matching of the model parameters onto the variables entering in the triangular barrier bounce action of Sec.~\ref{sec:approxFOPT}. In the following, we will compare our analytical expectations with the full numerical analysis of the FOPT from the origin to the 
$\langle x \rangle_{\text{true}} $ vacuum.

\subsubsection{First order phase transition dynamics}
We now study the model at finite temperature with an eye towards the dynamics of the phase transition associated to $R$-symmetry breaking. The thermal corrections to the $X$ potential are equivalent to the ones that we studied in the simplest O'Raifeartaigh model, up to small corrections proportional to $\epsilon$. The main difference is that the thermal effects are added on top of a zero-temperature potential described in the previous section, where the global minimum is far away from the origin. If we restrict to the parameter space where $\lambda F/m^2\sim1$ (the green region of Fig.~\ref{fig:NoamOR}),  the role of the thermal corrections is to stabilize the origin at high temperature. Lowering the temperature, the thermal history is very similar to the one described in section \ref{sec:anatomy}: the negative thermal contributions at the origin decrease in absolute value until we reach $T_c$, where the minimum at $x=0$ is degenerate with the minimum at $x_{\text{true}}$. An analytic estimate of this temperature can be obtained following Eq.~\eqref{Tcrit}. By further lowering the temperature, the thermal corrections become more and more negligible and one recovers the zero-temperature potential with a local minimum at the origin separated from the true vacuum by a loop-induced barrier.

\paragraph{Bounce action and nucleation temperature}

\begin{figure}[t!]
  \centering
\includegraphics[width=1.0\textwidth]{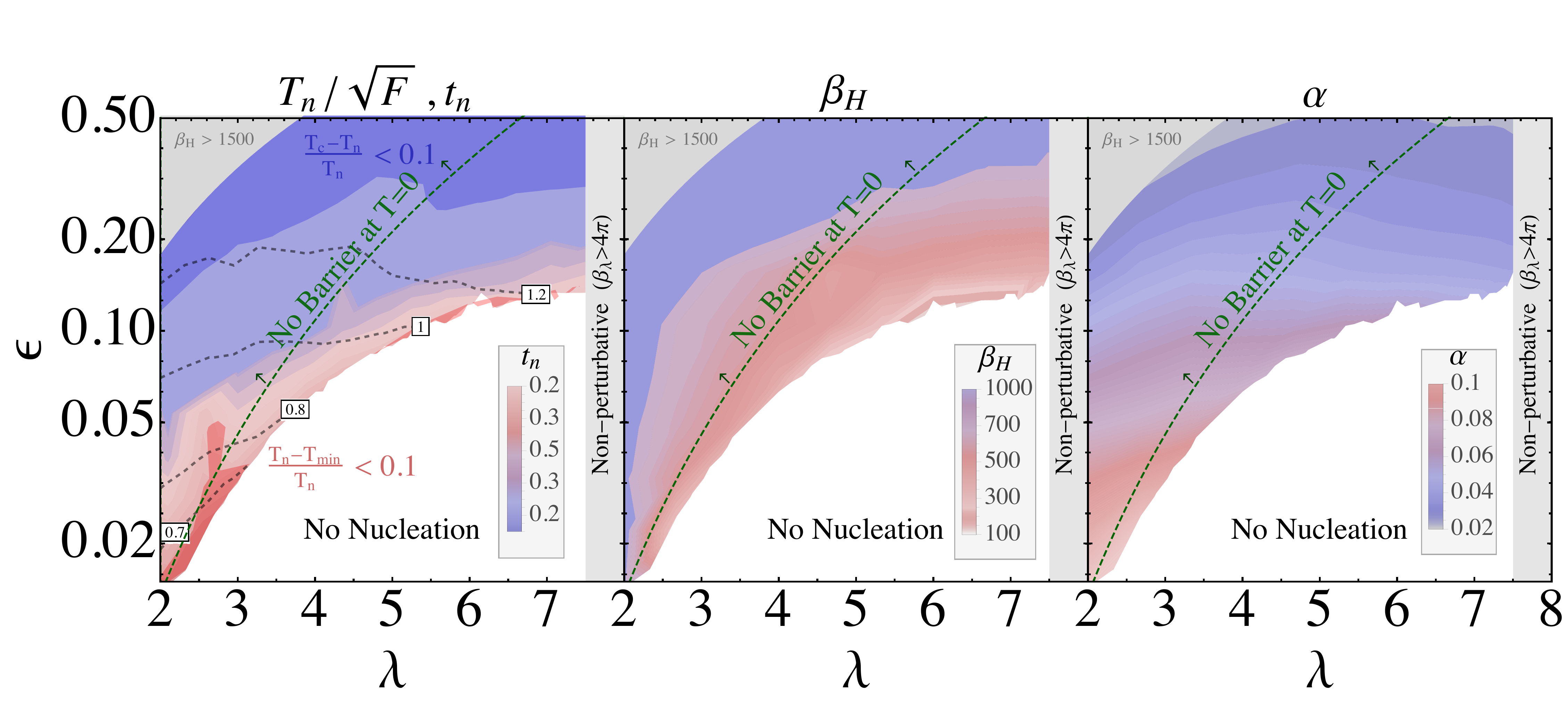}
\caption{From the left to the right, we show the behavior of $T_n/\sqrt{F}$, $\beta_H$ and $\alpha$ in the O'Raifeartaigh model with explicit $R$-symmetry breaking described in Eq.~\eqref{eq:X3def}. We fix $y_F=3/4$ and $F=30\text{ PeV}$ so that the entire parameter space of the model can be shown in the $(\lambda,\epsilon)$ plane. The {\bf black dashed contours} in the left plot show $T_n/\sqrt{F}$. The {\bf red-to-blue} gradients show contours of $t_n$ as defined in Eq.~\eqref{eq:tndef} (left), of $\beta_H$ as defined in Eq.~\eqref{eq:beta} (center) and of $\alpha$ as defined in Eq.~\eqref{eq:alpha} (right). The GW signal weakens going from red to blue.  Above the {\bf green dashed} line, the barrier separating the false and true vacua disappears at zero temperature. The {\bf grey} regions are not considered in our numerical scan because $\beta_H$ is too large (top left) or $\lambda$ is non-perturbative. In the {\bf white} region the nucleation condition in Eq.~\eqref{eq:Tn} cannot be satisfied.}\label{fig:X3AlphaPlot30PeV}
\end{figure}

The next step in determining the phase transition dynamics is to compute the bounce action and the nucleation temperature.  In the left panel of Fig.~\ref{fig:X3AlphaPlot30PeV}, we show the numerical result for the nucleation temperature $T_{n}$
as a function of the two dimensionless couplings of the model, having fixed $y_F \equiv \lambda F/m^2 =3/4$ and the scale of SUSY breaking to 
$\sqrt{F} = 30$ PeV for concreteness. Numerically we see that $T_n\sim\sqrt{F}$ as assumed in the general discussion around Eq.~\ref{eq:general}. As can be seen from the explicit formula in Eq.~\ref{Tnucl_1}, this feature is a consequence of the fact that $F$ cannot be arbitrarily decoupled from $m^2$ if we require $y_F\sim 1$ and perturbativity of $\lambda$.

Keeping $\epsilon$ fixed, we see that the nucleation temperature $T_n$ decreases when $\lambda$ increases, up until reaching the no-nucleation zone.
Indeed, increasing $\lambda$ makes the barrier between the two vacua higher, and the bounce action larger, decreasing the likelihood that the phase transition completes.
When the bounce action increases, the nucleation temperature lowers, approaching $T_{\text{min}}$ at the border of the no-nucleation region. Conversely, decreasing $\lambda$ at fixed $\epsilon$ shifts the nucleation temperature towards $T_c$. In Fig.~\ref{fig:X3AlphaPlot30PeV} we plot the quantity 
\begin{equation}
t_n\defeq\text{min}\left(\frac{T_c-T_n}{T_c},\frac{T_n-T_{\text{min}}}{T_{\text{min}}}\right)\ ,\label{eq:tndef}
\end{equation}
which indicates whether $T_n$ is closer to $T_c$ or $T_{\text{min}}$. As we will see this quantity is strongly correlated with the strength of the signal. 

The main effect of $\epsilon$ on the pseudomodulus potential is to set the distance in field space between the origin and the true vacuum. The barrier is also $\epsilon$-dependent, but away from the region where the effective mass in Eq.~\eqref{eq:X3potential} changes sign, the effect of varying $\epsilon$ is negligible. 
Decreasing $\epsilon$ makes $x_{\text{true}}=f_a$ larger. From Eq.~\eqref{S3oTsimp} we see that the bounce action grows, making it more difficult for the phase transition to occur. This explains why lowering $\epsilon$ at fixed $\lambda$ causes the nucleation temperature to decrease until the no-nucleation region is reached.

The interesting area of the parameter space is the sliver between the no-nucleation region and the region where there is not a barrier at $T=0$. Within this sliver, $\frac{\lambda^3}{16 \pi^2} \sim \epsilon$ and the effective mass of the pseudomodulus in Eq.~\eqref{eq:X3potential} is small and positive (in units of $\sqrt{F} \sim \Delta V^{1/4}$). The whole region shrinks for small $\epsilon$ because the true vacuum is pushed to large field values, and there is no nucleation unless the effective mass at the origin is tuned to be small.

The scaling of $T_n/\sqrt{F}$ with the Lagrangian parameters can be captured by the analytic approximations presented in Sec.~\ref{sec:anatomy}.
We match the generic parameterization of Sec.~\ref{sec:anatomy} using the expressions in \eqref{eq:X3potential} and \eqref{eq:VpeakX3}, giving
\be
\label{Tnucl_anal_X3}
T_{n} \sim \frac{40 \sqrt{\lambda F}}{63}\frac{1}{ \log\left(1 +0.76 \,\frac{ \lambda ^{6/7}}{\mathcal{C}^{4/21} \epsilon ^{2/7}}  \right)}
\left(
1\, -0.015 \left( \frac{\lambda ^9}{\mathcal{C}^2 \epsilon ^3}
\right)^{1/5}
\right)\ ,
\ee
where we have approximated the radiative corrections for $y_F \sim 1$. This expression qualitatively reproduces the left panel of Fig.~\ref{fig:X3AlphaPlot30PeV}, up to an overall normalization of the bounce action (corresponding to a shift in $\mathcal{C}$).

For large $\epsilon$, the rightmost term in parentheses in \eqref{Tnucl_anal_X3} is always $\mathcal{O}(1)$, and hence the variation of $T_n$ is largely controlled by the prefactor.
The $\sim \sqrt{\lambda}$ scaling of the numerator is balanced by the $\log \lambda$ scaling in the denominator, and the resulting prefactor of $T_n$ is essentially flat in $\lambda$ and only
decreases with decreasing $\epsilon$.
In the small $\epsilon$ region, the rightmost term in parentheses in  \eqref{Tnucl_anal_X3} becomes smaller than $1$ and controls the shapes of the $T_n$ contours, 
in agreement with the left panel of Fig.~\ref{fig:X3AlphaPlot30PeV}.

\paragraph{$\alpha$, $\beta_H$ and fine-tuning} The microscopic properties of the FOPT dynamics are encoded in the two parameters $\alpha$ and $\beta_H$, which correspond to the energy release and the duration of the phase transition.  In the central and rightmost panels of Fig.~\ref{fig:X3AlphaPlot30PeV}, we show the behavior of $\beta_H$ and $\alpha$ in the $(\lambda, \epsilon)$ plane. Here the scaling of $\alpha$ is essentially dictated by the scaling of $T_{n}/\sqrt{F}$, since $\alpha \sim \frac{30}{g_*\pi^2} \frac{\Delta V}{T_n^4}$ and $\Delta V \sim F^2$. We see that in the parameter space explored here, we cannot reach large values of $\alpha$ except in the thin sliver towards small $\epsilon$ and $\lambda$ where $T_{n}\sim T_{\text{min}}$ and $T_{\text{min}}$ is minimized with respect to $\sqrt{F}$. This unfortunate feature is a generic prediction of a single-scale SUSY-breaking hidden sector, as discussed in Sec.~\ref{sec:toy}. 

\begin{figure}[t!]
  \centering
\includegraphics[width=0.46\textwidth]{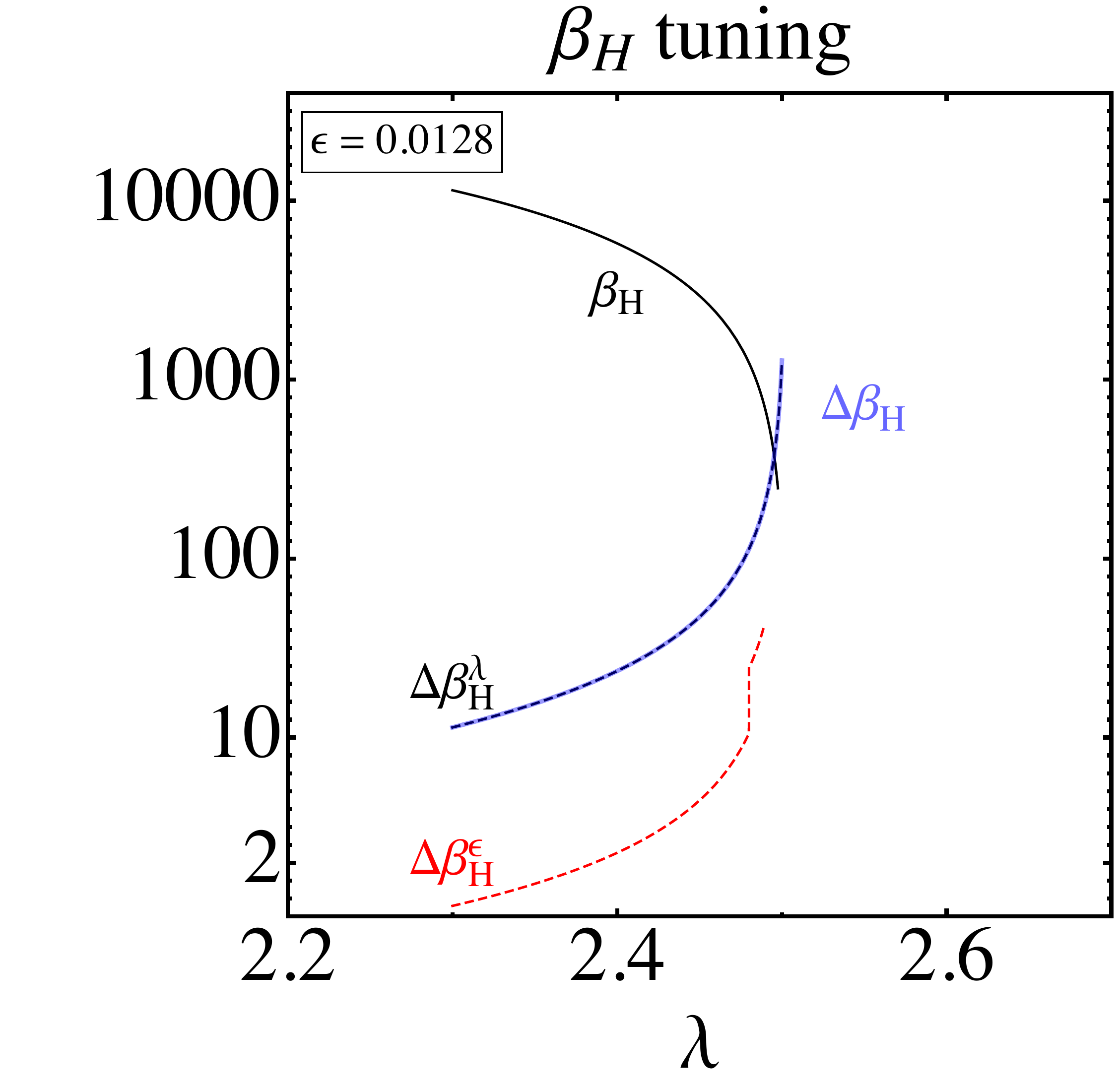}\hfill  
\includegraphics[width=0.48\textwidth]{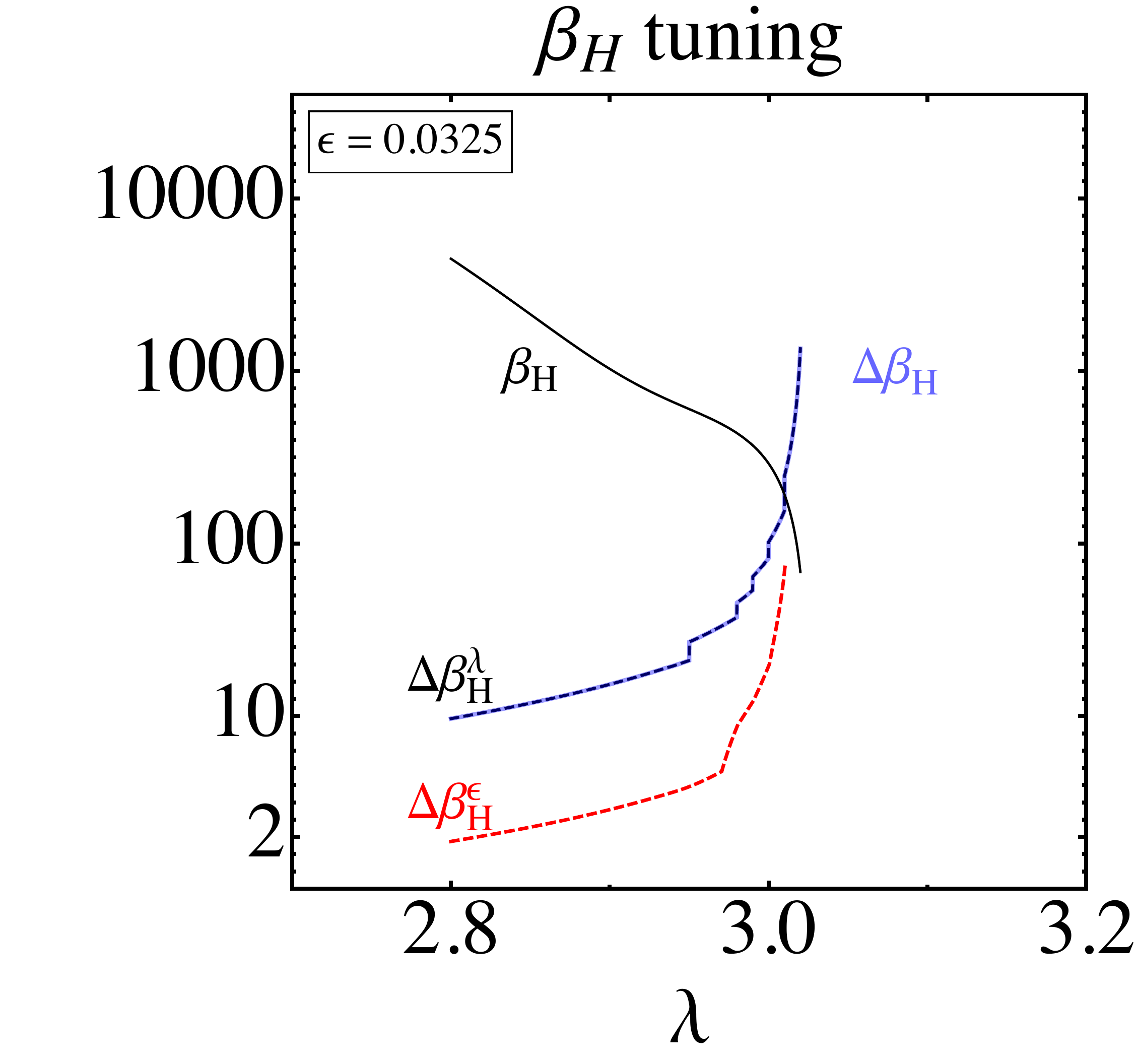}
\caption{Behavior of $\beta_H$ ({\bf black solid} curve) and the associated fine-tunings defined in Eq.~\eqref{eq:betaFT} as a function of $\lambda$ for $m^2/F=4$ and $\sqrt{F}=30\text{ PeV}$. The {\bf left} and {\bf right} plots correspond to two different values of $\epsilon$. The {\bf black dashed} curve shows the fine tuning w.r.t. $\lambda$, which dominates over the fine tuning w.r.t $\epsilon$ shown as a {\bf red dashed} curve. The $\lambda$ fine-tuning corresponds to the total fine-tuning $\Delta\beta_H$ shown as a {\bf light blue solid} line.\label{fig:X3lambdaTuning}}
\end{figure}

As shown in Fig.~\ref{fig:X3AlphaPlot30PeV}, $\beta_H$ is small in the regions of the parameter space at the border of the no-nucleation zone where $T_{n}\sim T_{\text{min}}$. Getting closer and closer to this boundary, one can achieve $\beta_H\lesssim 100$ at the price of a large tuning of the model parameters as discussed in Sec.~\ref{sec:FOPTandGWs}. The fine-tuning is dominated by the tuning of the barrier $V_P$ between the two minima. Substituting the dependence of $V_P$ on the Lagrangian parameters, we can estimate the the tuning of $\beta_H$ with respect to $\lambda$ as
\be
\frac{\partial \log \beta_H}{\partial \log \lambda} \gtrsim 
 8 \left( \frac{\mathcal{C}(T_n)}{\beta_H}  \right)\ .
\ee
The same fine-tuning can be computed numerically using the prescription of Eq.~\eqref{beta_tune2}. We show the results in Fig.~\ref{fig:X3lambdaTuning}, where we see that $\beta_H\lesssim 100$ corresponds to $\Delta_{\beta_H}\sim 10^3$, which is larger than the estimate derived above. The numerical results also confirm our expectation that 
the tuning associated with the $\lambda$ parameter (setting the height of the barrier) dominates relative to the tuning associated with the $\epsilon$ parameter (setting the location of the true vacuum). Comparing the left and right panels in Fig.~\ref{fig:X3lambdaTuning}, it is apparent that the tuning grows in the region of small $\epsilon$, where $\alpha$ is larger.

\subsubsection{Phenomenological Challenges}\label{sec:X3GWpheno}

As discussed in the previous section, the model presented here is not optimal for generating a sizable SGWB signal. Indeed, by comparing the resulting values of $\alpha$
and $\beta_H$ in Figure \ref{fig:X3AlphaPlot30PeV} to the values displayed in Figure \ref{fig:alphabeta}, it is clear that the typical value of $\alpha$ is too small to lead to a detectable signal. Of course, this issue can be resolved by going in a tuned region of the parameter space where a very small $\epsilon$ and an appropriately fine-tuned $\lambda$ give $T\sim T_{\text{min}}$ and a suppressed $T_{\text{min}}$ compared to $\sqrt{F}$. However, it is fair to say that, in general, a perturbative single-scale SUSY-breaking hidden sector cannot lead to strong SGWB signals. For this reason, we do not display the SGWB for this model, although it may easily be inferred from the $\alpha, \beta_H, T_{n}$ plots in Figure \ref{fig:X3AlphaPlot30PeV}.

As shown in the simple toy model of Sec.~\ref{sec:toy}, the suppression of $\alpha$ is a consequence of the fact that in a single-scale SUSY-breaking hidden sector one cannot significantly separate $\Delta V$ from $T_n^4$. Notice that this conclusion hinges on requiring $y_F\sim1$, which is necessary to avoid the region shown in Fig.~\ref{fig:NoamOR} where thermal corrections at high temperatures induce new minima. A more careful study of the dynamics of single scale models for $y_F\ll1$ is left for future work. 

In addition to the $\alpha$ suppression, the true vacuum in this model restores SUSY, so that the phase transition is not genuinely a ``SUSY-breaking phase transition''; this sector, as presented, cannot be responsible for the SUSY breaking transmitted to the MSSM. However, this is not a fatal obstruction, as the vacuum energy in the true vacuum far from the origin can be easily lifted by coupling to another source of SUSY breaking, very much in the spirit of \cite{McCullough:2010wf}.  If we require the new source of SUSY-breaking to not significantly affect the dynamics of the phase transition, the parametrics of the model will not significantly deviate from those presented here and the resulting $\alpha$ will be still suppressed. Interestingly, our analysis seems to point towards SUSY-breaking hidden sectors with multiple dynamical field directions and scales. In the next section, we will exhibit the simplest model of this type, leaving a more thorough exploration of the different possibilities for future study.

\subsection{O'Raifeartaigh model with gauge interactions}
\label{sec:FD}
In the previous subsection we analyzed a simple model displaying a first order phase transition associated with the breaking of the $R$-symmetry.
However, there were two aspects that were not completely satisfactory: i) SUSY breaking in the global minimum had to be added as a further deformation, and ii) the phase transition was generically not strong enough to generate a sizable signal.
Both issues were related to the fact that there was only one SUSY breaking scale in the problem. In this subsection we resolve these issues in a hidden sector where the global minimum breaks both SUSY and $R$-symmetry spontaneously and the presence of two SUSY breaking scales leads to a strong FOPT from the origin to the true minimum. 

It is well-known that adding gauge interactions to SUSY breaking models with chiral superfields modifies the potential and typically leads to a new SUSY- and $R$-symmetry-breaking vacuum at 
large field values (see e.g.~\cite{Intriligator:2007py}). As a prototype of this class of models we consider the simplest realization, which consists of the vector-like O'Raifeartaigh model of the previous sections where the anomaly-free $U(1)_D$ flavor symmetry defined in the right panel of Fig.~\ref{fig:OR_spectrum} is gauged. The model we consider has been studied at zero temperature in \cite{Vaknin:2014fxa}. The qualitative features that we find here are generic to models where SUSY is broken through the interplay of $F$- and $D$-term effects. 

The field content and superpotential are the same as those introduced in \eqref{Orafe_W}. The gauging of the $U(1)_D$ symmetry contributes new terms in the scalar potential from the $D$-term contribution. The $F$- and $D$-term contributions to the potential together give
\bea 
V_F+V_D &=&
|F-\lambda \phi_1 \tilde \phi_2|^2+|\lambda X \tilde \phi_2+m \tilde \phi_1|^2+ 
|\lambda X \phi_1+m  \phi_2|^2 + |m \phi_1|^2+ |m\tilde \phi_2|^2+ \nonumber \\
&&+\frac{g^2}{2} \left(\frac{D}{g} + |\phi_1|^2-|\tilde \phi_1|^2+|\phi_2|^2-|\tilde \phi_2|^2 \right)^2\ ,
\eea
where $g$ is the gauge coupling of the $U(1)_D$ symmetry and we have also included a UV Fayet-Iliopoulos (FI) term $ D/g$. This FI term contributes a second source of SUSY breaking that will strengthen the GW signal. Note that the model contains, in addition to the O'Raifeartaigh degrees of freedom, a gauge boson and gaugino associated with the $U(1)_D$ vector multiplet. We focus on the regime where $ y_F\lesssim 1 $
and we do not discuss the origin of the FI term here.\footnote{The inclusion of a fundamental FI term is not strictly required to obtain a strong FOPT. A very similar potential for the pseudomodulus can be obtained by considering two different masses for the messengers and working in the regime where $\lambda F > m_1 m_2$. Models with multiple $F$-terms would also lead to similar conclusions.\label{footnote}}


\paragraph{The scalar potential at zero temperature}
As a first step, we analyze the zero-temperature vacuum structure and map it onto the parameterization of Section \ref{sec:anatomy}. Neglecting the gauge dynamics, the tree-level potential has a minimum at $\phi_i = \tilde \phi_i = 0$ where SUSY is broken everywhere along the $F$-flat pseudomoduli space parameterized by $X$. Including the gauge interactions, the minimization of the $D$-term part of potential favors configurations where the $\tilde \phi_i$ fields acquire a VEV to compensate for the FI term $D/g$. This results in a tension between the minimization of the $F$-term and $D$-term contributions to the potential. While the $F$-term can never be set to zero, one can find a runaway direction in field space which leads, asymptotically, to the vanishing of the $D$-term. 

\begin{figure}[t!]
  \centering
  \includegraphics[width=0.48\textwidth]{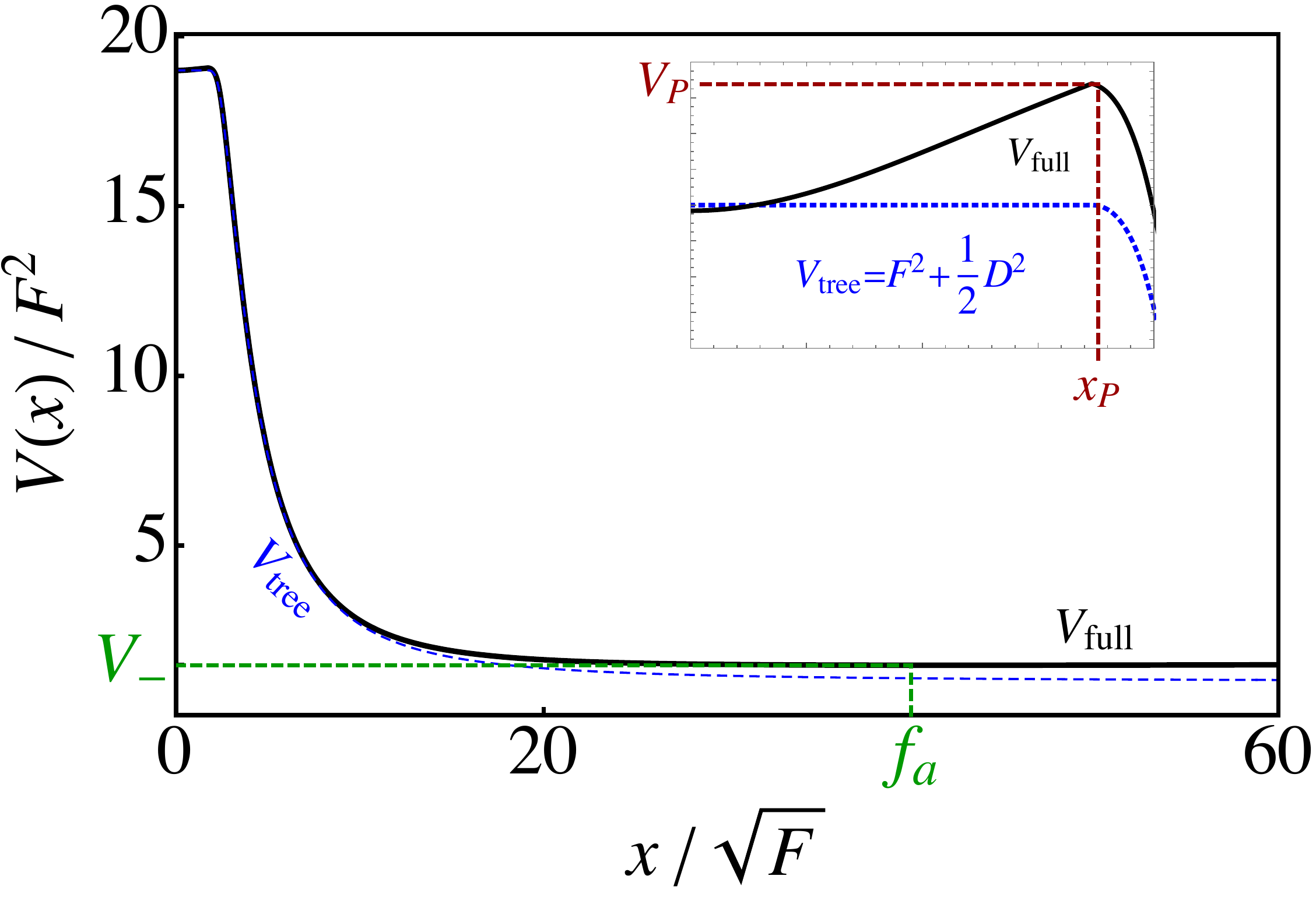} ~~~
 \includegraphics[width=0.48\textwidth]{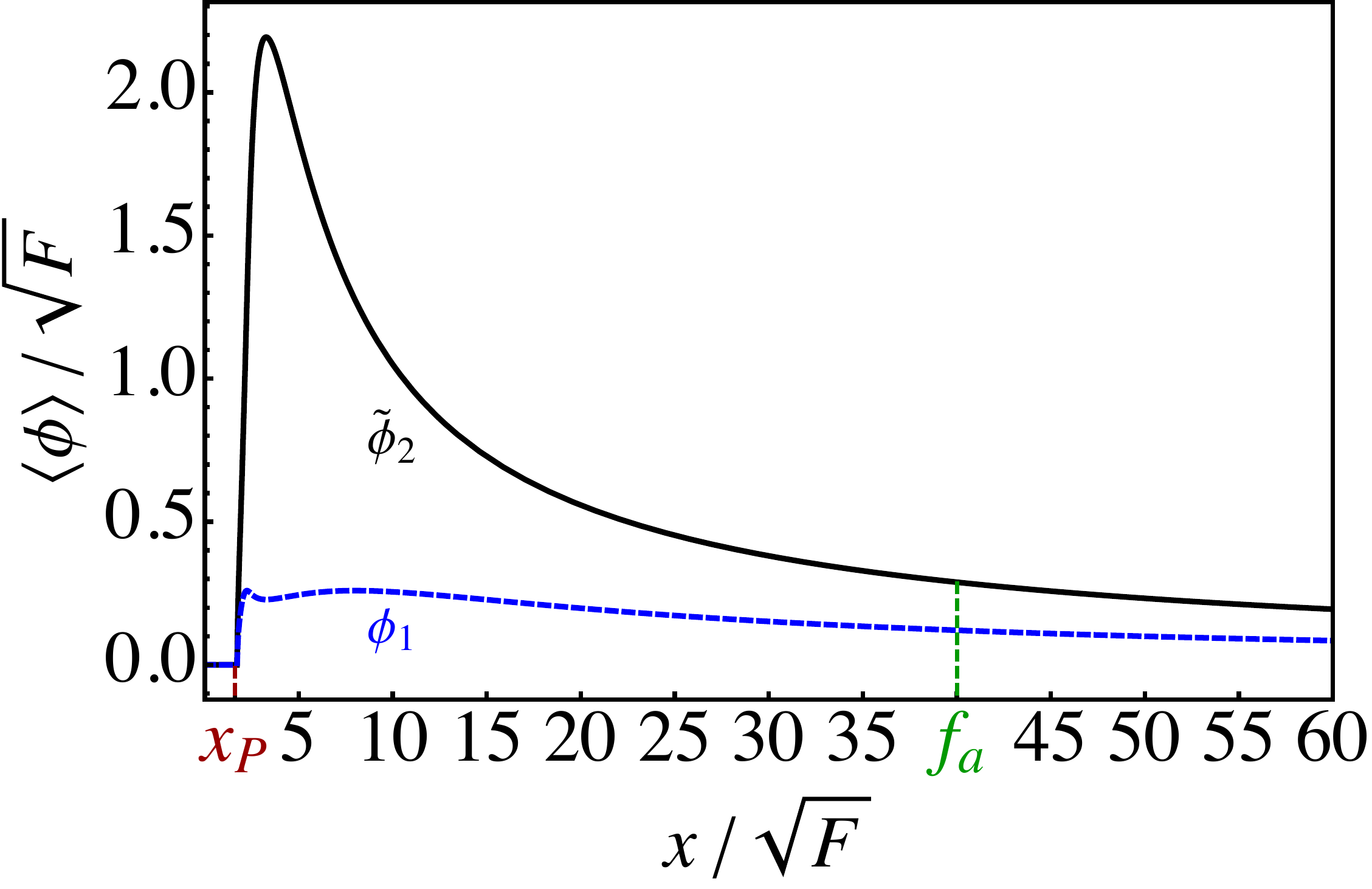}
\caption{
{\bf Left:} Tree level and one-loop scalar potential as a function of the pseudomodulus direction $x$ minimizing the directions $\phi_1$ and $\tilde \phi_2$. The {\bf dashed blue} line shows the tree level potential which is flat around the origin and develops a runaway at $x_P\simeq x_{\text{trans}}$ (see Eq,~\eqref{eq:VEVtrans}). Quantum corrections generate a local minimum at the origin as shown by the {\bf black solid} line in the small quadrant and a global minimum far away in field space indicated with a {\bf green dashed} line. The difference in energy density is $\Delta V\simeq \frac{1}{2}D^2$. {\bf Right:} The VEVs of the fields $\phi_1$ and $\tilde \phi_2$ while moving along the $x$-direction. Interestingly, both VEVS increase only at the barrier and they are otherwise quite small compared to $\sqrt{F}$.  For reference, the benchmark used in both plots has $(F=1\, , m=2\, , D=6\, , \lambda=2.9\, , g=1)$.
\label{fig:Vtree_one}
}
\end{figure}

First, we can solve for the $F$-terms of $\Phi_1$ and $\tilde \Phi_2$ by taking 
\be
\tilde \phi_1 = -\frac{\lambda}{m} X \tilde \phi_2\quad , \quad \phi_2 = -\frac{\lambda}{m} X \phi_1\ .
\ee
On this solution the scalar potential simplifies to 
\begin{align}
V &= 
|F^2-\lambda \phi_1 \tilde \phi_2|^2
+|m \phi_1|^2+|m \tilde \phi_2|^2\\
&+\frac{g^2}{2}  \left[
|\phi_1|^2 \left(\frac{\lambda^2 |X|^2}{m^2}+1\right)
-|\tilde \phi_2|^2 \left(\frac{\lambda^2 |X|^2}{m^2}+1\right) + \frac{D}{g} \right]^2\ .\label{eq:VtreeFD}
\end{align}
Note that $\phi_1,\tilde \phi_2$ have vanishing $R$-charge, so the only direction where the $R$-symmetry is spontaneously broken is along $x$. In order to visualize the shape of the scalar potential and the approach to the runaway, we show in Fig.~\ref{fig:Vtree_one} the tree level scalar potential as a function of $x$, as well as the values of $\phi_1$ and $\tilde \phi_2$ as a function of $x$. 

The scalar potential is flat around the origin and then turns to the runaway direction along which the $D$-term diminishes. The turning point along $x$ is where
 the fields $\phi_1$ and $\tilde \phi_2$ acquire a non-vanishing VEV. The VEV of $\phi_1$ is different from zero since the potential energy is most efficiently minimized if $\phi_1$ partially cancels the first term in \eqref{eq:VtreeFD} as well as minimizing the $D$-term. 
The VEV of $\phi_1$ is suppressed by a factor $\sim \frac{g F}{\lambda D}$ with respect to the VEV of $\tilde \phi_2$. An analytic estimate of the scalar potential can then be captured by working at zeroth order in the VEV of $\phi_1$.
In this approximation, and focusing on the parameter region where $g D/m^2<1$, the effective mass-squared for the $\tilde \phi_2$ field is $X$-dependent, 
\be
m_{\tilde \phi_2}^2= 
m^2 - g D -\frac{ \lambda^2 g D }{2m^2} x^2 
\qquad \Rightarrow \qquad x_{\text{trans}}^2 \simeq 
\frac{2 m^2 (m^2-g D)}{ \lambda^2 g D }\label{eq:VEVtrans}
\ee
and turns negative at the transition point $x_{\text{trans}}$ where the field $\tilde \phi_2$ develops a VEV.


The potential for $x$ is flat for $x \leq x_{\text{trans}}$ , while for  $x \geq x_{\text{trans}}$ it can be obtained by integrating out $\tilde \phi_2$,
\be
V_{\text{tree}}(x) \simeq 
\begin{cases} & 
F^2+\frac{1}{2}D^2= V_{+}
\qquad \qquad \qquad \qquad \qquad  x< x_{\text{trans}}\\
&
F^2+\frac{1}{2}D^2
-\frac{\lambda^4 D^2 \left(  x^2-x_{\text{trans}}^2 \right)^2}{2(2m^2 + \lambda^2 x^2)^2}
\qquad \qquad~~
x> x_{\text{trans}}
\end{cases}\ . \label{eq:Vtreeapprox}
\ee

Since we work in the small-$g$ regime, radiative corrections from the gauge sector may be neglected, such that the 1-loop corrections are the same as the ones discussed in the previous sections (see Eq.s~\eqref{V_1loop_approx}
and \eqref{V_1loop_large}).
They have two effects, namely i) they create a local minimum at the origin, and ii) they generate a global minimum at large $x$ values along the $D$-flat direction.

The barrier between the two vacua is approximately at $x\simeq x_{\text{trans}}$ where we can estimate the one-loop potential simply 
by the large field behaviour in \eqref{V_1loop_large}, giving
\be
V_P -V_{+} \simeq \frac{\lambda^2 F^2}{16 \pi ^2} \log\left( \frac{ x_{\text{trans}}^2}{m^2}\right) \qquad , \qquad  x_{P} \simeq x_{\text{trans}}\ .\label{eq:VPFD}
\ee
Combining the approximate tree level potential in \eqref{eq:Vtreeapprox} with the loop corrections in \eqref{V_1loop_large}, we find that the true vacuum at large field values
lies at
\be
\langle x \rangle_{\text{true}} =f_a \simeq \frac{4 \sqrt{2} \pi}{\lambda y_F} \sqrt{\frac{D}{g}}\quad ,\quad \Delta V \simeq \frac{1}{2}D^2\ ,\label{eq:truevacFD}
\ee
where the difference in potential energy between the two minima is dominated by the $D$-term contribution. This completes the matching of the potential of this model to the general discussion of Section \ref{sec:anatomy}. Note that here the SUSY-breaking $F$-term controls the height of the barrier in Eq.~\eqref{eq:VPFD}, while the SUSY-breaking $D$-term sets the potential energy difference as in Eq.~\eqref{eq:truevacFD} . This implies that the phase transition can have sizable values of $\alpha$, as we will see in the numerical analysis.

\begin{figure}[htbp]
	\centering
		\includegraphics[width=1\textwidth]{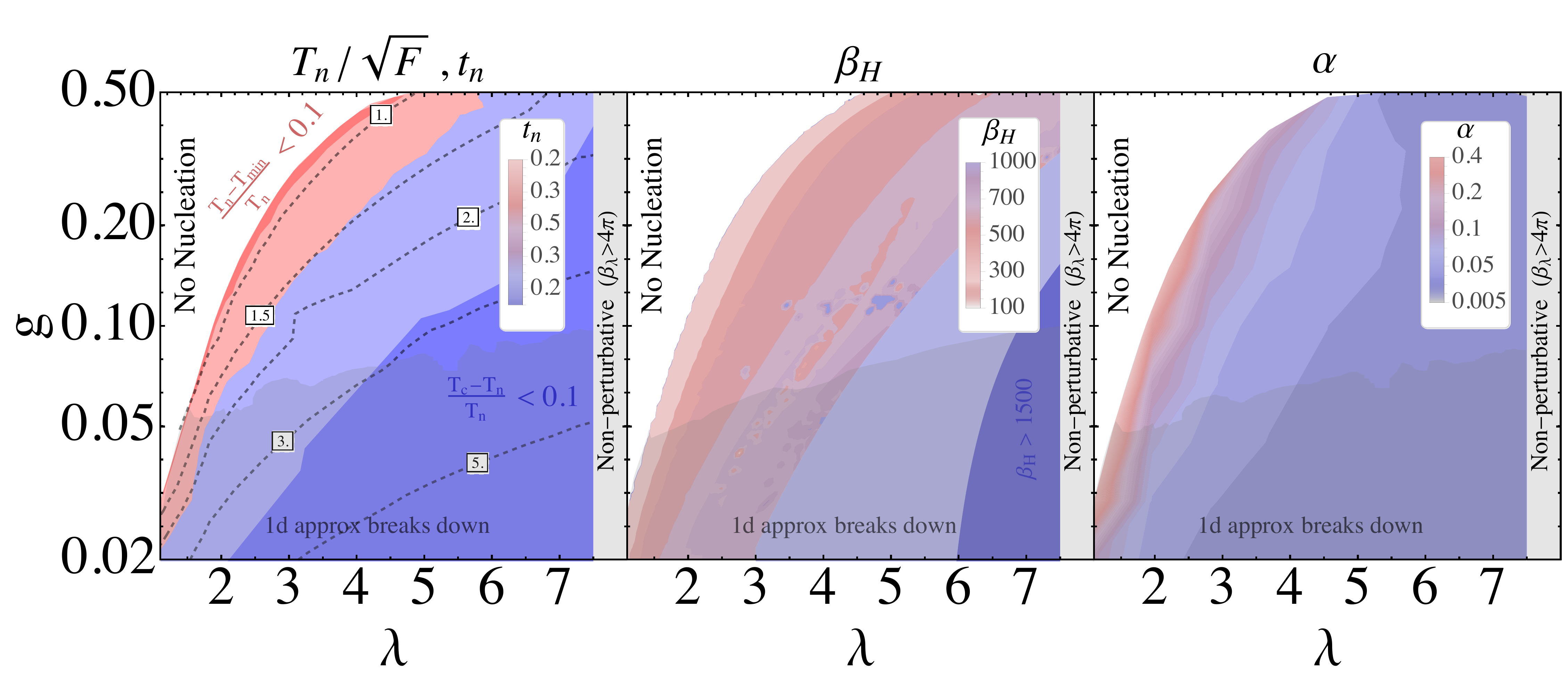}
		\caption{From the left to the right we show the behavior of $T_n/\sqrt{F}$, $\beta_H$ and $\alpha$ in the O'Raifeartaigh model with gauge interactions described in Eq.~\eqref{eq:VtreeFD}. We fix $F=30\text{ PeV}$, $y_F=3/4$ and $y_D=1/5$ so that the entire parameter space of the model can be shown in the $(\lambda,g)$ plane. The {\bf black dashed contours} in the left plot show $T_n/\sqrt{F}$. The {\bf red-to-blue} gradients show contours of $t_n$ as defined in Eq.~\eqref{eq:tndef} (left), of $\beta_H$ as defined in Eq.~\eqref{eq:beta} (center) and of $\alpha$ as defined in Eq.~\eqref{eq:alpha} (right). The GW signal weakens going from red to blue.  In the {\bf gray shaded} region at the bottom $R_{1d/3d}>0.5$ and as described in Eq.~\eqref{eq:1dvs3d} our $1d$ approximation is expected to break down. The {\bf grey} region on the left is excluded by the perturbativity of $\lambda$ below $m$. In the {\bf white} region the nucleation condition in Eq.~\eqref{eq:Tn} cannot be satisfied.}\label{fig:FDgvslambda}
			\end{figure}

\subsubsection{First order phase transition dynamics}

We now turn to the finite-temperature corrections and compute the parameters associated with the phase transition.
Note that the spectrum is similar to the O'Raifeartaigh model with the addition of the gauge boson and the gaugino of the $U(1)_D$ symmetry.
These additional states are massless in the false vacuum and massive in the true vacuum, so they contribute to making the origin the global minimum at high temperatures. 

We numerically evaluate the one-loop and the thermal corrections to the scalar potential, and then 
compute the bounce action for tunneling from the false vacuum to the true vacuum. In Appendix~\ref{ssub:triangular_bounce_for_the_} we present the triangular barrier approximation for this model and compare it with the full numerics. Even though the bounce profile in field space involves three different fields, i.e. $(x, \phi_1,  \tilde \phi_2)$, in our numerical scan, we approximate the bounce as one-dimensional, neglecting the contribution from the $\phi_1,  \tilde \phi_2$ directions. As detailed in Appendix~\ref{sub:single_field_approximation_for_the_multi_field_problem_}, we checked the single field approximation against the full $3d$ bounce action computed numerically with both FindBounce~\cite{Guada:2020xnz} and CosmoTransitions~\cite{Wainwright:2011kj}. As a result, the single field approximation gives a good description of the bounce as long as $\phi_1,  \tilde \phi_2$ are smaller than $X$ at the bounce release point (defined as the starting point of the tunneling set at $r=0$, where the kinetic terms of all the fields are exactly zero). In order to estimate where we expect sizable deviations from the multidimensional contribution, we borrow some intuition from the triangular barrier approximation, where $S_3/T$ scales as $\sim X^3$, and define
\begin{equation}\label{eq:1dvs3d}
R_{1d/3d}
\defeq
\left.\frac{X^3(r)}
{\left(X^2(r)+\phi_{1}^2(r)+\tilde \phi_{2}^2(r)\right)^{3/2}}\right\vert_{r=0}\, ,
\end{equation}
where $X(0)$, $\phi_{1}(0)$ and $\tilde \phi_{2}(0)$ are the field distances from the origin computed at the release point $r=0$.  In Fig.~\ref{fig:FDgvslambda} we show the region where $R_{1d/3d}>0.5$ and we expect deviations of $50\%$ or more from our one-dimensional estimate of the bounce action. As we can see, this region is not phenomenologically relevant since it is quite far from the interesting region for GW signals.

\begin{figure}[htbp]
	\centering
\includegraphics[width=1\textwidth]{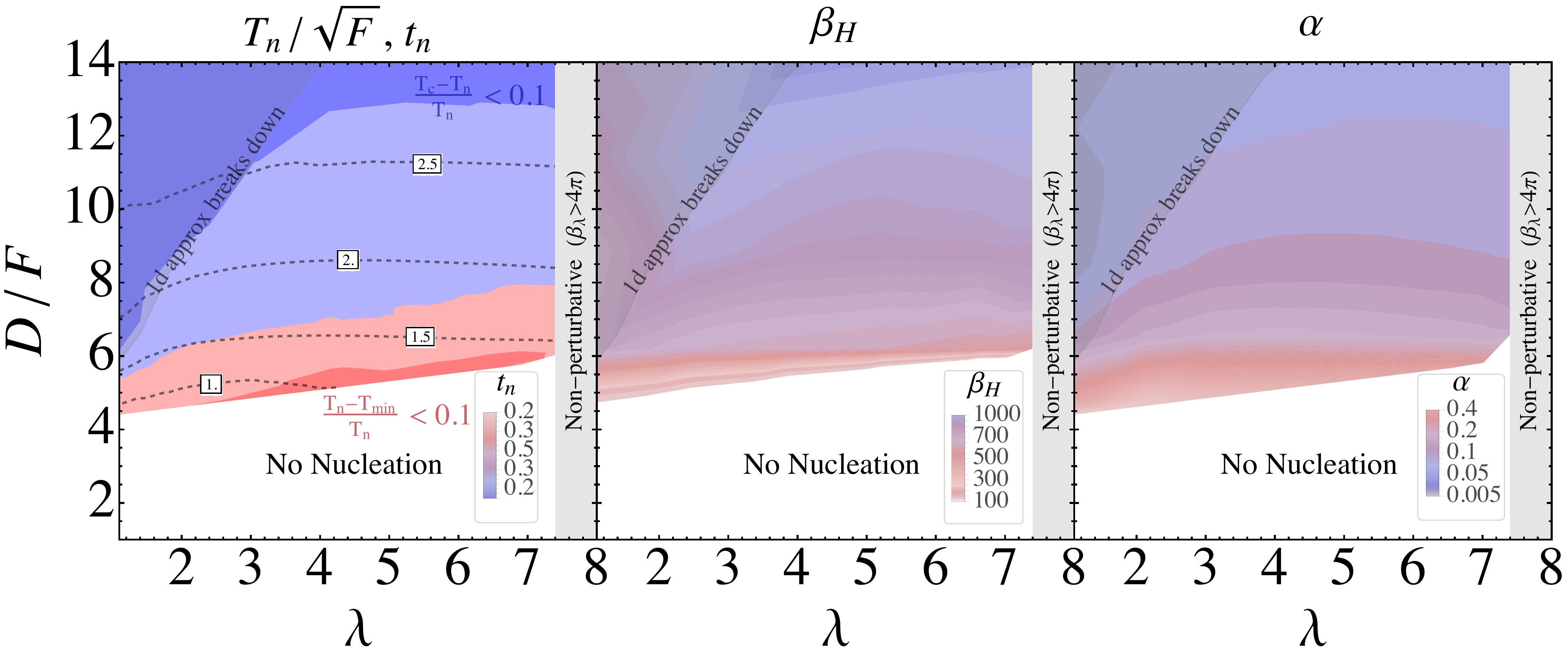}
\caption{Same as in Fig.~\ref{fig:FDgvslambda} but in the $(\lambda, D/F)$ plane, fixing $F=30\text{ PeV}$, $y_F=3/4$ and $g=0.1$.}\label{fig:FDDvslambda} 
\end{figure}
The parameter space of this model can be explored at fixed $F$, after fixing the two ratios 
\begin{equation}
y_F \defeq \frac{\lambda F}{m^2}\quad ,\quad y_D \defeq\frac{g D}{m^2}\ .\label{eq:ratiosFD}
\end{equation}
In Fig.~\ref{fig:FDgvslambda} we show the behavior of $T_n$, $\alpha$, and $\beta_H$ in the $(\lambda, g)$ plane, having fixed $\sqrt{F}=30$ PeV and $y_F=3/4$ as in the previous model and set $y_D=1/5$. Keeping fixed the ratios in Eq.~\eqref{eq:ratiosFD}, the triangular barrier parameters scale as 
\begin{equation}
f_a \sim \frac{1}{g \sqrt{\lambda}}\quad ,\quad\frac{\Delta V}{F^2} \sim \frac{\lambda^2}{g^2}\quad ,\quad \frac{V_P}{F^2} \sim \lambda^2\quad ,\quad \frac{m_{*}}{\sqrt{F}} \sim \sqrt{\lambda} \ .
\end{equation}
As a consequence of these scalings, using Eq.~\eqref{S3oTsimp} it is straightforward to see that for fixed $\lambda$ the boundary of the nucleation region is reached for large $g$, while for fixed $g$ the boundary lies at small $\lambda$. The shape of the nucleation temperature $T_n$ can be captured by a simple analytic formula after rewriting Eq.~\eqref{Tnucl_1} in terms of the theory parameters:
\be
T_{n} \sim 0.73 \sqrt{\lambda F} \frac{1}{\log\left( 1 + \frac{22.5}{\mathcal{C}^{4/21}} \left(\frac{g}{\lambda }\right)^{4/7} \right)} \left(1 - 20.5 \left(\frac{g^6}{\mathcal{C}^2 \lambda^6} \right)^{1/5} \right)\ .
\ee
This expression reproduces the contours in Fig.~\ref{fig:FDgvslambda} (left) up to overall normalization.

In the middle and right panels of Fig.\ref{fig:FDgvslambda} we show the contours for $\beta_H$ and $\alpha$, respectively. The main difference compared to the model in Sec.~\ref{sec:X3} is that even if $T_n \sim \sqrt{F}$, it is possible to obtain sizable values of $\alpha$ because $\Delta V$ is here controlled by the $D$-term. Approaching the boundary of the nucleation zone without fine-tuning the theory parameters by more than $\mathcal{O}(1)$ we can reach $\beta_H\sim100$ and $\alpha\sim 0.3-0.4$, which we use as a benchmark for our summary plot in Fig.~\ref{fig:moneyplot}. 

The interplay of the two SUSY-breaking scales $\sqrt{F}$ and $\sqrt{D}$ is an essential ingredient for a strong FOPT. This is illustrated in Fig.~\ref{fig:FDDvslambda}, where we show the behavior of $T_n$, $\alpha$, and $\beta_H$ in the $(\lambda, D/F)$ plane, having again set $\sqrt{F}=30$ PeV and $y_F=3/4$, and now fixing $g=0.1$. In this scaling the FOPT is essentially independent of $\lambda$, and one can see clearly that the separation of $D$ from $F$ is the crucial ingredient for a sufficiently strong phase transition. Notice that the required separation is $\mathcal{O}(1)$ and therefore not obviously in tension with theoretical bounds on large $D$-terms \cite{Dumitrescu:2010ca}.  Strictly speaking, these bounds do not  apply to our simple model, where a tree level Fayet-Iliopoulos term makes the Ferrara-Zumino multiplet not gauge invariant~\cite{Komargodski:2009pc}. However they would have applied if we were to UV complete this model to a full-fledged model of dynamical SUSY-breaking or for instance if we were to explore the second branch of the model with two different messengers masses and $\lambda F > m_1 m_2$.  

\subsubsection{Gravitational Wave spectrum and phenomenology}

\begin{figure}[htbp]
	\centering
\includegraphics[width=0.95\textwidth]{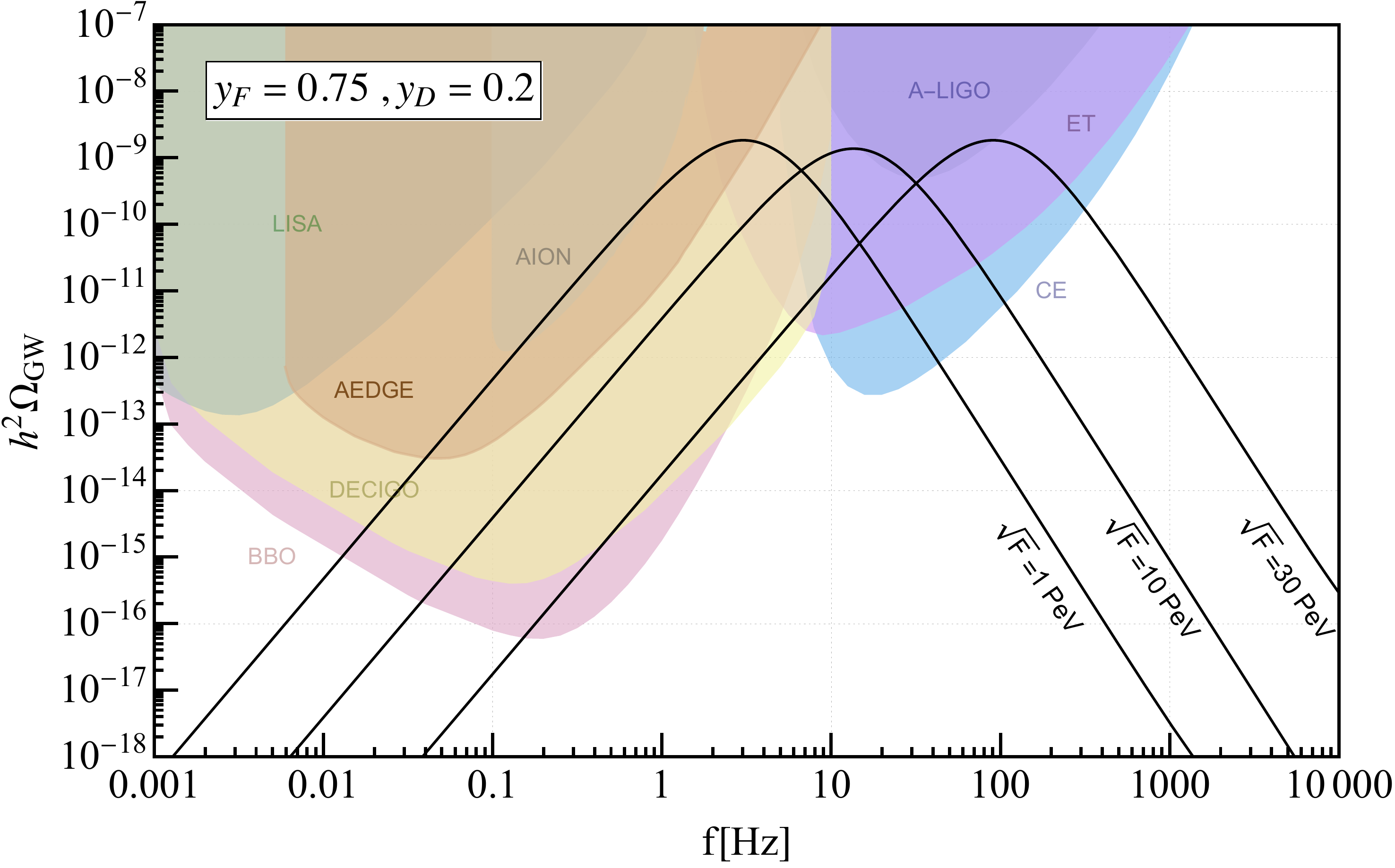}
	\caption{Predicted SGWB in the O'Raifeartaigh model with gauge interactions. We show the prediction for various values of the SUSY-breaking scale $\sqrt{F}=1,10,30\text{ PeV}$ and choose the theory parameters such that $\alpha\simeq0.3$ and $\beta_H\simeq50$. As discussed in the text, achieving these values does not require any tuning in this model. The SUSY-breaking scale correlates with the peak frequency of the GW spectrum, which is always dominated by sound waves as shown in Eq.~\eqref{eq:alwaysstop}.}
	\label{fig:everybodyFD}
\end{figure}

Having shown that a strong FOPT can be achieved without fine-tuning in a SUSY-breaking hidden sector with at least two SUSY-breaking scales, we now turn to the gravitational wave signal itself, again using the simple model presented in the previous section as a benchmark. As discussed in Sec.~\ref{sec:FOPTandGWs}, computing the SGWB signal requires understanding the macroscopic dynamics of the vacuum bubbles expanding in the plasma. This is essentially determined by the balance of the energy $\Delta V$ released in the FOPT and the pressure effects from the plasma (see Eq.~\eqref{eq:pressure}). If pressure effects stop the bubbles before they collide, most of the SGWB signal will be sourced by the energy released in the plasma.  

In Sec.~\ref{sec:FOPTandGWs} we described a quite unique friction mechanism at work in our class of models. This mechanism is a direct consequence of two peculiar features of the pseudomodulus potential: i) the nucleation temperature $T_n\sim\sqrt{F}$ is set by exponentially suppressed temperature corrections to be smaller than the typical scale of the heavy states in the theory $m$, and ii) the true vacuum VEV is typically the larger scale in the problem and controls the mass variation $\Delta m^2/m^2\sim \lambda^2f_a^2/m^2\gg1$ of the heavy states from the false to the true vacuum. These two properties together imply that when the vacuum bubbles accelerate enough, $\gamma T_n>m_{\text{true}}$ and the heavy states can cross the bubble wall. Their crossing switches on a new pressure effect which is generically larger than $\Delta V$ and immediately stops the bubble runaway. 

This last statement can be checked explicitly with the parametric dependence of the simple model described here. The heavy state pressure term in Eq.~\eqref{eq:heavypressure} scales as
\begin{equation}
\Delta P_{\text{LO}}^{\text{heavy}}\sim\frac{4\pi^2}{3 y_F^2 g} F D e^{-m_{\text{false}}/\sqrt{F}}\ ,
\end{equation}
where we used the scaling of the true vacuum as a function of the theory parameters in Eq.~\eqref{eq:truevacFD} and approximated $T_n\simeq \sqrt{F}$ for simplicity (this approximation is numerically correct up to an $\mathcal{O}(1)$ factor as shown by the dashed contours in Fig.~\ref{fig:FDgvslambda}). Inside the exponential, we should take the lightest heavy states in the plasma $m_{\text{false}}\sim\sqrt{m^2-\lambda F}$ which are of course less Boltzmann suppressed and dominate the friction. Comparing this quantity with the energy released in the phase transition $\Delta V= D^2/2$ we can get the range of the gauge coupling $g$ such that this friction prevents the bubble runaway,
\begin{equation}
g\lesssim \frac{8\pi^2}{3 y_F}\frac{F}{D} e^{-m_{\text{false}}/\sqrt{F}}\ .\label{eq:alwaysstop}
\end{equation} 
Plugging in the typical numbers for our phase transition ($F/D\sim 1/5$, $y_F\sim3/4$ and $m_{\text{false}}/\sqrt{F}\lesssim\sqrt{\lambda}\lesssim 2.5 $) indicates that the vacuum bubbles are always stopped in the range of interest for the gauge coupling $g$ for perturbative values of $\lambda$. The predicted boost factor at equilibrium in this case is
\begin{equation}
\gamma_{\text{eq}}^{\text{heavy}}= \frac{m_{\text{true}}}{T_n}\sim \frac{\lambda f_a}{\sqrt{F}}\ ,
\end{equation}
where again $f_a$ is defined in this model by Eq.~\eqref{eq:truevacFD}. As a final remark, we notice that the NLO friction induced by gauge degrees of freedom radiated through the wall never dominates over the one from heavy states in the interesting range of the gauge coupling~$g$. 

Given that the bubble runaway is always prevented, the dominant SGWB comes from sound waves in the plasma. The predicted energy fraction as a function of frequency at GW interferometers has been discussed in Eq.~\eqref{eq:swsignal} and below. Putting everything together, in Fig.~\ref{fig:everybodyFD} we compare our model predictions with the PLI curves for future GW interferometers derived in Appendix~\ref{ssub:pli_curves}. This clearly demonstrates that SUSY-breaking hidden sectors with multiple SUSY-breaking scales can generate stochastic signals detectable at future GW interferometers. Moreover, it makes explicit the expected correlation between the SUSY-breaking scale and the peak frequency of the resulting SBGW. All that remains is to explore the full range of viable SUSY-breaking scales (and hence signal frequencies), as well as the correlation between signals at GW interferometers and other experiments. In the next section, we will bound the SUSY-breaking scale from above around  $\sim$few tens of PeV by computing the gravitino cosmological abundance. By specifying a mediation mechanism, we will also use the explicit hidden sector presented here to show how the SUSY-breaking scale determines the spectrum of MSSM superpartners, thereby correlating signals at GW interferometers and future colliders.

 \section{Phenomenology}\label{sec:pheno}
Having demonstrated that the first-order phase transition in a SUSY-breaking hidden sector can generate an observable GW signal, we now turn to complementary aspects of hidden sector phenomenology that shape the motivated parameter space and suggest additional experimental tests in the event of a signal at GW interferometers. We begin with universal features that are intrinsic to the hidden sector itself and independent of the mediation mechanism that connects the hidden sector to the MSSM. This includes key aspects of gravitino cosmology, where we will see that the requirement $T_{\text{r.h.}}=\sqrt{F}$ implies an upper bound on $\sqrt{F}$ even if $m_{3/2}$ receives extra contributiosn from other SUSY-breaking sectors as in Eq.~\eqref{eq:m32}. We also explore the prospects for collider searches for the gravitino (independent of the MSSM spectrum), finding that future high energy lepton colliders could probe almost the entirety of the light gravitino window (i.e. $m_{3/2}<16\text{ eV}$) by directly producing gravitino pairs. We then relate the parameters of the hidden sector to the spectrum of the MSSM, which requires specifying details of the mediation mechanism. Here we consider the prototypical example of gauge mediation via vector-like messengers, where the parameter space for observable GW signals generates a superpartner spectrum within reach of future proton-proton colliders such as FCC-hh.

\begin{figure}[t!]
  \centering
 \includegraphics[width=1\textwidth]{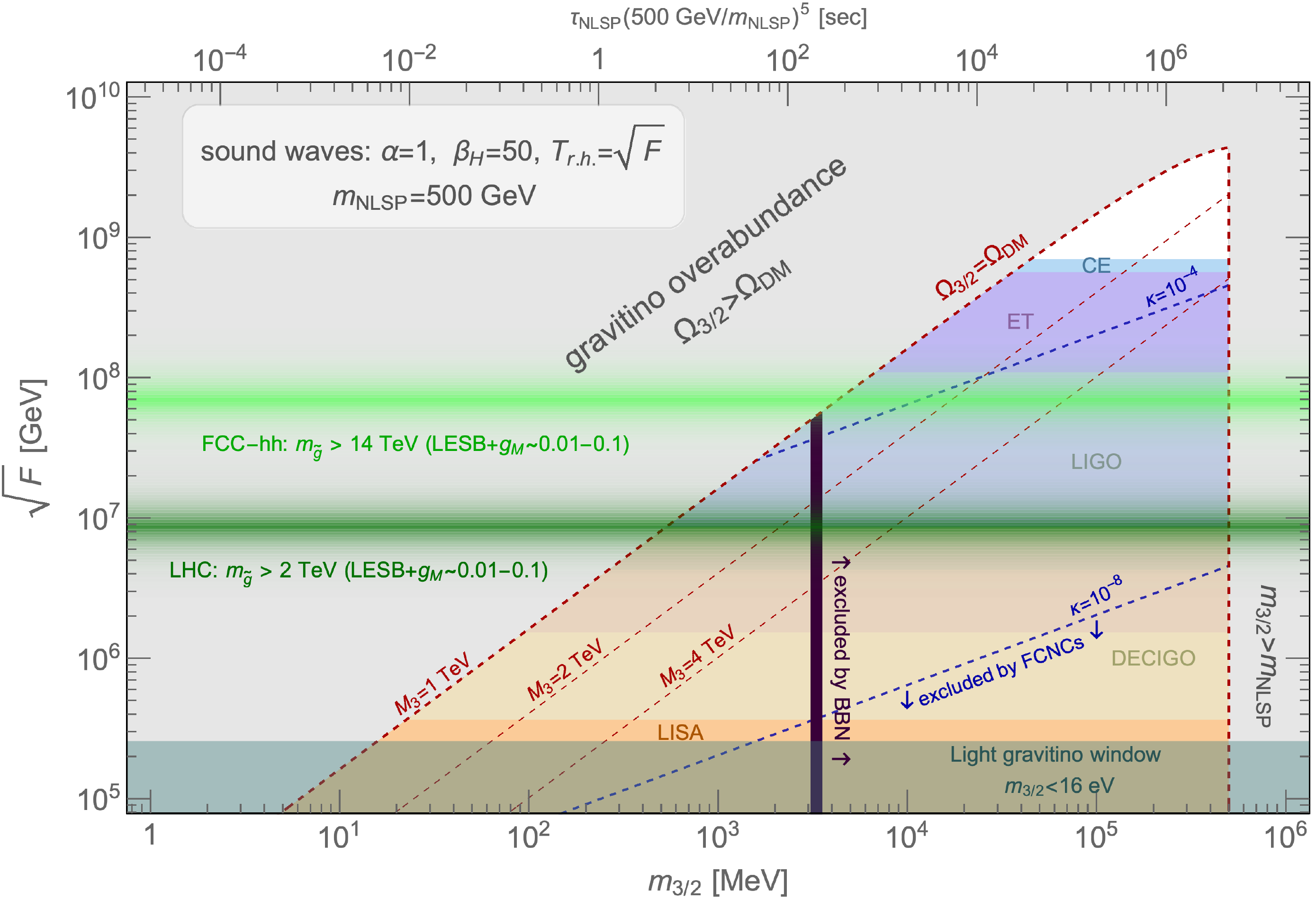}
\caption{Parameter space of low energy SUSY-breaking in the $(m_{3/2},\sqrt{F})$ plane. The {\bf gray shaded} region is excluded by gravitino overabundance and the requirement $m_{3/2}<m_{\text{NLSP}}$, having fixed $m_{\text{NLSP}}=500\text{ GeV}$. The red dashed line shows the region where $\Omega_{3/2}=\Omega_{\text{DM}}$ for different values of the gluino soft mass. The colored regions show the sensitivities of different GW interferometers to signals with fixed $\alpha=1$ and $\beta_H=50$. The two {\bf dark green} and {\bf light green} bands show the impact of the present LHC bounds~\cite{Aaboud:2018doq,Aaboud:2018mna,ATLAS:2019vcq,ATLAS-CONF-2020-047} and the future FCC-hh reach on gluinos~\cite{Arkani-Hamed:2015vfh} for perturbative messenger sectors with $g_M\in(0.01,0.1)$ (see Eq.~\eqref{eq:gluinomass} for a definition of $g_M$). The {\bf dark blue dashed} lines show the values of $\kappa=F/F_0$, the ratio between the total SUSY-breaking scale $F_0$ and one controlling the soft masses (see Eq.~\eqref{eq:m32}). As discussed in Eq.~\eqref{eq:FCNCvskappa}, we expect constraints on flavor changing neutral currents to exclude $\kappa\lesssim10^{-8}$  as indicated by the {\bf dark blue arrows}. The {\bf dark magenta thick} line indicates the BBN bound on the higgsino NLSP decaying to gravitino plus hadrons as obtained in~\cite{Jedamzik:2006xz}.}\label{fig:gravitinoproblem}
\end{figure}

\subsection{Gravitino cosmology vs future colliders} 
The gravitino overabundance is a well known problem of LESB scenarios~\cite{Moroi:1993mb,Kawasaki:1994af,Moroi:1995fs}. This problem is exacerbated in our setup, because having sizeable GW signals from the SUSY-breaking hidden sector requires the reheating temperature $T_{\text{r.h.}}$ to be at least as high as the SUSY-breaking scale, enhancing the gravitino production from scattering as detailed in Eq.~\eqref{eq:m32abundance}. In light of this tension, here we delve into further detail about the two viable scenarios sketched Sec.~\ref{sec:summaryparspace}. Since $T_{\text{r.h.}}\sim \sqrt{F}$, which is much larger than the scale of the soft masses, the main player in determining the final gravitino abundance is the production from UV scattering computed in~\cite{Bolz:2000fu,Pradler:2006qh,Pradler:2006hh,Rychkov:2007uq}. The final yield can be written as
\begin{equation}
Y_{3/2}^{\text{UV}}=C_{\text{UV}}\frac{M_3^2 \sqrt{F}}{m_{3/2}^2 M_{\text{Pl}}}\quad , \quad C_{UV}=\frac{45 \sqrt{5}f_3}{8\pi^{13/2}g_*^{3/2}}\simeq 4\times 10^{-5}\ ,
\end{equation}
where the production through gluon-gluino scattering dominates over the other channels and we have substituted $T_{\text{r.h.}}\simeq \sqrt{F}$, which is the lowest reheating temperature compatible with our scenario. Following Eq.~\eqref{eq:m32}, we assume that the gravitino mass $m_{3/2}$ is set by an independent SUSY-breaking scale $F_0=F/\kappa $, possibly higher than the one setting the soft spectrum (i.e. $\kappa\ll1$).

\paragraph{Ultralight gravitino window vs.~pair production at future colliders}
If the gravitino mass and the soft spectrum are set by the same SUSY-breaking scale $\sqrt{F}$, the yield scales as $Y_{3/2}^{\text{UV}}\sim M_{\text{Pl}}/\sqrt{F}$. For sufficiently low SUSY-breaking scales, the yield becomes just the equilibrium one, $Y_{3/2}^{\text{UV}}>Y_{\text{eq}}$, where $Y_{\text{eq}}=n_{3/2}^{\text{eq}}/s=1.8\times 10^{-3}$. The gravitino is a thermal relic as long as $\sqrt{F}\lesssim\left(\frac{3}{45} M_{\text{Pl}}M_3^2\right)^{1/3}$, which corresponds to $\sqrt{F}\lesssim8.6\times 10^7\text{ GeV}$ for $M_3=2\text{ TeV}$. Moreover, since $\sqrt{F}\gg m_{3/2}$, the gravitino is relativistic at freeze-out and its abundance today is constrained by measurements of the matter power spectrum at short scales~\cite{Pierpaoli:1997im,Viel:2005qj}.  The current bounds imply 
\begin{equation}
m_{3/2}\lesssim16\text{ eV}\quad ,\quad F\lesssim260\text{ TeV}\ .
\end{equation}
The above requirement identifies the \emph{ultralight gravitino window}. Although it is unquestionably challenging to decouple the soft spectrum from the LHC in this window (see \cite{Hook:2015tra,Hook:2018sai} for attempts in this direction), it is interesting to ask whether future colliders can test this window in a model-independent fashion through direct pair production of the longitudinal component of the gravitino, the goldstino. This production rate depends directly on $\sqrt{F}$ even when the MSSM superpartners are decoupled, and so provides a direct experimental test of the SUSY-breaking sector.   

The projected sensitivity to gravitino pair production at both hadron and lepton colliders is displayed in Figure \ref{fig:moneyplot}. For the bound at future lepton colliders, we consider a high energy lepton collider operating at $\sqrt{s} = 30$ TeV. Assuming minimal cuts on the photon kinematics ($E_{\gamma}> 50$ GeV, $|\eta_{\gamma}|<2.4$),
the signal cross section from Eq.~\eqref{eq:directpairxsec} is
\be
\sigma_{\text{30TeV}} (\ell^+ \ell^- \to \tilde G \tilde G \gamma) \simeq 487 ~\text{fb} \left(\frac{10 \, \text{TeV}}{\sqrt{F}} \right)^8\ .
\ee
Applying the same minimal cuts on the photon, the SM background (estimated with MadGraph5 \cite{Alwall:2011uj,Alwall:2014hca}) 
is $\sigma_{SM}\simeq 2$pb. In contrast to LEP,  at high energy lepton colliders the SM background is dominated by $WW$ fusion while the Drell-Yan process with an ISR photon is negligible. We can then derive a lower bound on the scale of SUSY breaking as displayed in Figure \ref{fig:moneyplot} given an assumed integrated luminosity, namely
\be
\sqrt{F} \gtrsim 25 \, \text{TeV} \left( \frac{\mathcal{L}}{100 \, \text{ab}^{-1}} \right)^{1/16}\ ,
\ee
which is still an order of magnitude away from entirely closing the ultralight gravitino window. However, improved analyises and new cosmological data could strengthen the gravitino mass bound by an order of magnitude, potentially closing the ultralight gravitino window completely. For instance, Ref.~\cite{Osato:2016ixc} already claims a bound on the gravitino mass of $m_{3/2}<4.7\text{ eV}$; although the robustness of this bound is subject to interpretation, improved limits from Planck data are likely to be comparable.  
 
In order to estimate the reach of future hadron colliders, we perform a rescaling of the limits discussed in \cite{Maltoni:2015twa}, based on the 
mono-photon search of ATLAS \cite{Aad:2014tda},
which constrain $\sqrt{F} \gtrsim 850 $GeV with $20.3 $fb$^{-1}$ at $\sqrt{s}=8$ TeV.\footnote{The bounds from mono-jet searches are comparable \cite{Maltoni:2015twa}, but involve backgrounds from a mix of both quark- and gluon-initiated processes that are less amenable to simple rescaling. Thus we focus on the mono-photon signal for simplicity.}
As at lepton colliders, the signal cross section for gravitino pair production in association with a photon $\sigma(p p \to \tilde G \tilde G \gamma)$
scales as $s^3/F^4$ at hadron colliders \cite{Brignole:1998me}. To estimate the limit attainable at $\sqrt{s} = 100$ TeV, we first compute the ratio of the signal cross sections at $\sqrt{s} = 100$ TeV and $\sqrt{s} = 8$ TeV, taking the partonic signal cross section to scale as $\sigma_{\rm sig} \sim \hat s^3 / F^4$ and assuming that the $p_{T,\gamma} \geq 125$ GeV cut at $\sqrt{s} = 8$ TeV is increased to $p_{T,\gamma} \geq 1$ TeV at $\sqrt{s} = 100$ TeV. We additionally compute the ratio of background cross sections, assuming the partonic background cross section scales as $\sigma_{\rm bkg} \sim 1/\hat s$. Using the $\sqrt{s} = 8$ TeV signal and background predictions in \cite{Aad:2014tda} and the above ratios, we find the expected limit at $\sqrt{s} = 100$ TeV to be
\be
\sqrt{F} \gtrsim 12 \, \text{TeV} \left( \frac{\sqrt{s}}{100 \, \text{TeV}} \right)^{3/4} \left( \frac{\mathcal{L}}{30 \, \text{ab}^{-1}} \right)^{1/16}\ ,
\ee
which is the one displayed in Figure \ref{fig:moneyplot}.  Even with this aggressive estimate, the reach of high energy hadron colliders is limited compared to the reach of high energy lepton colliders because the signal cross section at the former only grows with $\hat s^3$, while at the latter it grows as $s^3$.

\paragraph{Gravitino Dark Matter window} If the SUSY-breaking scale $\sqrt{F_0}$ setting the gravitino mass exceeds the scale $\sqrt{F}$ of the hidden sector, we can treat $m_{3/2}$ as a free parameter and access an interesting region where the gravitino is never in thermal equilibrium with the SM. This could arise naturally from additional sequestered sectors that break supersymmetry at higher scales. As shown in Fig.~\ref{fig:gravitinoproblem}, we also require this new source of SUSY-breaking to not spoil the defining phenomenological features of LESB, namely i) the gravitino is still the LSP, and ii) the soft masses are dominated by the flavor-diagonal contribution from gauge mediation. 

Requiring the gravitino avoid thermalization, $Y_{3/2}^{\text{UV}}<Y_{\text{eq}}$, we obtain an upper bound on $\sqrt{F}$ at fixed gravitino mass which  can be cast as an upper bound on the ratio between the two SUSY-breaking scales, $\kappa=F/F_0$: 
\begin{equation}
Y_{3/2}^{\text{UV}}<Y_{\text{eq}}\quad \Rightarrow\quad \kappa< 0.02 \left(\frac{F}{10^7\text{ GeV}}\right)^{1/4}\left(\frac{0.1}{g_M}\right)\ ,
\end{equation}
Here we have used the expression for the gravtino mass in Eq.~\eqref{eq:m32} and the one for the gluinos in Eq.~\eqref{eq:gluinomass}, where the parameter $g_M$ encodes the model-dependence of the latter.  If the gravitino is never in thermal equilibrium, we can assume (as usual in freeze-in scenarios) that the gravitino sector is not directly reheated after inflation and the gravitino abundance is  frozen-in through scattering of SM states with their superpartners.  Setting the gravitino abundance to explain the DM abundance today, we can predict the gluino mass in the $(m_{3/2}, \sqrt{F})$ plane,  
\begin{equation}
M_{3}\simeq  2\text{ TeV}\left(\frac{10^7\text{ GeV}}{\sqrt{F}}\right)^{1/2} \left(\frac{m_{3/2}}{2.5\text{ GeV}}\right)^{1/2}\simeq  2\text{ TeV}\left(\frac{\sqrt{F}}{10^7\text{ GeV}}\right)^{1/2}\left(\frac{10^{-5}}{\kappa}\right)\ ,\label{eq:boundary}
\end{equation}
which corresponds to the red lines of Fig.~\ref{fig:gravitinoproblem} where the gravitino accounts for the total DM abundance today at fixed gluino mass. The current LHC bounds on the gluino mass set a boundary of our parameter space, which is shown in Fig.~\ref{fig:gravitinoproblem}.\footnote{Strictly speaking, the gluino mass here is the soft mass at computed at the high scale; since the low-scale pole mass will be larger, we generously show the parameter space up to $M_3=1\text{ TeV}$.} The second scaling in Eq.~\eqref{eq:boundary} shows the value of $\kappa$ required to achieve a given gluino mass. As shown in Fig.~\ref{fig:gravitinoproblem}, the parameter space of interest has $\kappa$ between $(10^{-8},10^{-4})$, where smaller values of $\kappa$ would not open up more parameter space and in any event would be in tension with FCNC constraints as discussed in Eq.~\eqref{eq:FCNCvskappa}. 

The bound at larger gravitino masses (the gray band on the r.h.s. on Fig.~\ref{fig:gravitinoproblem}) is given by the requirement that the gravitino be the LSP. A stronger bound is derived from BBN constraints on the freeze-out abundance of the NLSP decaying into gravitinos. We have computed the NLSP freeze-out abundance assuming the NLSP is a pure higgsino NLSP and applied the BBN bound of Ref.~\cite{Jedamzik:2006xz} given the NLSP lifetime in Eq.~\eqref{eq:NLSPdecay}. The triangular-shaped region where the gravitino could be DM can be probed by both GW interferometers and future colliders, as shown in Fig.~\ref{fig:gravitinoproblem}. This highlights the potential for future colliders to determine whether a SUSY-breaking phase transition is the source of a SGWB signal observed at GW interferometers.
    
\subsection{A complete model of gauge mediation}\label{sec:gaugemediation}  
Finally, we can correlate the GW signals of the SUSY-breaking hidden sector with the superpartner spectrum of the MSSM by specifying a mediation mechanism. In order to embed the model of Sec.~\ref{sec:FD} into a successful model of gauge mediation, we work in terms of a simple generalization of the gauged O'Raifeartaigh model which allows a natural embedding of both the gauged $U(1)_D$ symmetry and the SM gauge group into the flavor symmetry of the messengers. This requires $M$ copies of the vector-like messengers $\Phi$ and $\tilde \Phi$ coupled to the singlet $X$, so that the superpotential is identical to the one of the O'Raifeartaigh model in Eq.~\eqref{Orafe_W}, but where now the fields are intended as vectors with $M$ components. The minimal setup requires $M=6$ so that the superpotential enjoys an $SU(6)$ symmetry, where a $U(1)$ subgroup of the $SU(6)$ is the gauged $U(1)_D$ with a non vanishing Fayet-Ilipoulos term, while the SM gauge group lies inside the remaining global $SU(5)$ such that the messengers can be taken to transform in the $5 + \bar 5$ representation of $SU(5)$ as in standard gauge mediation scenarios~\cite{Giudice:1998bp}.   

The mass matrix of the messenger fields is 
\be
\mathcal{M}_{\text{mess}} = 
\left(
 \begin{array}{cc}
  \frac{\lambda f_a}{\sqrt{2}} & m \\
 m & 0 
\end{array}
\right)\ ,
\ee
where $f_a$ is the VEV of the pseudomodulus given in Eq.~\eqref{eq:truevacFD}. Integrating out the messengers, one can compute the soft masses for the MSSM following the general formulas in~\cite{Martin:1996zb}. The scalar masses follow the standard gauge mediation scaling discussed in Eq.~\eqref{eq:softmasses}, while it is worth explicitly writing the parametric dependence of the gluino soft mass in the notation of Eq.~\eqref{eq:softmasses}:   
\be
M_3= \frac{\alpha_3}{4 \pi} \frac{\sqrt{2}F}{f_a} s_M\quad ,\quad s_M=\frac{y_F^2}{6}\ .
\ee
Here we have expanded in $\lambda f_a \gg m \gtrsim F$ and identified the gaugino screening factor $s_M$ in this model. Since we typically have $y_F\sim 1$ in our scenarios (in order to remain in the green region of Fig.~\ref{fig:NoamOR}) the gaugino screening factor does not provide significant suppression, but interestingly it is generic for models like ours where the messengers mass matrix is never singular along the pseudomodulus direction~\cite{Komargodski:2009jf,Cohen:2011aa}. Abandoning this requirement, one could avoid gaugino screening at the price of opening up  messenger field directions where the SM gauge group is spontaneously broken in the UV~\cite{Riotto:1995am}.   

Substituting the value of $f_a$ in Eq.~\eqref{eq:truevacFD} and taking as benchmark values a typical point with $\alpha\sim 0.3$ and $\beta_H\sim 100$ from Fig.~\ref{fig:FDgvslambda}, the gaugino pole mass is   
\begin{equation}
m_{\tilde{g}}\simeq 2\text{ TeV}\left(\frac{F}{30\text{ PeV}}\right)^{1/2}\left(\frac{y_F}{0.75}\right)^3\left(\frac{F}{2.5 D}\right)^{1/2}\left(\frac{\lambda}{4}\right)\left(\frac{g}{0.4}\right)\ .
\end{equation}  
This shows that the band between the the present exclusion at the LHC and the future reach of FCC-hh can be populated with simple, concrete models featuring strong SGWB signals within the reach of future high-frequency interferometers such as A-LIGO, ET and CE.

\section{Conclusions}\label{sec:conclusions}

We began by asking if future gravity wave detectors could provide a new window into supersymmetry by probing SUSY-breaking hidden sectors in a region not yet excluded by LHC searches. The answer to this question is well summarized in Fig.~\ref{fig:moneyplot}, which shows the complementarity of future gravitational wave interferometers and colliders in probing scenarios of low-energy supersymmetry breaking (LESB). Fortuitously, the cosmological history of the gravitino -- a key degree of freedom in LESB scenarios -- bounds the SUSY-breaking scale from above, so that the viable parameter space lies within reach of both high-frequency GW interferometers and high-energy colliders. 

The underlying assumption in Fig.~\ref{fig:moneyplot} is that the SUSY-breaking hidden sector actually undergoes a strong first-order phase transition. The remainder of the paper has been devoted to demonstrating, on general grounds, the circumstances under which strong FOPTs can be produced in SUSY-breaking hidden sectors. We have focused on phase transitions along the pseudomodulus direction, which as a universal feature of spontaneous SUSY-breaking is guaranteed to exist in a vast class of SUSY-breaking hidden sectors. Remarkably, the generic features of the pseudomodulus potential gave rise to a parametrically new way of realizing strong first-order phase transitions in field theory. 

The novelty of the pseudomodulus FOPT is a consequence of the flatness of the tree-level potential accompanied by the presence of a mass gap for the heavy states, which makes the theory calculable everywhere in field space.  Since the mass gap is supersymmetric, it does not destroy the flatness of the potential at large field values. The two resulting features of this setup are that i) the nucleation temperature is well below the scale of the heavy states, so that the low-$T$ expansion applies, and ii) the pressure from Boltzmann-suppressed heavy states  in the plasma is responsible for stopping the vacuum bubble runaway. The dominant GW signal then comes from the energy released in the plasma during the phase transition. 

The strength of the GW signal depends on finer details of the hidden sector dynamics. However, we found that multiple SUSY-breaking scales in the hidden sector are a necessary condition for generating strong GW signals without fine tuning of the theory parameters. This result is quite general and can be obtained analytically without reference to a specific model. For the sake of concreteness, we presented an explicit model for a hidden sector generating a strong GW signal, where SUSY is broken by both an $F$-term and a $D$-term. Detailed predictions for the GW signal and superpartner spectrum in this model substantiate the general phenomenological observations of Fig.~\ref{fig:moneyplot}.

\section*{Acknowledgements} 
We thank Tomer Volansky for collaboration at the initial stage of this project. We thank Andrea Tesi for asking if there was a non-SUSY realization of our phase transitions. In trying to answer this question we realized how special these SUSY-breaking phase-transitions were. We also thank Daniele Barducci, Toby Opferkuch for interesting discussions. We thank Victor Guada for assistance with the FindBounce package. 

AM is supported by the Strategic Research Program High-Energy Physics and the Research Council of the Vrije Universiteit Brussel, and by the ``Excellence of Science - EOS" - be.h project n.30820817. NC is supported in part by the Department of Energy under the award DE-SC0011702. NL would like to thank the Milner Foundation for the award of a Milner Fellowship.

\appendix

\section{The effective potential}
\label{app:potentials}
The effective scalar potential is given by a sum of quantum and thermal contributions
\begin{equation}\label{eq:}
V_{\mathrm{eff}}(x,T)= V_0(x)+
V_{T}(x, T).
\end{equation}
The temperature independent potential can be written as $V_0(x)=V_{\mathrm{tree}}(x)+V_{\mathrm{CW}}(x)$, where $V_{\mathrm{CW}}(x)$ is the one loop Coleman-Weinberg potential at zero temperature which in the $\overline{\rm{MS}}$ scheme is given by
\begin{eqnarray}\label{eq:}
V_{\mathrm{CW}}(x)
&=&
\sum_{i}
(-1)^F
\frac{g_i m_i^4(x) }{64\pi^2} 
\left(
\log{\frac{m_i^2(x)}{m_0^2}-c_i}
\right),
\end{eqnarray}
where $F=1\,(0)$ for fermions (bosons), the number of degrees of freedom associated with the particle $i$ is $g_i=1/2/3$ for real scalars, fermions and vectors, respectively,
and $c_i=\frac{3}{2}\,(\frac{5}{2})$ for scalars/fermions (vectors).
The thermal one-loop potential is given by
\begin{eqnarray}\label{eq:}
V_{T}\left(x, T\right)
&=&
\frac{T^{4}}{2 \pi^{2}}
\sum_{i}
(-1)^{F} g_{i}  J_{\mathrm{B} / \mathrm{F}}\left(\frac{m_{i}^{2}(x)}{T^{2}}\right),
\end{eqnarray}
where the thermal functions for both species are
\begin{eqnarray}\label{eq:}
J_{\mathrm{B} / \mathrm{F}}\left(z^{2}\right)=\int_{0}^{\infty} d x x^{2} \log [1 \mp \exp (-\sqrt{x^{2}+z^{2}})],
\end{eqnarray}
with $z_i\equiv m_i/T$.
These functions can only be fully evaluated numerically, but admit analytical approximations for large and small $|z^2|$. 
%
In the high-temperature limit, $\left|z^{2}\right| \ll 1$ and the thermal functions are
\begin{eqnarray}\label{eq:} 
J_{B}\left(z^{2}\right) 
& \approx&
J_{B}^{\mathrm{high}-T}\left(z^{2}\right)
=
-\frac{\pi^{4}}{45}+\frac{\pi^{2}}{12} z^{2}-\frac{\pi}{6} y^{3}-\frac{1}{32} 
z^{4} \log \left(\frac{z^{2}}{a_{b}}\right) ,
\\ \nn
 J_{F}\left(z^{2}\right) 
 & \approx&
  J_{F}^{\mathrm{high}-T}\left(z^{2}\right)
  =
  \frac{7 \pi^{4}}{360}-\frac{\pi^{2}}{24} z^{2}-\frac{1}{32} z^{4} \log \left(\frac{z^{2}}{a_{f}}\right), \quad \text { for }\left|z^{2}\right| \ll 1 ,
\end{eqnarray}
where $a_{b}=\pi^{2} \exp \left(3 / 2-2 \gamma_{E}\right)$ and $a_{f}=16\pi^{2} \exp \left(3 / 2-2 \gamma_{E}\right)$.
The low temperature limit (i.e. $\left|z^{2}\right| \gg 1$) can be approximated in terms of modified Bessel functions of the second kind
\begin{eqnarray}\label{eq:lowT}
J_{B}\left(z^{2}\right)
&=&
\tilde{J}_{B}^{(m)}\left(z^{2}\right)
=
-\sum_{n=1}^{m} \frac{1}{n^{2}} z^{2} K_{2}(z n),
\\  \nn
J_{F}\left(z^{2}\right)
&=&
\tilde{J}_{F}^{(m)}\left(z^{2}\right)
=
-\sum_{n=1}^{m} \frac{(-1)^{n}}{n^{2}} z^{2} K_{2}(z n),
 \quad \text { for }\left|z^{2}\right| \gg 1\ ,
\end{eqnarray}
where $m$ is high enough such that the series converge. For $T$ low enough, we can take only the first term in the series and further expand the modified Bessel function to the leading order $K_{\nu}(z)\underset{z\to\infty}{\simeq}\sqrt{\frac{\pi}{2 z}} e^{-z}$ to get
\begin{equation}
J_{B}\left(z^{2}\right)=-J_{F}\left(z^{2}\right)\underset{z\to\infty}{\simeq}-\left(\frac{\pi z^3}{2}\right)^{1/2}e^{-z}\ .
\end{equation}
Within this approximation we can obtain a simple expression for $V_{\mathrm{th}}(x,T)$ which is valid at the leading order in the low-$T$ expansion:
\begin{eqnarray}\label{eq:approxefflowT}
V_{T}(x,T)
 &\simeq&
  -T^{4}
  \sum_\mathrm{B/F}
  g_i
  \left(\frac{m_i(x)}{2\pi T} \right) ^{3/2}
e^{-m_i(x) /T},
\end{eqnarray}
where it is important to notice that bosons and fermions contribute with the same (negative) sign to the effective potential. The approximation above is used in Section~\ref{sec:anatomy} to derive an analytical scaling of the dynamics of FOPTs.


\paragraph{Lambert function}
As a consequence of the low-$T$ expansion the equations we will be dealing with have the typical form 
\begin{equation}
A z^{-a}e^{-z}-B=0\quad , \quad z= a \mathcal{W}\left[\frac{1}{a}\left(\frac{A}{B}\right)^{1/a}\right]\ ,
\end{equation}
where $\mathcal{W}(z)$ is the Lambert function, which is defined such that $\mathcal{W}(z) e^{\mathcal{W}(z)}=z$.
Without entering into the details of the interesting properties of this function, we restrict our interest to finding 
a good approximation for it using simpler functions. 
First, we will consider $\mathcal{W}(z)$ for strictly positive arguments.
Second, we note that $\mathcal{W} \to 0$ for $z \to 0$ and that $\mathcal{W}(z) \sim \log z $ for $z \to \infty$.
By inspection one finds that a good approximation of $\mathcal{W}(z)$ is given simply by
\be
\label{approx_Lamb}
\mathcal{W}(z) \simeq \frac{3}{4} \log\left(1+z \right)\ . 
\ee
The relative difference between the Lambert function and the approximation with the logarithm in \eqref{approx_Lamb} 
is at most $\sim 1/4$ (for $z \to 0$ and $z \to \infty$) and smaller (in absolute value) in intermediate regions. 
For practical purposes in analytic estimations of relevant quantities, we will hence often consider the approximation in \eqref{approx_Lamb}.

%
%
%

\section{Bounce action computation schemes}\label{app:bounce}
The transition of a quantum system from a meta-stable vacuum state to the true vacuum can be driven either by quantum tunneling or by thermal fluctuations. In the FOPTs describe in this paper the latter are always dominant. The probability of thermal tunnelling is described semi-classically by Eq.~\ref{eq:decay} and it is exponentially dependent on the classical $O(3)$-symmetric bounce solution~\cite{Linde:1980tt,Linde:1981zj}. In this appendix we review both the analytical and the numerical approaches we used to study the behavior of the $O(3)$-symmetric bounce in our FOPTs. 

First, in Sec.~\ref{sub:single_field_triangular_vs_numerical_solution}, we describe in detail the triangular barrier approximation introduced in Ref.~\cite{Duncan:1992ai} and its generalization to the  $O(3)$-symmetric case~\cite{Amariti:2009kb}. We compare this approximation with the behavior of the full bounce action computed numerically with the FindBounce package~\cite{Guada:2018jek,Guada:2020xnz} and the CosmoTransitions code~\cite{Wainwright:2011kj}. In Sec.~\ref{sec:optimal} we discuss an optimization of the triangular barrier approximation which leads to excellent agreement with the full numerical computation. In Sec.~\ref{ssub:triangular_bounce_for_the_}  we consider the O'Raifeartaigh model with gauge interactions of Sec.~\ref{sec:FD} as our main case study. 

Second, in Sec.~\ref{sub:single_field_approximation_for_the_multi_field_problem_} we discuss the single field approximation of the bounce action in the model of Sec.~\ref{sec:FD}. We study numerically the behavior of the bounce action in the full three-dimensional field space and compare it with the single field approximation, deriving where we expect the latter to deviate sensibly from the full solution.

\subsection{Triangular barrier approximation of the bounce action} 
\label{sub:single_field_triangular_vs_numerical_solution}

The $d$-dimensional Euclidean action for $n$ scalar fields $\phi_i$ is
\begin{eqnarray}\label{eq:}
S_{d}=
\Omega_{d} \sum_i \!\!\int_{0}^{\infty} r^{d-1} d r\left[\frac{1}{2} \dot{\phi_i}^2+V(\phi_i)\right]\ ,
\end{eqnarray}
where $\Omega_d=2 \pi^{d / 2}/\Gamma(d / 2)$ and the equations of motion for $\phi_i$ are 
\begin{eqnarray}\label{eq:}
\ddot{\phi_i}+\frac{(d-1)}{r} \dot{\phi_i}=V^{\prime}(\phi_i)\, .
\end{eqnarray}
The boundary conditions defining the bounce solution are 
\begin{eqnarray}\label{eq:}
\dot{\phi_i}(r=0)=0, \quad \phi_i(r \rightarrow \infty)=\phi_{i_{f}} \quad \text { and } \quad \dot{\phi_i}(r \rightarrow \infty)=0\, ,
\end{eqnarray}
where the conditions at $r\to\infty$ ensures that the solution starts with zero kinetic energy from the false vacuum and stops at $r=0$ with zero kinetic energy. Here, we are interested in solving this equation for a single field $(n=1)$ and in $d=3$.   

The idea behind the triangular barrier approximation is to approximate the potential as a piecewise linear function anchored at three points: the false vacuum, the top of the barrier and the true vacuum. 
In the following, we review in some details the derivation of the 3d bounce action in the triangular barrier approximation, and
we refer to the original paper \cite{Duncan:1992ai} 
for more details and to \cite{Amariti:2020ntv} for the derivation in arbitrary dimensions.

Following the notation of \cite{Duncan:1992ai}, we define the false vacuum to be at the field position $\phi_+$ with potential $V_+$; the peak of the triangular barrier to be at $\phi_P$ with potential $V_P$; and the true vacuum to be at $\phi_-$ with potential $V_-$. 
It is then convenient to define the magnitudes of the gradients of the potential by
\begin{eqnarray}\label{eq:}
\lambda_{\pm}\defeq\frac{\Delta V_{\pm}}{\Delta \phi_{\pm}}\,,
\end{eqnarray}
so that $V^{\prime}(\phi)=\pm \lambda_{\pm}$ on either side of the barrier, precisely
\bea
V(\phi) = \left\{
\begin{array}{l}
V_+ +\lambda_+ (\phi-\phi_+) \quad \text{for} \quad  \phi < \phi_P\\
V_-  + \lambda_- (\phi_- -\phi) \quad \text{for} \quad \phi> \phi_P
 \end{array}
 \label{eq:piecewise}
\right.
\eea

In order to solve the equation of motions we have to specify the boundary conditions.
At a large radius $R_+$ the field attains the false vacuum, so we have
\begin{eqnarray}\label{eq:}
\phi\left(R_{+}\right)=\phi_{+}\quad \quad \dot{\phi}\left(R_{+}\right)=0\,,
\end{eqnarray}
Then, we have to specify the boundary conditions at the start of the tunneling.
There are two possibilities:
\begin{enumerate}
\item The field immediately start to roll at $r=0$ and hence we impose
\begin{eqnarray}\label{eq:}
\begin{array}{c}\phi(0)=\phi_{0} \qquad \dot{\phi}(0)=0, \end{array}
\end{eqnarray}
where the initial field value $\phi_0$ is the undetermined release point. This is the valid regime if one finds that $\phi_0 \leq \phi_-$.
\item Otherwise, the field sits in the true vacuum for $r<R_0$ and then starts rolling. In this second case the boundary conditions are
\begin{eqnarray}\label{eq:}
&&\phi(r)=\phi_{-} \qquad   ~~0<r <R_0\\
&&\phi(R_0)=\phi_- \qquad \dot{\phi}(R_0)=0, 
\end{eqnarray}
\end{enumerate}

We begin with the analysis of the first case.
Imposing the previous boundary conditions one finds the following solutions for the equations of motion in the different radius intervals
\be
\phi(r) = 
\left\{ 
\begin{array}{l}
\phi_R(r) = \phi_0 -\frac{\lambda_-}{6} r^2 \qquad \qquad  \qquad \qquad \qquad  0 < r < R_P\\
\phi_L(r) = \phi_+ + \frac{\lambda_+}{6} \left(r^2 -3 R_+^2+2 \frac{R_+^3}{r} \right) \qquad R_P < r < R_+
 \end{array}
\right.
\label{eq:TBA_1}
\ee
Then we impose that the two solutions match at $R_P$ and also that their first derivatives match at $R_P$, obtaining the following conditions 
\bea
&& \phi_0 = \phi_P +\frac{\lambda_+ R_P^2}{6} \\
&& R_F^3 =(1+r_{\lambda}) R_{P}^3 \\
&& R_P^2 = \frac{6 \Delta \phi_+^2 }{\left[3+2 r_\lambda-3\left(1+r_\lambda\right)^{2 / 3}\right]\Delta V_+ } 
\label{eq:variuos_TBA_1}
\eea
where we have introduced $r_{\lambda} \equiv \frac{\lambda_-}{\lambda_+}$.
Then we insert into the action the solutions \eqref{eq:TBA_1} and we integrate from $r=0$ to $r=R_+$ with the appropriate potential (see \eqref{eq:piecewise}).
From this computation we have to subtract the action for the case in which the field sits at the false vacuum from $r=0$ to $r=R_+$, that is we compute all in all
\be
S_3^{TBA} = S_3[\phi(r)] - S_3[\phi_+]
\ee
Using the equations in \eqref{eq:variuos_TBA_1} we can rewrite the result as a function of the parameters of the potential to obtain
\begin{eqnarray}\label{eq:triST}
\left(\frac{S_{3}}{T}\right)_{\text {TBA }}
=
\frac{16 \sqrt{6} \pi}{5} \frac{1}{T} \frac{\left(1+r_\lambda\right)}{\left[3+2 r_\lambda-3\left(1+r_\lambda\right)^{2 / 3}\right]^{3 / 2}}\left(\frac{\Delta \phi_{+}^{3}}{\sqrt{\Delta V_{+}}}\right),
\end{eqnarray}
The condition to select the first case (i.e. $\phi_0 > \phi_-$) can also be rewritten by employing again formula \eqref{eq:variuos_TBA_1} as
\begin{eqnarray}\label{eq:}
\frac{\Delta \phi_{-}}{\Delta \phi_{+}} 
\geq
 \frac{r_\lambda}{
 3+2 r_\lambda-3\left(1+r_\lambda\right)^{2 / 3}
}.
\end{eqnarray}

We then analyse the second case. Imposing the boundary conditions we get the solutions
\be
\phi(r) = 
\left\{ 
\begin{array}{l}
\phi_R(r) = \phi_-  \qquad \qquad  \qquad \qquad \qquad \qquad  \quad 0 < r < R_0\\
\phi_R(r) =\phi_- - \frac{\lambda_-}{6} \left(r^2 -3 R_0^2+2 \frac{R_0^3}{r} \right)  \qquad  R_0 < r < R_P\\
\phi_L(r) = \phi_+ + \frac{\lambda_+}{6} \left(r^2 -3 R_+^2+2 \frac{R_+^3}{r} \right) \qquad R_P < r < R_+
 \end{array}
\right.
\label{eq:TBA_2}
\ee
By imposing matching of the fields and the derivative at $R=R_P$ we get the following equations for the unknown parameters $R_0,R_P,R_+$
\bea
&& R_P^3 = \frac{r_{\lambda} R_0^3+R_+^3}{1+r_{\lambda}} \\
&& \Delta \phi_- = \frac{r_{\lambda} \lambda_+ (R_0 - R_P)^2 (2 R_0 +R_P)}{6 R_P}\\
&& \Delta \phi_+ = \frac{ \lambda_+ (R_+ - R_P)^2 (2 R_+ +R_P)}{6 R_P}
\label{eq:variuos_TBA_2}
\eea
We then compute the bounce action by inserting the solutions and integrating from $r$ to $R_+$, and after some rearrangements we get 
\be
\left(\frac{S_{3}}{T}\right)_{\text {TBA }}
= \frac{8 \pi}{15} \frac{1}{T} \left(R_+^3 \Delta \phi_+ - r_{\lambda} R_0^3 \Delta \phi_- \right) \frac{\Delta V_+}{\Delta \phi_+ }
\ee
where $R_+$ and $R_0$ are functions of parameters of the scalar potential through the implicit equations \eqref{eq:variuos_TBA_2}.

So we conclude that in both possible cases, the computation of the bounce action in the TBA approximation only needs to specify the \emph{critical} points
of the potential, that is the metastable vacuum, the peak of the barrier and the true vacuum.
For the scenarios studied in this paper the first case \eqref{eq:triST} is the relevant one, and the validity condition is reported also in  \eqref{S3overTfull}
(note that in the conventions in the main text we always choose the metastable vacuum to be at the origin of the field space).
In this Appendix we have nevertheless reviewed both cases for completeness.

In the following we will compute the TBA bounce action by both evaluating the potential numerically and approximating it analytically.  These approximations can be compared with the results of the full-fledged numerical bounce action computation.

\subsubsection{Optimized triangular bounce}\label{sec:optimal}

Studying the evolution of the bubble profiles for the actions computed numerically we note that the release point $\phi_0$ is typically closer to the potential barrier than to the true vacuum at $\phi_-$ in our setups. Therefore, we introduce here a modified version of the TBA that takes into account this feature, leading to a better agreement with the 
full numerics than the analytic formulas presented above.  The idea is to allow the minimum of the potential to be a free parameter rather than to fix it at the true vacuum, allowing the TBA to more closely represent the shape of the potential, where the slope is closer to linear. 

The TBA bounce action is computed by replacing $\phi_-\to \phi_0$ and $V_-(\phi_-)\to V_-(\phi_0)$, where $\phi_0$ is now an arbitrary point along the potential
in the interval $\phi_\mathrm{eq}<\phi_0<\phi_-$ (here $\phi_\mathrm{eq}>\phi_P$ is the point after the potential barrier where $V(\phi_\mathrm{eq})=V_+$). This procedure defines a function $S_3/T(\phi_0)$ that we can minimize over $\phi_0$ in the allowed interval. The resulting minimum is the sought bounce action, that 
we have dubbed the optimal TBA.
\subsubsection{Triangular bounce for the O'Raifeartaigh model with gauge interactions} 
\label{ssub:triangular_bounce_for_the_}

In this section, we specify the discussion to the  triangular barrier approximation for the model of Sec.~\ref{sec:FD} and compute the inputs needed for the TBA bounce action. We denote the usual combination of parameters as $y_F \equiv \frac{\lambda F}{m^2}$ and $y_D \equiv \frac{g D}{m^2}$.

The local minimum at the origin is approximated as 
\begin{eqnarray}\label{eq:}
\phi_+
&=& 0\ , \\
V_+ (T)
&=& 
F^2+\frac{1}{2} D^2-\frac{3 T^{5/2} e^{-\frac{m}{T}} \left(128 m^2+240 m T\right)}{128 \sqrt{2} \pi ^{3/2} \sqrt{m}}\\ \nn
&+&\frac{m^4}{32 \pi ^2}
\left[
\left(1+y_F\right)^2 \log \left(1+y_F\right)+
\left(1-y_F\right)^2 \log \left(1-y_F\right)
-3 y_F^2
\right]\ ,
\end{eqnarray}
The peak of the barrier is located at
\begin{eqnarray}\label{eq:}
\phi_{P}
&\simeq&
\frac{F}{m}
\frac{\sqrt{2-2y_D}}{y_F  \sqrt{y_D}}\ ,
\\
V_{P}(T)
&\simeq&
F^2+\frac{1}{2}D^2
+\frac{m^4}{16 \pi ^2}
y_F^2
 \log \left(\frac{1-y_D}{y_D}\right)
-
\frac{T^{5/2} e^{-\frac{m}{T}} \left(128 m^2+240 m T\right)}{32 \sqrt{2} \pi ^{3/2} \sqrt{m}}\ ,
\end{eqnarray}
Finally, the true vacuum location and energy are given by
\begin{eqnarray}\label{eq:}
\phi_-
&\simeq& 
\frac{F}{m} 
\frac{4\pi }{g}
\frac{\sqrt{2 y_D}}{ y_F^2}\ ,
\\
V_{-}
&\simeq& 
F^2+
\frac{m^4}{16 \pi ^2} y_F^2 \log \left(\frac{16 \pi ^2}{g^2}\frac{ y_D  }{ y_F^2} \right)\ , 
\end{eqnarray}
where we set the renormalization scale $\mu = m$.
Within our approximation the thermal effects only enter at the origin, and at the top of the barrier, where they act to lower the potential relative to the true vacuum, and the potential difference between the top of the barrier and origin, respectively.
 
\begin{figure}[t!]
  \centering
\includegraphics[width=0.48\textwidth]{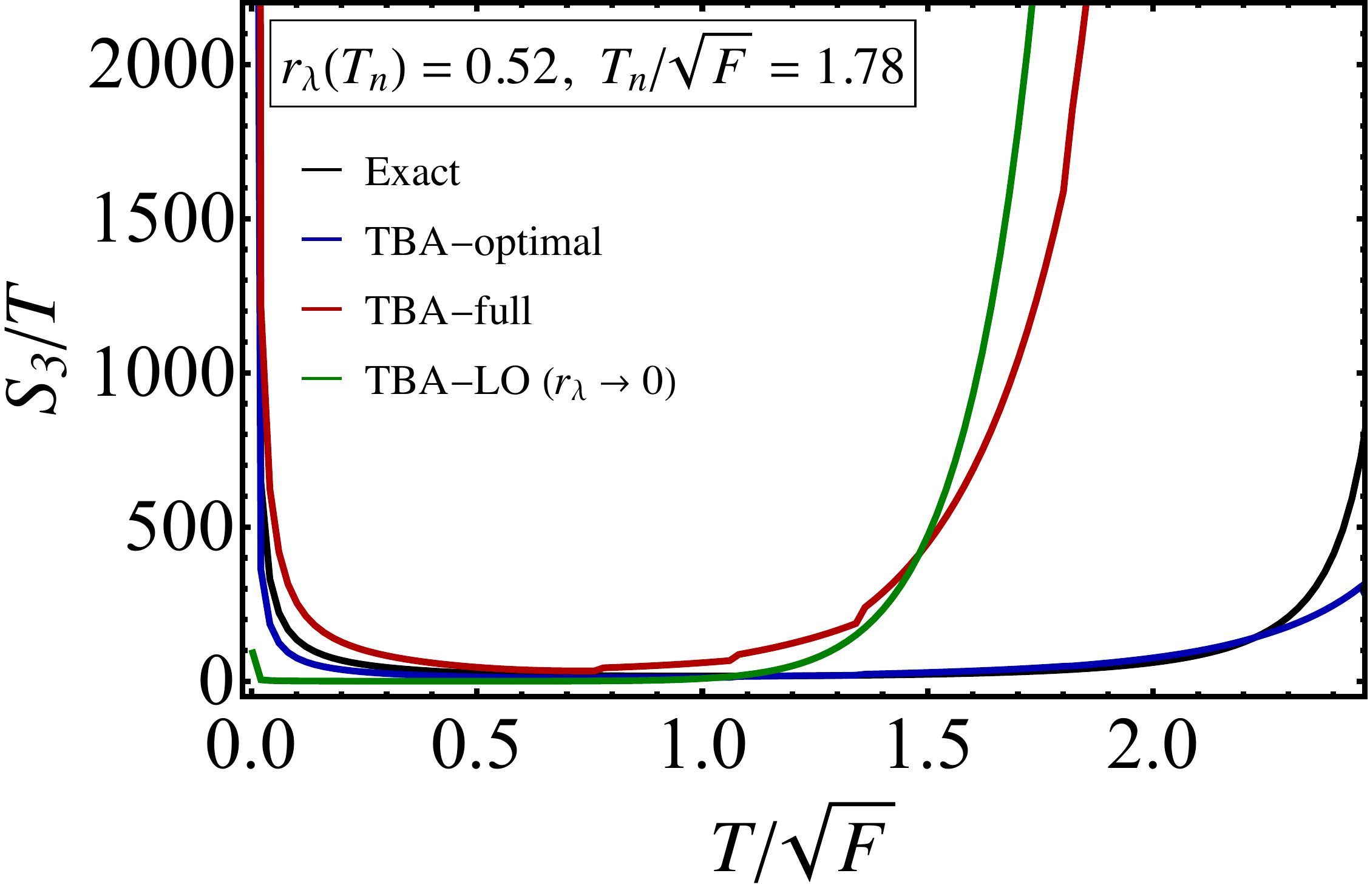}~~~
\includegraphics[width=0.48\textwidth]{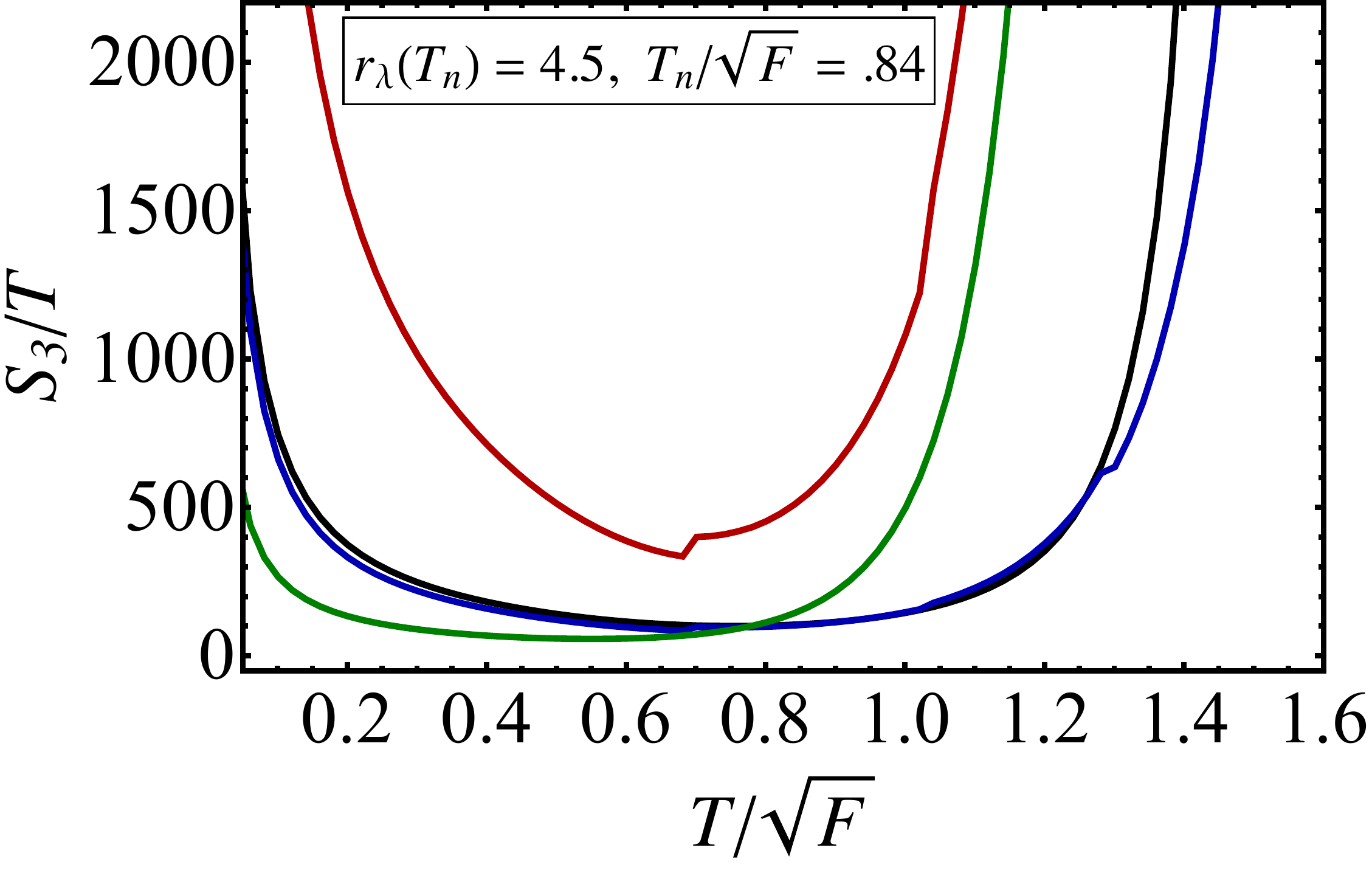}
	\caption{Bounce action computed using different approximations in the O'Raifeartaigh model with gauge interactions of Sec.~\ref{sec:FD}. The benchmarks on the left  and on the right are distinguished by the size of $r_\lambda$. The {\bf black lines} are the full numerical computation of the bounce action. In {\bf blue} we show the TBA computed using the numerical scalar potential and optimized with the procedures explained in the text. The {\bf red lines} are the standard TBA approximations as described in Appendix \ref{sub:single_field_triangular_vs_numerical_solution}. The {\bf green line} is the standard TBA evaluated on the analytic approximation of the scalar potential (as detailed in the text) and taking only the zeroth order term in the expansion for small $r_{\lambda}$ (as in Eq.~\eqref{S3overTfull}). This last approximation is the one used in Section \ref{sec:anatomy} to derive analytic estimates.}\label{fig:triFD}
\end{figure}

In Figure \ref{fig:triFD} we consider two benchmarks with very different $r_{\lambda}$ at $T_n$ and show the bounce action $S_3/T$ as a function of the temperature, computed in different approximations. 
The black line is computed using the fully numerical thermal effective potential and the mathematica package ``FindBounce''~\cite{Guada:2020xnz}.
The blue line is obtained with the TBA evaluated on the full-numerical scalar potential and optimized with the procedure explained above.
The red line is the TBA (as computed in Appendix \ref{sub:single_field_triangular_vs_numerical_solution}) evaluated on the full-numerical scalar potential. Finally, the green is the TBA evaluated on the analytical approximation of the critical points of the scalar potential as explained above, and moreover keeping only the leading order term in the small $r_{\lambda}$ expansion in Eq.~\eqref{S3overTfull}. This is the approximation used to derive the analytic formulas in Section \ref{sec:anatomy}.

First, we see that the optimal TBA reproduces almost perfectly the numerical bounce computation. 
The standard TBA can predict well the location of $T_{\text{min}}$ but the overall normalization can be off up by a factor of $\sim$few,
and in particular it does not agree with the numerical result in the vicinity of $T_c$.
However, the different trend in the overall shape of the bounce action in the two benchmarks (e.g. very flat around $T_{\text{min}}$ in the left one) is also captured in the standard TBA approximation.

Then, we note that the simplest approximation of the standard TBA reproduces very well the TBA when $r_{\lambda} $ is small. This was not obvious a priori since, besides expanding the TBA at leading order in small $r_{\lambda}$, we have: i) assumed that the only temperature dependence is in the height of the potential at the origin; ii) employed the low-$T$ approximation of the scalar potential to estimate it.
The agreement between the red and green curve in the left panel of Fig.~\ref{fig:triFD} hence confirms the fact that low-$T$ is the correct approximation to employ,
as discussed at length in Section \ref{sec:anatomy}.
When $r_{\lambda} $ is larger (right plot) the simplest approximation (green line) clearly deviates from the standard TBA (red line), but nevertheless capture approximately 
the location of $T_{\text{min}}$ and the shape of the numerical results.
It is important to observe that even if the normalization of the bounce action and its raising towards $T_c$ are not exactly reproduced by the approximations employed, 
they can still track the changes of the bounce action (shape deformations and overall size) as a function of the fundamental parameters of the model.
This elucidates why the analytic estimates obtained in Section~\ref{sec:anatomy} can capture the scaling of $T_n$ 
in the different models as discussed in Section~\ref{sec:X3} and \ref{sec:FD}.


\subsection{Single field approximation of the multi-field bounce action}\label{sub:single_field_approximation_for_the_multi_field_problem_}
\begin{figure}[t!]
  \centering
\includegraphics[width=0.48\textwidth]{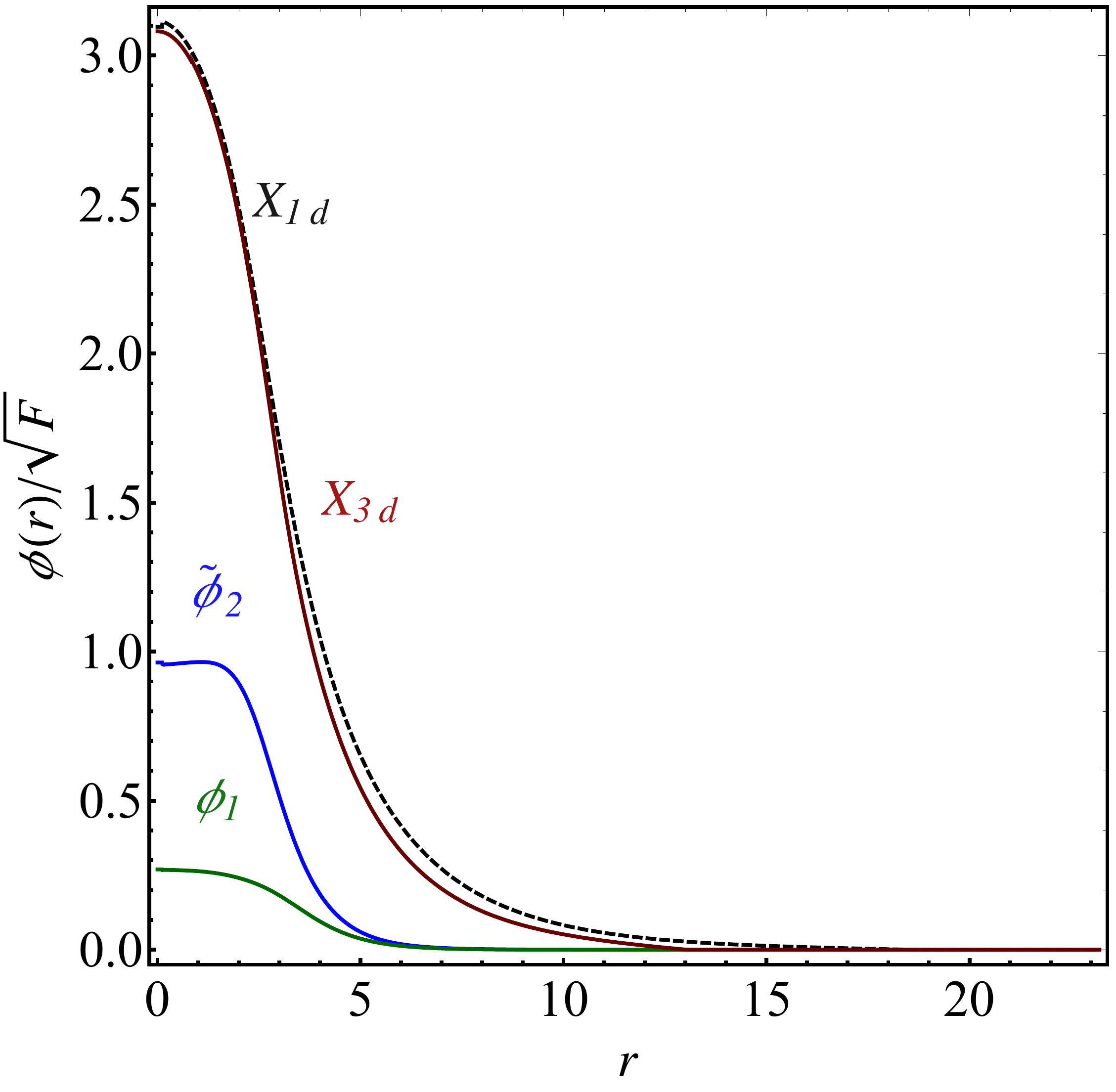}~
\includegraphics[width=0.48\textwidth]{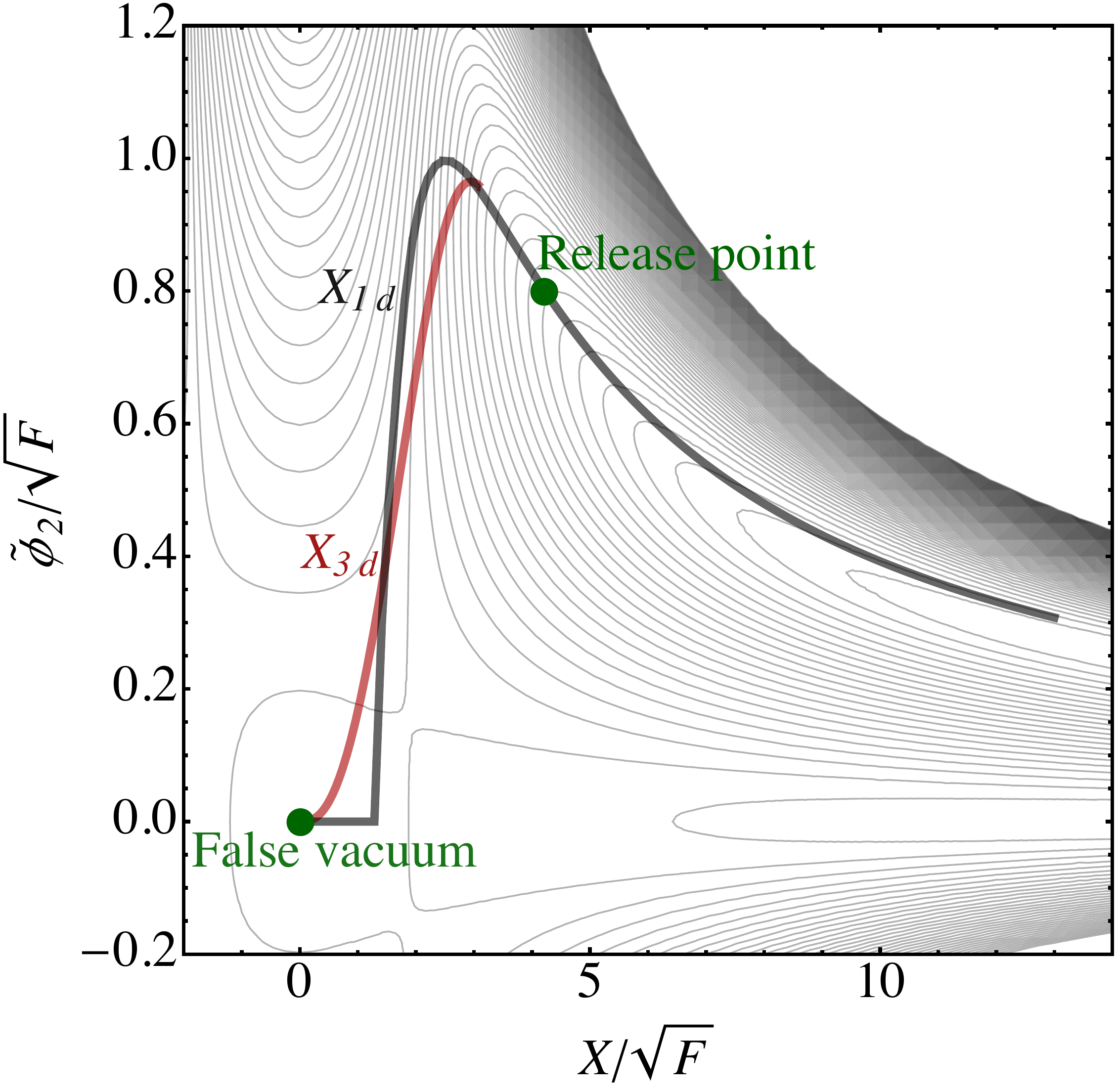}
\caption{\label{fig:singlevs3}
{\bf Left:} Bubble profiles for the single field and three field bounce actions as a function of the bubble radii. The release point for the single field (pseudo flat) direction is practically identical in both schemes, while the contribution coming from motion along the  other field directions is generally small, but non-zero. {\bf Right:} The bubble trajectory in field space for the single field and three field bounce action. The x-axis represents motion along the pseudo flat direction while the y-axis represents motion along the larger of the other two field directions ($\tilde{\phi}_2$). It is clear from both plots that the motion along the pseudo flat direction is mostly unaffected by the motion along the other directions, and therefore the single field path approximation is viable.
}
\end{figure}
As discussed previously, the model presented in Sec.~\ref{sec:FD}, contains more than one dynamical degree of freedom that actually enters into the bounce action computation. 
Namely, the fields $X,\phi_1$ and $\tilde \phi_2$ vary along the minimal potential energy trajectory in field space which connects the two minima of the potential.
This implies that the full bounce solution is that of a three field problem, which is in general only solvable numerically. 
In this Appendix we explain why in the model under study the bounce action can be effectively approximated with a one field problem and what
is the regime of validity of such approximation.

The bounce action involving the three fields is
\begin{eqnarray}\label{eq:}
S_3
&=&
\frac{2\pi^{3/2} }{\Gamma(3/2)}
\sum_i\int_0^\infty
r^2
\left( T_i + V(\phi_i)\right)
\\ \nn
&=&
\frac{2\pi^{3/2} }{\Gamma(3/2)}
\int_0^\infty
r^2
\left( \frac{1}{2}\dot{X}^2
+\frac{1}{2}\dot{\phi_1}^2
+\frac{1}{2}\dot{\tilde{\phi}}_2^2
 + V(X,\phi_1,\tilde{\phi}_2)\right),
\end{eqnarray}
where $T_i$ is the kinetic energy associated with each field, which is an additive quantity.
In our analysis we approximate this action by minimizing $V(X,\phi_1,\tilde{\phi}_2)$ along all three directions and by solving the bounce equation only for $X$.
This corresponds to neglect the contribution from the kinetic energy of the other two fields $\phi_1$ and $\tilde \phi_2$.
Since the kinetic energy of these fields is related to the potential energy along the same directions by the equation of motion, we expect that the
kinetic energy contributions of $\phi_1$ and $\tilde \phi_2$ can be consistently neglected 
if their VEVs along the bounce trajectory are small compared to the one of $X$.

This is typically what happens in this model as we show in the left panel of Figure \ref{fig:singlevs3}, 
where we plot the bubble profiles of the full three field problem computed using FindBounce (red, blue and green line).
We also show for comparison (in dashed black) the bubble profile that we obtain for the single-field bounce solution for $X$,
which is essentially identical to the one in the three-field solution.
In the right panel of Figure \ref{fig:singlevs3} we show a two dimensional slice of the field path along the bounce solution, in the $X,\tilde \phi_2$ plane (we show $\tilde \phi_2$ since the $\phi_1$ vev is smaller and hence it has even a smaller impact on the value of the bounce action).
We see that the $X$ trajectory is almost unchanged by the addition of the second field.
We hence conclude that it is typically a robust approximation in this model to neglect the kinetic 
contribution of the fields $\phi_1$ and $\tilde \phi_2$ and to solve the one-field problem.

Nevertheless, we would like to estimate the range of 
validity of our approximation.
As mentioned, we expect the difference to come from the kinetic terms of $\phi_1$ and $\tilde \phi_2$, which will be non negligible
if the size of the $\phi_1$ and $\tilde \phi_2$ VEV's compared to the one of $X$ is not negligible.
In particular, we would like to estimate the impact of this approximation in the overall bounce action.
We hence use intuition from the TBA where the size of the bounce action is proportional to the cubic power of the field displacement.
We define the following ratio 
\begin{equation}\label{eq:1dvs3d}
R_{1d/3d}
\defeq
\left.\frac{X^3(r)}
{\left(X^2(r)+\phi_{1}^2(r)+\tilde \phi_{2}^2(r)\right)^{3/2}}\right\vert_{r=0}\, ,
\end{equation}
where $X(0)$, $\phi_{1}(0)$ and $\tilde \phi_{2}(0)$ are the field distances from the origin computed at the release point $r=0$. 
The ratio $R_{1d/3d}$ provides a measure of the relative error of the single-field bounce action computation against the full three-field one.
Indeed, we have also cross-checked numerically in several benchmarks that the difference in the values of the bounce action is negligible when 
$R_{1d/3d}$ is small.
In the main body of the paper, we will define the region where the 1-field approximation breaks down as when the quantity $R_{1d/3d} > 0.5$, 
corresponding approximately to a relative error of $\sim 50\%$ of the single field approximation compared to the full three field solution.

\section{Sensitivity of GW interferometers} 
\label{sub:signal_detection}
In this section we briefly discuss the interpretation and generation of the sensetivity curves used to define detection of a GW signal. We follow standard definitions and conclusions obtained in \cite{Allen:1996vm,Allen:1997ad,Maggiore:1999vm,Romano:2016dpx}, see also \cite{Schmitz:2020syl} and references therein.
The detection sensitivity for GW background for a given experimental setup, is given by the integrated signal-to-noise ratio (SNR) over an observation time interval $t_{obs}$ as
\begin{eqnarray}\label{eq:rho}
\rho=
\frac{\langle S \rangle}{\langle N^{1/2} \rangle}=
 \left[n_{\mathrm{det}} t_{\mathrm{obs}} \int_{f_{\mathrm{min}}}^{f_{\mathrm{max}}} d f\left(\frac{S_{\mathrm{S}}(f)}{S_{\mathrm{N}}^{\mathrm{eff}}(f)}\right)^{2}\right]^{1 / 2}\quad ,\quad S_{\mathrm{N}}^{\mathrm{eff}}(f)=\frac{D_{\mathrm{N}}(f)}{\mathcal{R}(f)}\ ,
\end{eqnarray}
where $\langle S \rangle$ is the mean signal, $\langle N^{1/2} \rangle=\sqrt{\left\langle S_{I J}^{2}\right\rangle-\left\langle S_{I J}\right\rangle^{2}}$ is the average noise, $I,J=1,2$ indicate coupled detectors, $n_{\rm det}$ distinguishes between experiments that aim at detecting the signal by means of an auto-correlation of a single detector ($n_{det}=1$) or a cross-correlation of a couple of detectors ($n_{\rm det}=2$) measurement. The effective noise strain can be written in terms of the noise strain power spectrum   $D_\mathrm{N}(f)$ and the frequency dependent detector response function $\mathcal{R}(f)$. The latter quantities are the ones typically reported by the experimental collaborations.  The detector response becomes more intricate in the case of correlated detectors~\cite{Romano:2016dpx}, where an overlap detection function must be computed. We perform the appropriate procedure when computing the relevant SNR.
The signal/noise strains can be rewritten in terms of the signal/noise spectrum density as 
\begin{equation}
\Omega_{\mathrm{S/N}}(f)=\frac{2 \pi^{2}f^{3}}{3 H_{0}^{2}}  S_{\mathrm{S/N}}(f)\ ,
\end{equation}
so that Eq.~\eqref{eq:rho} becomes
\begin{eqnarray}\label{eq:sensitivitygood}
\rho=\left[n_{\mathrm{det}} t_{\mathrm{obs}} \int_{f_{\mathrm{min}}}^{f_{\mathrm{max}}} d f\left(\frac{\Omega_{\mathrm{S}}(f)}{\Omega_{\mathrm{N}}(f)}\right)^{2}\right]^{1 / 2},
\end{eqnarray}
where in this paper $\Omega_{\rm S}(f)$ will be the energy density of a SGWB produced by the SUSY-breaking FOPT in the early universe redshifted till today (see Eq.~\eqref{eq:GWspectrumtoday}). The frequency interval $(f_{\min},f_{\max})$ is determined by the bandwidth of each experiment, typically related to the length scale of each detector. This integrated quantity will imply detection if it surpasses a predefined threshold value $\rho^{2}\geq \rho_{\text{thr}}^2$ set by baysian probabilistic measures~\cite{Allen:1996vm}, usually taken between 3 and 10.

By defining a frequency dependent shape function which models the expected signal one can define the sensitivity of a given experiment even though this procedure does not provide a precise statistical indication of the expected sensitivity as the signal to noise ratio in Eq.~\eqref{eq:sensitivitygood}. Depending on how well the signal shape is approximated, sensitivity curve based on the shape function approximation can provide an instructive visual tool for estimating detection in a frequency dependent way. In the following section we discuss the most common method of generating sensitivity curves for 1OPT GW signal detection. 

\subsection{PLI curves} 
\label{ssub:pli_curves}
\begin{figure}[t]
	\centering
	\includegraphics[width=0.9\textwidth]{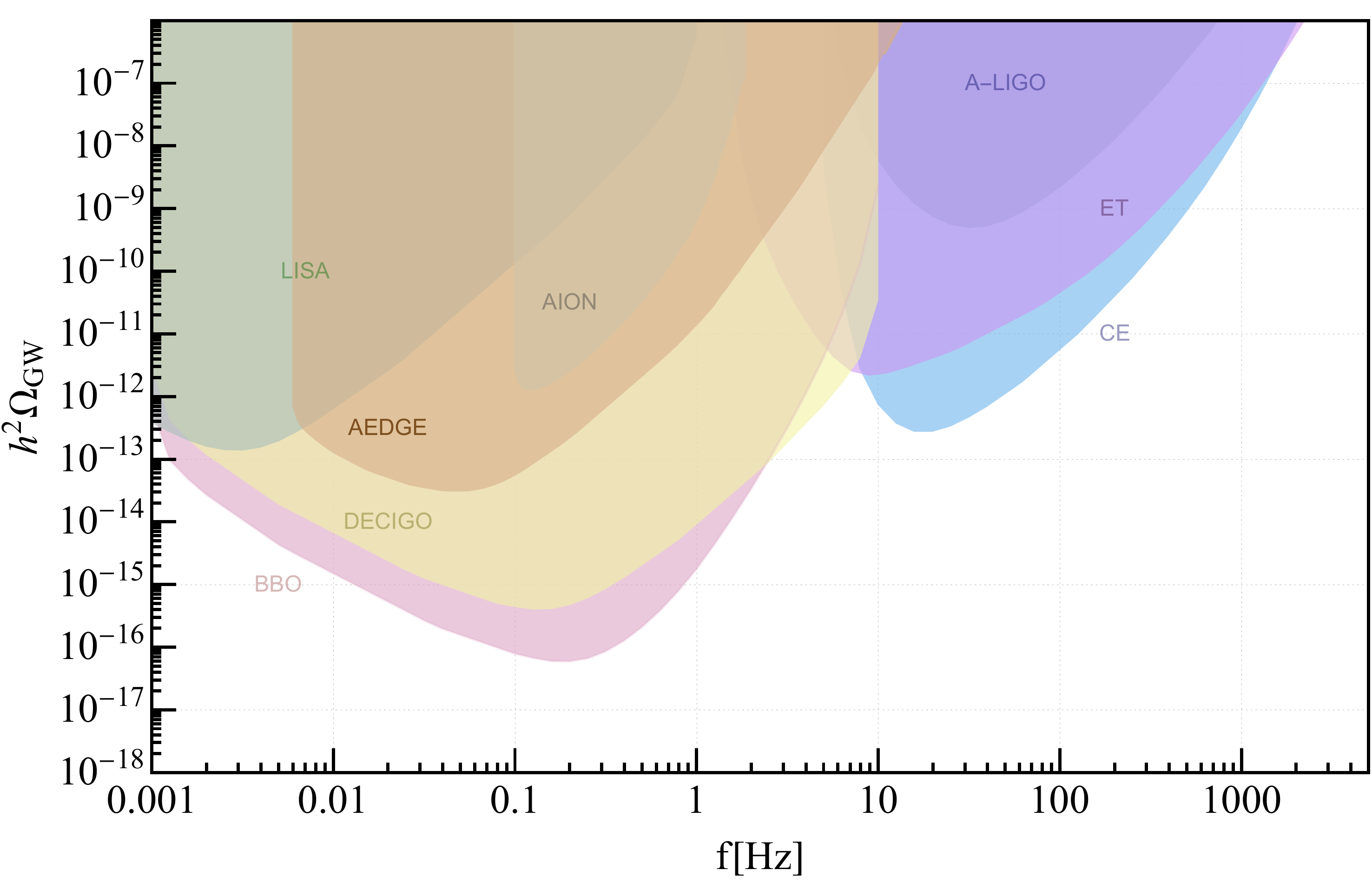}
	\caption{PLI curves for the different experiments considered here with the threshold value for the signal to noise ratio is conservatively fixed to  $\rho_{\text{thr}}=10$. The required data to derive a PLI curve for every experiment are collected in Table~\ref{tab:summary}.}
	\label{fig:pli}
\end{figure}

Power Law Integrated (PLI) curves are generated by considering a power law function of the frequency $f$ for the GW signal shape. The most common assumption, is a power law of the form 
\begin{eqnarray}\label{eq:power}
h^{2}{\Omega}_{\rm S}({f})
=
h^{2}\tilde{\Omega}_{\rm S, b}(\tilde{f})
\left(\frac{f}{\tilde{f}}\right)^b,
\end{eqnarray}
where $b$ is known as the spectral index. Taking this assumption we obtain the integrated SNR as~\cite{Cornish:2001bb,Thrane:2013oya}
\begin{eqnarray}\label{eq:intSNR}
h^{2} \tilde{\Omega}_{\rm S, b}(\tilde{f})
>h^{2} \tilde{\Omega}_{\rm G W, b}^{\text{thr}}(\tilde{f}) 
\equiv \frac{\rho_{\text{thr}}}{\sqrt{ t_{\text{obs}} n_{\text{det}}}}\left[\int_{f_{\min }}^{f_{\max }} d f\left(\frac{(f / \tilde{f})^{b}}{h^{2} \Omega_{\text{N}}(f)}\right)^{2}\right]^{-1 / 2}.
\end{eqnarray}
Finally, we can define the PLI sensitivity curve by maximizing the integrated SNR over the spectral index $b$, that is
\begin{eqnarray}\label{eq:PLI}
h^{2} \Omega_{\text{PLI}}(\tilde{f}) \equiv h^{2} \max _{b}
\left[
 \tilde{\Omega}_{\rm S, b}^{\text{thr}}(\tilde{f})
\right],
\end{eqnarray}
which gives the threshold value for the signal at each frequency. A curve which crosses the threshold value at a given frequency will therefore represent a detectable signal, assuming an approximate power law behavior. Given that GW signals from FOPTs have a broken power law shape, the method described here is typically appropriate to visualize the sensitivity of a given experiment. In Fig.~\ref{fig:pli} we summarize the PLI curves for the different GW interferometers considered here.

\begin{table}[htbp]
\centering
\begin{tabular}{|c|c|c|c|c|c|}
\hline
 & $n_{\mathrm{det}}$ & $t_{obs}$(months) & $\Delta f$ (Hz) & $\mathcal{R}(f), \Gamma_{IJ}(f), D_{\mathrm{noise}}(f),S_\mathrm{noise}(f)$ 
 \\ \hline
LISA   & 1 & 48 & $[10^{-5},1]$    
&\cite{Cornish:2018dyw, Caprini:2015zlo, Caprini:2019egz} 
\\ \hline
DECIGO & 2 & 48 & $[10^{-3},10]$    
& \cite{Yagi:2013du, Kuroyanagi:2014qza} 
\\ \hline
BBO    & 2 & 48 & $[10^{-3},10]$    
& \cite{Yagi:2011yu,Thrane:2013oya} 
\\ \hline
CE     & 1 & 60 & $[4.98, 5000]$    
& \cite{Evans:2016mbw} 
\\ \hline
ET     & 2 & 60 & $[1.12, 7066.72]$ 
& \cite{Sathyaprakash:2012jk} 
\\ \hline
LIGO   & 2 & 20 & $[4.98, 4978]$    
& \cite{TheLIGOScientific:2014jea,Aasi:2013wya, Thrane:2013oya,Nishizawa:2009bf, Himemoto:2017gnw} 
\\ \hline
AEDGE   & 2 & 60 & $[0.006, 14.83]$    
& \cite{Bertoldi:2019tck} 
\\ \hline
AION   & 1 & 60 & $[0.1, 1.84]$    
& \cite{Badurina:2019hst} 
\\ \hline
\end{tabular}
\caption{Summary of the experimental parameters used in generating PLI curves for GW detection in this work. Auto/cross-correlation measurement is indicated by $n_\mathrm{det}=1(2)$, the observation time and bandwidth for each experiment are presented above. The detector response $\mathcal{R}(f)$, multiple detector overlap function $\Gamma_{IJ}(f)$ and noise strain spectrum $ D_{\mathrm{noise}}(f),S_\mathrm{noise}(f)$ are extracted from the references herein.  \label{tab:summary}}	
\end{table}

\subsubsection{Experimental parameters} 
\label{ssub:current_and_future_gw_experiments}

For completeness we recompute here the PLI curves as described in the previous section for ground based interferometers such as The Laser Interferometer Gravitational Wave Observatory (LIGO)~\cite{TheLIGOScientific:2014jea}, the Einstein Telescope (ET)~\cite{Sathyaprakash:2012jk}, the Cosmic Explorer (CE)~\cite{Evans:2016mbw,Reitze:2019iox} and the Atom Interferometer Observatory and Network (AION)~\cite{Badurina:2019hst}, as well as for space based detectors such as The Laser Interferometer Space Antenna (LISA)~\cite{Cornish:2018dyw}, the Big Bang Observer (BBO)~\cite{Yagi:2011yu}, the Deci-Hertz Interferometer Gravitational-Wave Observatory (DECIGO)~\cite{Seto:2001qf, Yagi:2011wg, Isoyama:2018rjb} and the Atomic Experiment for Dark Matter and Gravity Exploration (AEDGE)~\cite{Bertoldi:2019tck}. While many more experiments are planned in the future, such as AIGSO~\cite{Gao:2017rgh,Wang:2019oeu}, AMIGO~\cite{Ni:2019nau}, Taiji~\cite{Hu:2017mde}, TianGO~\cite{Kuns:2019upi}, TianQin~\cite{Luo:2015ght,Hu:2018yqb} and more, we focus on the ones cited above as representatives of the potential detection range in coming years.

In order to determine the reach of a given experiment, we need to know the frequency band, the response function of the detector $\mathcal{R}(f)$ within this band, the measured noise at every accessible frequency $D_{\text {noise }}^{I}(f)$, the time of observation $t_{\text{obs}}$ and the number of coupled detectors $n_{\text{det}}$. We report the extracted parameters for the various experiments in the Table~\ref{tab:summary}.

\bibliography{GWSUSY}

\providecommand{\href}[2]{#2}\begingroup\raggedright\begin{thebibliography}{100}

\bibitem{Baumann:2011nk}
D.~Baumann and D.~Green, \emph{{Signatures of Supersymmetry from the Early
  Universe}}, \href{https://doi.org/10.1103/PhysRevD.85.103520}{\emph{Phys.
  Rev. D} {\bfseries 85} (2012) 103520}
  [\href{https://arxiv.org/abs/1109.0292}{{\ttfamily 1109.0292}}].

\bibitem{Craig:2014rta}
N.~Craig and D.~Green, \emph{{Testing Split Supersymmetry with Inflation}},
  \href{https://doi.org/10.1007/JHEP07(2014)102}{\emph{JHEP} {\bfseries 07}
  (2014) 102} [\href{https://arxiv.org/abs/1403.7193}{{\ttfamily 1403.7193}}].

\bibitem{Craig:2009zx}
N.~J. Craig, \emph{{Gravitational Waves from Supersymmetry Breaking}},
  \href{https://arxiv.org/abs/0902.1990}{{\ttfamily 0902.1990}}.

\bibitem{Abbott:2016blz}
{\scshape LIGO Scientific, Virgo} collaboration, \emph{{Observation of
  Gravitational Waves from a Binary Black Hole Merger}},
  \href{https://doi.org/10.1103/PhysRevLett.116.061102}{\emph{Phys. Rev. Lett.}
  {\bfseries 116} (2016) 061102}
  [\href{https://arxiv.org/abs/1602.03837}{{\ttfamily 1602.03837}}].

\bibitem{Abbott:2017mem}
{\scshape LIGO Scientific, Virgo} collaboration, \emph{{Constraints on cosmic
  strings using data from the first Advanced LIGO observing run}},
  \href{https://doi.org/10.1103/PhysRevD.97.102002}{\emph{Phys. Rev. D}
  {\bfseries 97} (2018) 102002}
  [\href{https://arxiv.org/abs/1712.01168}{{\ttfamily 1712.01168}}].

\bibitem{LIGOScientific:2019vic}
{\scshape LIGO Scientific, Virgo} collaboration, \emph{{Search for the
  isotropic stochastic background using data from Advanced LIGO's second
  observing run}},
  \href{https://doi.org/10.1103/PhysRevD.100.061101}{\emph{Phys. Rev. D}
  {\bfseries 100} (2019) 061101}
  [\href{https://arxiv.org/abs/1903.02886}{{\ttfamily 1903.02886}}].

\bibitem{Witten:1984rs}
E.~Witten, \emph{{Cosmic Separation of Phases}},
  \href{https://doi.org/10.1103/PhysRevD.30.272}{\emph{Phys. Rev. D} {\bfseries
  30} (1984) 272}.

\bibitem{Hogan:1984hx}
C.~Hogan, \emph{{NUCLEATION OF COSMOLOGICAL PHASE TRANSITIONS}},
  \href{https://doi.org/10.1016/0370-2693(83)90553-1}{\emph{Phys. Lett. B}
  {\bfseries 133} (1983) 172}.

\bibitem{Hogan:1986qda}
C.~Hogan, \emph{{Gravitational radiation from cosmological phase transitions}},
  {\emph{Mon. Not. Roy. Astron. Soc.} {\bfseries 218} (1986) 629}.

\bibitem{Turner:1990rc}
M.~S. Turner and F.~Wilczek, \emph{{Relic gravitational waves and extended
  inflation}}, \href{https://doi.org/10.1103/PhysRevLett.65.3080}{\emph{Phys.
  Rev. Lett.} {\bfseries 65} (1990) 3080}.

\bibitem{Caprini:2015zlo}
C.~Caprini et~al., \emph{{Science with the space-based interferometer eLISA.
  II: Gravitational waves from cosmological phase transitions}},
  \href{https://doi.org/10.1088/1475-7516/2016/04/001}{\emph{JCAP} {\bfseries
  04} (2016) 001} [\href{https://arxiv.org/abs/1512.06239}{{\ttfamily
  1512.06239}}].

\bibitem{Mazumdar:2018dfl}
A.~Mazumdar and G.~White, \emph{{Review of cosmic phase transitions: their
  significance and experimental signatures}},
  \href{https://doi.org/10.1088/1361-6633/ab1f55}{\emph{Rept. Prog. Phys.}
  {\bfseries 82} (2019) 076901}
  [\href{https://arxiv.org/abs/1811.01948}{{\ttfamily 1811.01948}}].

\bibitem{Nelson:1993nf}
A.~E. Nelson and N.~Seiberg, \emph{{R symmetry breaking versus supersymmetry
  breaking}}, \href{https://doi.org/10.1016/0550-3213(94)90577-0}{\emph{Nucl.
  Phys. B} {\bfseries 416} (1994) 46}
  [\href{https://arxiv.org/abs/hep-ph/9309299}{{\ttfamily hep-ph/9309299}}].

\bibitem{Intriligator:2007py}
K.~A. Intriligator, N.~Seiberg and D.~Shih, \emph{{Supersymmetry breaking,
  R-symmetry breaking and metastable vacua}},
  \href{https://doi.org/10.1088/1126-6708/2007/07/017}{\emph{JHEP} {\bfseries
  07} (2007) 017} [\href{https://arxiv.org/abs/hep-th/0703281}{{\ttfamily
  hep-th/0703281}}].

\bibitem{Komargodski:2009jf}
Z.~Komargodski and D.~Shih, \emph{{Notes on SUSY and R-Symmetry Breaking in
  Wess-Zumino Models}},
  \href{https://doi.org/10.1088/1126-6708/2009/04/093}{\emph{JHEP} {\bfseries
  04} (2009) 093} [\href{https://arxiv.org/abs/0902.0030}{{\ttfamily
  0902.0030}}].

\bibitem{TheLIGOScientific:2014jea}
{\scshape LIGO Scientific} collaboration, \emph{{Advanced Ligo}},
  \href{https://doi.org/10.1088/0264-9381/32/7/074001}{\emph{Class. Quant.
  Grav.} {\bfseries 32} (2015) 074001}
  [\href{https://arxiv.org/abs/1411.4547}{{\ttfamily 1411.4547}}].

\bibitem{Sathyaprakash:2012jk}
B.~Sathyaprakash et~al., \emph{{Scientific Objectives of Einstein Telescope}},
  \href{https://doi.org/10.1088/0264-9381/29/12/124013}{\emph{Class. Quant.
  Grav.} {\bfseries 29} (2012) 124013}
  [\href{https://arxiv.org/abs/1206.0331}{{\ttfamily 1206.0331}}].

\bibitem{Evans:2016mbw}
{\scshape LIGO Scientific} collaboration, \emph{{Exploring the Sensitivity of
  Next Generation Gravitational Wave Detectors}},
  \href{https://doi.org/10.1088/1361-6382/aa51f4}{\emph{Class. Quant. Grav.}
  {\bfseries 34} (2017) 044001}
  [\href{https://arxiv.org/abs/1607.08697}{{\ttfamily 1607.08697}}].

\bibitem{Reitze:2019iox}
D.~Reitze et~al., \emph{{Cosmic Explorer: the U.S. Contribution to
  Gravitational-Wave Astronomy Beyond Ligo}}, {\emph{Bull. Am. Astron. Soc.}
  {\bfseries 51} (2019) 035}
  [\href{https://arxiv.org/abs/1907.04833}{{\ttfamily 1907.04833}}].

\bibitem{Chung:2012vg}
D.~J. Chung, A.~J. Long and L.-T. Wang, \emph{{125~GeV Higgs boson and
  electroweak phase transition model classes}},
  \href{https://doi.org/10.1103/PhysRevD.87.023509}{\emph{Phys. Rev. D}
  {\bfseries 87} (2013) 023509}
  [\href{https://arxiv.org/abs/1209.1819}{{\ttfamily 1209.1819}}].

\bibitem{Brignole:1997sk}
A.~Brignole, F.~Feruglio and F.~Zwirner, \emph{{Signals of a superlight
  gravitino at $e^+ e^-$ colliders when the other superparticles are heavy}},
  \href{https://doi.org/10.1016/S0550-3213(97)00825-0}{\emph{Nucl. Phys. B}
  {\bfseries 516} (1998) 13}
  [\href{https://arxiv.org/abs/hep-ph/9711516}{{\ttfamily hep-ph/9711516}}].

\bibitem{Brignole:1998me}
A.~Brignole, F.~Feruglio, M.~L. Mangano and F.~Zwirner, \emph{{Signals of a
  superlight gravitino at hadron colliders when the other superparticles are
  heavy}}, \href{https://doi.org/10.1016/S0550-3213(98)00254-5}{\emph{Nucl.
  Phys. B} {\bfseries 526} (1998) 136}
  [\href{https://arxiv.org/abs/hep-ph/9801329}{{\ttfamily hep-ph/9801329}}].

\bibitem{Maltoni:2015twa}
F.~Maltoni, A.~Martini, K.~Mawatari and B.~Oexl, \emph{{Signals of a superlight
  gravitino at the LHC}},
  \href{https://doi.org/10.1007/JHEP04(2015)021}{\emph{JHEP} {\bfseries 04}
  (2015) 021} [\href{https://arxiv.org/abs/1502.01637}{{\ttfamily
  1502.01637}}].

\bibitem{Aaboud:2018doq}
{\scshape ATLAS} collaboration, \emph{{Search for photonic signatures of
  gauge-mediated supersymmetry in 13 TeV $pp$ collisions with the ATLAS
  detector}}, \href{https://doi.org/10.1103/PhysRevD.97.092006}{\emph{Phys.
  Rev. D} {\bfseries 97} (2018) 092006}
  [\href{https://arxiv.org/abs/1802.03158}{{\ttfamily 1802.03158}}].

\bibitem{Aaboud:2018mna}
{\scshape ATLAS} collaboration, \emph{{Search for squarks and gluinos in final
  states with hadronically decaying $\tau$-leptons, jets, and missing
  transverse momentum using $pp$ collisions at $\sqrt{s}$ = 13 TeV with the
  ATLAS detector}},
  \href{https://doi.org/10.1103/PhysRevD.99.012009}{\emph{Phys. Rev. D}
  {\bfseries 99} (2019) 012009}
  [\href{https://arxiv.org/abs/1808.06358}{{\ttfamily 1808.06358}}].

\bibitem{ATLAS:2019vcq}
{\scshape ATLAS} collaboration, \emph{{Search for squarks and gluinos in final
  states with jets and missing transverse momentum using 139 fb$^{-1}$ of
  $\sqrt{s}$ =13 TeV $pp$ collision data with the ATLAS detector}}, .

\bibitem{ATLAS-CONF-2020-047}
{\scshape ATLAS Collaboration} collaboration, \emph{{Search for squarks and
  gluinos in final states with an isolated lepton, jets, and missing transverse
  momentum at $\sqrt{s}=13$ TeV with the ATLAS detector}},  Tech. Rep.
  ATLAS-CONF-2020-047, CERN, Geneva, Aug, 2020.

\bibitem{Arkani-Hamed:2015vfh}
N.~Arkani-Hamed, T.~Han, M.~Mangano and L.-T. Wang, \emph{{Physics
  opportunities of a 100 TeV proton\textendash{}proton collider}},
  \href{https://doi.org/10.1016/j.physrep.2016.07.004}{\emph{Phys. Rept.}
  {\bfseries 652} (2016) 1} [\href{https://arxiv.org/abs/1511.06495}{{\ttfamily
  1511.06495}}].

\bibitem{Hall:2013uga}
L.~J. Hall, J.~T. Ruderman and T.~Volansky, \emph{{A Cosmological Upper Bound
  on Superpartner Masses}},
  \href{https://doi.org/10.1007/JHEP02(2015)094}{\emph{JHEP} {\bfseries 02}
  (2015) 094} [\href{https://arxiv.org/abs/1302.2620}{{\ttfamily 1302.2620}}].

\bibitem{Jedamzik:2006xz}
K.~Jedamzik, \emph{{Big bang nucleosynthesis constraints on hadronically and
  electromagnetically decaying relic neutral particles}},
  \href{https://doi.org/10.1103/PhysRevD.74.103509}{\emph{Phys. Rev. D}
  {\bfseries 74} (2006) 103509}
  [\href{https://arxiv.org/abs/hep-ph/0604251}{{\ttfamily hep-ph/0604251}}].

\bibitem{Deser:1977uq}
S.~Deser and B.~Zumino, \emph{{Broken Supersymmetry and Supergravity}},
  \href{https://doi.org/10.1103/PhysRevLett.38.1433}{\emph{Phys. Rev. Lett.}
  {\bfseries 38} (1977) 1433}.

\bibitem{Bagger:1994hh}
J.~Bagger, E.~Poppitz and L.~Randall, \emph{{The R axion from dynamical
  supersymmetry breaking}},
  \href{https://doi.org/10.1016/0550-3213(94)90123-6}{\emph{Nucl. Phys. B}
  {\bfseries 426} (1994) 3}
  [\href{https://arxiv.org/abs/hep-ph/9405345}{{\ttfamily hep-ph/9405345}}].

\bibitem{Bellazzini:2017neg}
B.~Bellazzini, A.~Mariotti, D.~Redigolo, F.~Sala and J.~Serra, \emph{{$R$-axion
  at colliders}},
  \href{https://doi.org/10.1103/PhysRevLett.119.141804}{\emph{Phys. Rev. Lett.}
  {\bfseries 119} (2017) 141804}
  [\href{https://arxiv.org/abs/1702.02152}{{\ttfamily 1702.02152}}].

\bibitem{Dine:2009sw}
M.~Dine, G.~Festuccia and Z.~Komargodski, \emph{{A Bound on the
  Superpotential}}, \href{https://doi.org/10.1007/JHEP03(2010)011}{\emph{JHEP}
  {\bfseries 03} (2010) 011} [\href{https://arxiv.org/abs/0910.2527}{{\ttfamily
  0910.2527}}].

\bibitem{Giudice:1998bp}
G.~Giudice and R.~Rattazzi, \emph{{Theories with gauge mediated supersymmetry
  breaking}}, \href{https://doi.org/10.1016/S0370-1573(99)00042-3}{\emph{Phys.
  Rept.} {\bfseries 322} (1999) 419}
  [\href{https://arxiv.org/abs/hep-ph/9801271}{{\ttfamily hep-ph/9801271}}].

\bibitem{Brignole:1997pe}
A.~Brignole, F.~Feruglio and F.~Zwirner, \emph{{On the effective interactions
  of a light gravitino with matter fermions}},
  \href{https://doi.org/10.1088/1126-6708/1997/11/001}{\emph{JHEP} {\bfseries
  11} (1997) 001} [\href{https://arxiv.org/abs/hep-th/9709111}{{\ttfamily
  hep-th/9709111}}].

\bibitem{Moroi:1993mb}
T.~Moroi, H.~Murayama and M.~Yamaguchi, \emph{{Cosmological constraints on the
  light stable gravitino}},
  \href{https://doi.org/10.1016/0370-2693(93)91434-O}{\emph{Phys. Lett. B}
  {\bfseries 303} (1993) 289}.

\bibitem{Kawasaki:1994af}
M.~Kawasaki and T.~Moroi, \emph{{Gravitino production in the inflationary
  universe and the effects on big bang nucleosynthesis}},
  \href{https://doi.org/10.1143/PTP.93.879}{\emph{Prog. Theor. Phys.}
  {\bfseries 93} (1995) 879}
  [\href{https://arxiv.org/abs/hep-ph/9403364}{{\ttfamily hep-ph/9403364}}].

\bibitem{Moroi:1995fs}
T.~Moroi, \emph{{Effects of the gravitino on the inflationary universe}},
  other thesis, 3, 1995.

\bibitem{Pierpaoli:1997im}
E.~Pierpaoli, S.~Borgani, A.~Masiero and M.~Yamaguchi, \emph{{The Formation of
  cosmic structures in a light gravitino dominated universe}},
  \href{https://doi.org/10.1103/PhysRevD.57.2089}{\emph{Phys. Rev. D}
  {\bfseries 57} (1998) 2089}
  [\href{https://arxiv.org/abs/astro-ph/9709047}{{\ttfamily
  astro-ph/9709047}}].

\bibitem{Viel:2005qj}
M.~Viel, J.~Lesgourgues, M.~G. Haehnelt, S.~Matarrese and A.~Riotto,
  \emph{{Constraining warm dark matter candidates including sterile neutrinos
  and light gravitinos with WMAP and the Lyman-alpha forest}},
  \href{https://doi.org/10.1103/PhysRevD.71.063534}{\emph{Phys. Rev. D}
  {\bfseries 71} (2005) 063534}
  [\href{https://arxiv.org/abs/astro-ph/0501562}{{\ttfamily
  astro-ph/0501562}}].

\bibitem{Bolz:2000fu}
M.~Bolz, A.~Brandenburg and W.~Buchmuller, \emph{{Thermal production of
  gravitinos}},
  \href{https://doi.org/10.1016/S0550-3213(01)00132-8}{\emph{Nucl. Phys. B}
  {\bfseries 606} (2001) 518}
  [\href{https://arxiv.org/abs/hep-ph/0012052}{{\ttfamily hep-ph/0012052}}].

\bibitem{Pradler:2006qh}
J.~Pradler and F.~D. Steffen, \emph{{Thermal gravitino production and collider
  tests of leptogenesis}},
  \href{https://doi.org/10.1103/PhysRevD.75.023509}{\emph{Phys. Rev. D}
  {\bfseries 75} (2007) 023509}
  [\href{https://arxiv.org/abs/hep-ph/0608344}{{\ttfamily hep-ph/0608344}}].

\bibitem{Pradler:2006hh}
J.~Pradler and F.~D. Steffen, \emph{{Constraints on the Reheating Temperature
  in Gravitino Dark Matter Scenarios}},
  \href{https://doi.org/10.1016/j.physletb.2007.02.072}{\emph{Phys. Lett. B}
  {\bfseries 648} (2007) 224}
  [\href{https://arxiv.org/abs/hep-ph/0612291}{{\ttfamily hep-ph/0612291}}].

\bibitem{Rychkov:2007uq}
V.~S. Rychkov and A.~Strumia, \emph{{Thermal production of gravitinos}},
  \href{https://doi.org/10.1103/PhysRevD.75.075011}{\emph{Phys. Rev. D}
  {\bfseries 75} (2007) 075011}
  [\href{https://arxiv.org/abs/hep-ph/0701104}{{\ttfamily hep-ph/0701104}}].

\bibitem{Cheung:2011nn}
C.~Cheung, G.~Elor and L.~Hall, \emph{{Gravitino Freeze-In}},
  \href{https://doi.org/10.1103/PhysRevD.84.115021}{\emph{Phys. Rev. D}
  {\bfseries 84} (2011) 115021}
  [\href{https://arxiv.org/abs/1103.4394}{{\ttfamily 1103.4394}}].

\bibitem{Feng:2003xh}
J.~L. Feng, A.~Rajaraman and F.~Takayama, \emph{{Superweakly interacting
  massive particles}},
  \href{https://doi.org/10.1103/PhysRevLett.91.011302}{\emph{Phys. Rev. Lett.}
  {\bfseries 91} (2003) 011302}
  [\href{https://arxiv.org/abs/hep-ph/0302215}{{\ttfamily hep-ph/0302215}}].

\bibitem{Feng:2003uy}
J.~L. Feng, A.~Rajaraman and F.~Takayama, \emph{{SuperWIMP dark matter signals
  from the early universe}},
  \href{https://doi.org/10.1103/PhysRevD.68.063504}{\emph{Phys. Rev. D}
  {\bfseries 68} (2003) 063504}
  [\href{https://arxiv.org/abs/hep-ph/0306024}{{\ttfamily hep-ph/0306024}}].

\bibitem{ArkaniHamed:2006mb}
N.~Arkani-Hamed, A.~Delgado and G.~Giudice, \emph{{The Well-tempered
  neutralino}},
  \href{https://doi.org/10.1016/j.nuclphysb.2006.02.010}{\emph{Nucl. Phys. B}
  {\bfseries 741} (2006) 108}
  [\href{https://arxiv.org/abs/hep-ph/0601041}{{\ttfamily hep-ph/0601041}}].

\bibitem{ArkaniHamed:1998kj}
N.~Arkani-Hamed, G.~F. Giudice, M.~A. Luty and R.~Rattazzi,
  \emph{{Supersymmetry breaking loops from analytic continuation into
  superspace}}, \href{https://doi.org/10.1103/PhysRevD.58.115005}{\emph{Phys.
  Rev. D} {\bfseries 58} (1998) 115005}
  [\href{https://arxiv.org/abs/hep-ph/9803290}{{\ttfamily hep-ph/9803290}}].

\bibitem{Cohen:2011aa}
T.~Cohen, A.~Hook and B.~Wecht, \emph{{Comments on Gaugino Screening}},
  \href{https://doi.org/10.1103/PhysRevD.85.115004}{\emph{Phys. Rev. D}
  {\bfseries 85} (2012) 115004}
  [\href{https://arxiv.org/abs/1112.1699}{{\ttfamily 1112.1699}}].

\bibitem{Martin:1993yx}
S.~P. Martin and M.~T. Vaughn, \emph{{Regularization dependence of running
  couplings in softly broken supersymmetry}},
  \href{https://doi.org/10.1016/0370-2693(93)90136-6}{\emph{Phys. Lett. B}
  {\bfseries 318} (1993) 331}
  [\href{https://arxiv.org/abs/hep-ph/9308222}{{\ttfamily hep-ph/9308222}}].

\bibitem{Martin:1997ns}
S.~P. Martin, \emph{{A Supersymmetry primer}},
  \href{https://arxiv.org/abs/hep-ph/9709356}{{\ttfamily hep-ph/9709356}}.

\bibitem{Barbieri:1995tw}
R.~Barbieri, L.~J. Hall and A.~Strumia, \emph{{Violations of lepton flavor and
  CP in supersymmetric unified theories}},
  \href{https://doi.org/10.1016/0550-3213(95)00208-A}{\emph{Nucl. Phys. B}
  {\bfseries 445} (1995) 219}
  [\href{https://arxiv.org/abs/hep-ph/9501334}{{\ttfamily hep-ph/9501334}}].

\bibitem{Hisano:1995nq}
J.~Hisano, T.~Moroi, K.~Tobe, M.~Yamaguchi and T.~Yanagida, \emph{{Lepton
  flavor violation in the supersymmetric standard model with seesaw induced
  neutrino masses}},
  \href{https://doi.org/10.1016/0370-2693(95)00954-J}{\emph{Phys. Lett. B}
  {\bfseries 357} (1995) 579}
  [\href{https://arxiv.org/abs/hep-ph/9501407}{{\ttfamily hep-ph/9501407}}].

\bibitem{Gabbiani:1996hi}
F.~Gabbiani, E.~Gabrielli, A.~Masiero and L.~Silvestrini, \emph{{A Complete
  analysis of FCNC and CP constraints in general SUSY extensions of the
  standard model}},
  \href{https://doi.org/10.1016/0550-3213(96)00390-2}{\emph{Nucl. Phys. B}
  {\bfseries 477} (1996) 321}
  [\href{https://arxiv.org/abs/hep-ph/9604387}{{\ttfamily hep-ph/9604387}}].

\bibitem{Ciuchini:1998ix}
M.~Ciuchini et~al., \emph{{Delta M(K) and epsilon(K) in SUSY at the
  next-to-leading order}},
  \href{https://doi.org/10.1088/1126-6708/1998/10/008}{\emph{JHEP} {\bfseries
  10} (1998) 008} [\href{https://arxiv.org/abs/hep-ph/9808328}{{\ttfamily
  hep-ph/9808328}}].

\bibitem{Baron:2013eja}
{\scshape ACME} collaboration, \emph{{Order of Magnitude Smaller Limit on the
  Electric Dipole Moment of the Electron}},
  \href{https://doi.org/10.1126/science.1248213}{\emph{Science} {\bfseries 343}
  (2014) 269} [\href{https://arxiv.org/abs/1310.7534}{{\ttfamily 1310.7534}}].

\bibitem{Andreev:2018ayy}
{\scshape ACME} collaboration, \emph{{Improved limit on the electric dipole
  moment of the electron}},
  \href{https://doi.org/10.1038/s41586-018-0599-8}{\emph{Nature} {\bfseries
  562} (2018) 355}.

\bibitem{Nakai:2016atk}
Y.~Nakai and M.~Reece, \emph{{Electric Dipole Moments in Natural
  Supersymmetry}}, \href{https://doi.org/10.1007/JHEP08(2017)031}{\emph{JHEP}
  {\bfseries 08} (2017) 031}
  [\href{https://arxiv.org/abs/1612.08090}{{\ttfamily 1612.08090}}].

\bibitem{Cesarotti:2018huy}
C.~Cesarotti, Q.~Lu, Y.~Nakai, A.~Parikh and M.~Reece, \emph{{Interpreting the
  Electron EDM Constraint}},
  \href{https://doi.org/10.1007/JHEP05(2019)059}{\emph{JHEP} {\bfseries 05}
  (2019) 059} [\href{https://arxiv.org/abs/1810.07736}{{\ttfamily
  1810.07736}}].

\bibitem{Coleman:1977py}
S.~R. Coleman, \emph{{The Fate of the False Vacuum. 1. Semiclassical Theory}},
  \href{https://doi.org/10.1103/PhysRevD.16.1248}{\emph{Phys. Rev. D}
  {\bfseries 15} (1977) 2929}.

\bibitem{Callan:1977pt}
J.~Callan, Curtis~G. and S.~R. Coleman, \emph{{The Fate of the False Vacuum. 2.
  First Quantum Corrections}},
  \href{https://doi.org/10.1103/PhysRevD.16.1762}{\emph{Phys. Rev. D}
  {\bfseries 16} (1977) 1762}.

\bibitem{Linde:1980tt}
A.~D. Linde, \emph{{Fate of the False Vacuum at Finite Temperature: Theory and
  Applications}},
  \href{https://doi.org/10.1016/0370-2693(81)90281-1}{\emph{Phys. Lett. B}
  {\bfseries 100} (1981) 37}.

\bibitem{Linde:1981zj}
A.~D. Linde, \emph{{Decay of the False Vacuum at Finite Temperature}},
  \href{https://doi.org/10.1016/0550-3213(83)90072-X}{\emph{Nucl. Phys. B}
  {\bfseries 216} (1983) 421}.

\bibitem{Kamionkowski:1993fg}
M.~Kamionkowski, A.~Kosowsky and M.~S. Turner, \emph{{Gravitational Radiation
  from First Order Phase Transitions}},
  \href{https://doi.org/10.1103/PhysRevD.49.2837}{\emph{Phys. Rev. D}
  {\bfseries 49} (1994) 2837}
  [\href{https://arxiv.org/abs/astro-ph/9310044}{{\ttfamily
  astro-ph/9310044}}].

\bibitem{Arnold:1993wc}
P.~B. Arnold, \emph{{One loop fluctuation - dissipation formula for bubble wall
  velocity}}, \href{https://doi.org/10.1103/PhysRevD.48.1539}{\emph{Phys. Rev.
  D} {\bfseries 48} (1993) 1539}
  [\href{https://arxiv.org/abs/hep-ph/9302258}{{\ttfamily hep-ph/9302258}}].

\bibitem{Mancha:2020fzw}
M.~Barroso~Mancha, T.~Prokopec and B.~Swiezewska, \emph{{Field theoretic
  derivation of bubble wall force}},
  \href{https://arxiv.org/abs/2005.10875}{{\ttfamily 2005.10875}}.

\bibitem{Bodeker:2009qy}
D.~Bodeker and G.~D. Moore, \emph{{Can Electroweak Bubble Walls Run Away?}},
  \href{https://doi.org/10.1088/1475-7516/2009/05/009}{\emph{JCAP} {\bfseries
  05} (2009) 009} [\href{https://arxiv.org/abs/0903.4099}{{\ttfamily
  0903.4099}}].

\bibitem{Bodeker:2017cim}
D.~Bodeker and G.~D. Moore, \emph{{Electroweak Bubble Wall Speed Limit}},
  \href{https://doi.org/10.1088/1475-7516/2017/05/025}{\emph{JCAP} {\bfseries
  05} (2017) 025} [\href{https://arxiv.org/abs/1703.08215}{{\ttfamily
  1703.08215}}].

\bibitem{Darme:2017wvu}
L.~Darm\'e, J.~Jaeckel and M.~Lewicki, \emph{{Towards the fate of the
  oscillating false vacuum}},
  \href{https://doi.org/10.1103/PhysRevD.96.056001}{\emph{Phys. Rev. D}
  {\bfseries 96} (2017) 056001}
  [\href{https://arxiv.org/abs/1704.06445}{{\ttfamily 1704.06445}}].

\bibitem{Ellis:2019oqb}
J.~Ellis, M.~Lewicki, J.~M. No and V.~Vaskonen, \emph{{Gravitational wave
  energy budget in strongly supercooled phase transitions}},
  \href{https://doi.org/10.1088/1475-7516/2019/06/024}{\emph{JCAP} {\bfseries
  06} (2019) 024} [\href{https://arxiv.org/abs/1903.09642}{{\ttfamily
  1903.09642}}].

\bibitem{Enqvist:1991xw}
K.~Enqvist, J.~Ignatius, K.~Kajantie and K.~Rummukainen, \emph{{Nucleation and
  bubble growth in a first order cosmological electroweak phase transition}},
  \href{https://doi.org/10.1103/PhysRevD.45.3415}{\emph{Phys. Rev. D}
  {\bfseries 45} (1992) 3415}.

\bibitem{Vanvlasselaer:2020niz}
A.~Azatov and M.~Vanvlasselaer, \emph{{Bubble wall velocity: heavy physics
  effects}},  \href{https://arxiv.org/abs/2010.02590}{{\ttfamily 2010.02590}}.

\bibitem{Hindmarsh:2013xza}
M.~Hindmarsh, S.~J. Huber, K.~Rummukainen and D.~J. Weir, \emph{{Gravitational
  waves from the sound of a first order phase transition}},
  \href{https://doi.org/10.1103/PhysRevLett.112.041301}{\emph{Phys. Rev. Lett.}
  {\bfseries 112} (2014) 041301}
  [\href{https://arxiv.org/abs/1304.2433}{{\ttfamily 1304.2433}}].

\bibitem{Hindmarsh:2015qta}
M.~Hindmarsh, S.~J. Huber, K.~Rummukainen and D.~J. Weir, \emph{{Numerical
  Simulations of Acoustically Generated Gravitational Waves at a First Order
  Phase Transition}},
  \href{https://doi.org/10.1103/PhysRevD.92.123009}{\emph{Phys. Rev. D}
  {\bfseries 92} (2015) 123009}
  [\href{https://arxiv.org/abs/1504.03291}{{\ttfamily 1504.03291}}].

\bibitem{Caprini:2019egz}
C.~Caprini et~al., \emph{{Detecting gravitational waves from cosmological phase
  transitions with LISA: an update}},
  \href{https://doi.org/10.1088/1475-7516/2020/03/024}{\emph{JCAP} {\bfseries
  03} (2020) 024} [\href{https://arxiv.org/abs/1910.13125}{{\ttfamily
  1910.13125}}].

\bibitem{Hook:2015tra}
A.~Hook and H.~Murayama, \emph{{Low-energy Supersymmetry Breaking Without the
  Gravitino Problem}},
  \href{https://doi.org/10.1103/PhysRevD.92.015004}{\emph{Phys. Rev. D}
  {\bfseries 92} (2015) 015004}
  [\href{https://arxiv.org/abs/1503.04880}{{\ttfamily 1503.04880}}].

\bibitem{Hook:2018sai}
A.~Hook, R.~McGehee and H.~Murayama, \emph{{Cosmologically Viable Low-energy
  Supersymmetry Breaking}},
  \href{https://doi.org/10.1103/PhysRevD.98.115036}{\emph{Phys. Rev. D}
  {\bfseries 98} (2018) 115036}
  [\href{https://arxiv.org/abs/1801.10160}{{\ttfamily 1801.10160}}].

\bibitem{Duncan:1992ai}
M.~J. Duncan and L.~G. Jensen, \emph{{Exact Tunneling Solutions in Scalar Field
  Theory}}, \href{https://doi.org/10.1016/0370-2693(92)90128-Q}{\emph{Phys.
  Lett.} {\bfseries B291} (1992) 109}.

\bibitem{Amariti:2009kb}
A.~Amariti and M.~Siani, \emph{{R-symmetry and supersymmetry breaking in 3D WZ
  models}}, \href{https://doi.org/10.1088/1126-6708/2009/08/055}{\emph{JHEP}
  {\bfseries 08} (2009) 055} [\href{https://arxiv.org/abs/0905.4725}{{\ttfamily
  0905.4725}}].

\bibitem{Curtin:2012yu}
D.~Curtin, Z.~Komargodski, D.~Shih and Y.~Tsai, \emph{{Spontaneous R-symmetry
  Breaking with Multiple Pseudomoduli}},
  \href{https://doi.org/10.1103/PhysRevD.85.125031}{\emph{Phys. Rev. D}
  {\bfseries 85} (2012) 125031}
  [\href{https://arxiv.org/abs/1202.5331}{{\ttfamily 1202.5331}}].

\bibitem{Witten:1981kv}
E.~Witten, \emph{{Mass Hierarchies in Supersymmetric Theories}},
  \href{https://doi.org/10.1016/0370-2693(81)90885-6}{\emph{Phys. Lett. B}
  {\bfseries 105} (1981) 267}.

\bibitem{Rattazzi:2000hs}
R.~Rattazzi and A.~Zaffaroni, \emph{{Comments on the holographic picture of the
  Randall-Sundrum model}},
  \href{https://doi.org/10.1088/1126-6708/2001/04/021}{\emph{JHEP} {\bfseries
  04} (2001) 021} [\href{https://arxiv.org/abs/hep-th/0012248}{{\ttfamily
  hep-th/0012248}}].

\bibitem{Guada:2020xnz}
V.~Guada, M.~Nemev\v{s}ek and M.~Pintar, \emph{{FindBounce: Package for
  multi-field bounce actions}},
  \href{https://doi.org/10.1016/j.cpc.2020.107480}{\emph{Comput. Phys. Commun.}
  {\bfseries 256} (2020) 107480}
  [\href{https://arxiv.org/abs/2002.00881}{{\ttfamily 2002.00881}}].

\bibitem{Wainwright:2011kj}
C.~L. Wainwright, \emph{{CosmoTransitions: Computing Cosmological Phase
  Transition Temperatures and Bubble Profiles with Multiple Fields}},
  \href{https://doi.org/10.1016/j.cpc.2012.04.004}{\emph{Comput. Phys. Commun.}
  {\bfseries 183} (2012) 2006}
  [\href{https://arxiv.org/abs/1109.4189}{{\ttfamily 1109.4189}}].

\bibitem{Intriligator:2006dd}
K.~A. Intriligator, N.~Seiberg and D.~Shih, \emph{{Dynamical SUSY breaking in
  meta-stable vacua}},
  \href{https://doi.org/10.1088/1126-6708/2006/04/021}{\emph{JHEP} {\bfseries
  04} (2006) 021} [\href{https://arxiv.org/abs/hep-th/0602239}{{\ttfamily
  hep-th/0602239}}].

\bibitem{ORaifeartaigh:1975nky}
L.~O'Raifeartaigh, \emph{{Spontaneous Symmetry Breaking for Chiral Scalar
  Superfields}},
  \href{https://doi.org/10.1016/0550-3213(75)90585-4}{\emph{Nucl. Phys. B}
  {\bfseries 96} (1975) 331}.

\bibitem{Craig:2006kx}
N.~J. Craig, P.~J. Fox and J.~G. Wacker, \emph{{Reheating Metastable
  O'Raifeartaigh Models}},
  \href{https://doi.org/10.1103/PhysRevD.75.085006}{\emph{Phys. Rev. D}
  {\bfseries 75} (2007) 085006}
  [\href{https://arxiv.org/abs/hep-th/0611006}{{\ttfamily hep-th/0611006}}].

\bibitem{Katz:2009gh}
A.~Katz, \emph{{On the Thermal History of Calculable Gauge Mediation}},
  \href{https://doi.org/10.1088/1126-6708/2009/10/054}{\emph{JHEP} {\bfseries
  10} (2009) 054} [\href{https://arxiv.org/abs/0907.3930}{{\ttfamily
  0907.3930}}].

\bibitem{McCullough:2010wf}
M.~McCullough, \emph{{Stimulated Supersymmetry Breaking}},
  \href{https://doi.org/10.1103/PhysRevD.82.115016}{\emph{Phys. Rev. D}
  {\bfseries 82} (2010) 115016}
  [\href{https://arxiv.org/abs/1010.3203}{{\ttfamily 1010.3203}}].

\bibitem{Vaknin:2014fxa}
T.~Vaknin, \emph{{New Phases in O`raifeartaigh-Like Models and R-Symmetry
  Breaking}}, \href{https://doi.org/10.1007/JHEP09(2014)004}{\emph{JHEP}
  {\bfseries 09} (2014) 004} [\href{https://arxiv.org/abs/1402.5851}{{\ttfamily
  1402.5851}}].

\bibitem{Dumitrescu:2010ca}
T.~T. Dumitrescu, Z.~Komargodski and M.~Sudano, \emph{{Global Symmetries and
  D-Terms in Supersymmetric Field Theories}},
  \href{https://doi.org/10.1007/JHEP11(2010)052}{\emph{JHEP} {\bfseries 11}
  (2010) 052} [\href{https://arxiv.org/abs/1007.5352}{{\ttfamily 1007.5352}}].

\bibitem{Komargodski:2009pc}
Z.~Komargodski and N.~Seiberg, \emph{{Comments on the Fayet-Iliopoulos Term in
  Field Theory and Supergravity}},
  \href{https://doi.org/10.1088/1126-6708/2009/06/007}{\emph{JHEP} {\bfseries
  06} (2009) 007} [\href{https://arxiv.org/abs/0904.1159}{{\ttfamily
  0904.1159}}].

\bibitem{Alwall:2011uj}
J.~Alwall, M.~Herquet, F.~Maltoni, O.~Mattelaer and T.~Stelzer, \emph{{MadGraph
  5 : Going Beyond}},
  \href{https://doi.org/10.1007/JHEP06(2011)128}{\emph{JHEP} {\bfseries 06}
  (2011) 128} [\href{https://arxiv.org/abs/1106.0522}{{\ttfamily 1106.0522}}].

\bibitem{Alwall:2014hca}
J.~Alwall, R.~Frederix, S.~Frixione, V.~Hirschi, F.~Maltoni, O.~Mattelaer
  et~al., \emph{{The automated computation of tree-level and next-to-leading
  order differential cross sections, and their matching to parton shower
  simulations}}, \href{https://doi.org/10.1007/JHEP07(2014)079}{\emph{JHEP}
  {\bfseries 07} (2014) 079} [\href{https://arxiv.org/abs/1405.0301}{{\ttfamily
  1405.0301}}].

\bibitem{Osato:2016ixc}
K.~Osato, T.~Sekiguchi, M.~Shirasaki, A.~Kamada and N.~Yoshida,
  \emph{{Cosmological Constraint on the Light Gravitino Mass from CMB Lensing
  and Cosmic Shear}},
  \href{https://doi.org/10.1088/1475-7516/2016/06/004}{\emph{JCAP} {\bfseries
  06} (2016) 004} [\href{https://arxiv.org/abs/1601.07386}{{\ttfamily
  1601.07386}}].

\bibitem{Aad:2014tda}
{\scshape ATLAS} collaboration, \emph{{Search for new phenomena in events with
  a photon and missing transverse momentum in $pp$ collisions at $\sqrt{s}=8$
  TeV with the ATLAS detector}},
  \href{https://doi.org/10.1103/PhysRevD.91.012008}{\emph{Phys. Rev. D}
  {\bfseries 91} (2015) 012008}
  [\href{https://arxiv.org/abs/1411.1559}{{\ttfamily 1411.1559}}].

\bibitem{Martin:1996zb}
S.~P. Martin, \emph{{Generalized Messengers of Supersymmetry Breaking and the
  Sparticle Mass Spectrum}},
  \href{https://doi.org/10.1103/PhysRevD.55.3177}{\emph{Phys. Rev.} {\bfseries
  D55} (1997) 3177} [\href{https://arxiv.org/abs/hep-ph/9608224}{{\ttfamily
  hep-ph/9608224}}].

\bibitem{Riotto:1995am}
A.~Riotto and E.~Roulet, \emph{{Vacuum decay along supersymmetric flat
  directions}}, \href{https://doi.org/10.1016/0370-2693(96)00313-9}{\emph{Phys.
  Lett. B} {\bfseries 377} (1996) 60}
  [\href{https://arxiv.org/abs/hep-ph/9512401}{{\ttfamily hep-ph/9512401}}].

\bibitem{Guada:2018jek}
V.~Guada, A.~Maiezza and M.~Nemev\v{s}ek, \emph{{Multifield Polygonal
  Bounces}}, \href{https://doi.org/10.1103/PhysRevD.99.056020}{\emph{Phys. Rev.
  D} {\bfseries 99} (2019) 056020}
  [\href{https://arxiv.org/abs/1803.02227}{{\ttfamily 1803.02227}}].

\bibitem{Amariti:2020ntv}
A.~Amariti, \emph{{Analytic bounces in d dimensions}},
  \href{https://arxiv.org/abs/2009.14102}{{\ttfamily 2009.14102}}.

\bibitem{Allen:1996vm}
B.~Allen, \emph{{The Stochastic Gravity Wave Background: Sources and
  Detection}},  in \emph{{Les Houches School of Physics: Astrophysical Sources
  of Gravitational Radiation}}, pp.~373--417, 4, 1996,
  \href{https://arxiv.org/abs/gr-qc/9604033}{{\ttfamily gr-qc/9604033}}.

\bibitem{Allen:1997ad}
B.~Allen and J.~D. Romano, \emph{{Detecting a Stochastic Background of
  Gravitational Radiation: Signal Processing Strategies and Sensitivities}},
  \href{https://doi.org/10.1103/PhysRevD.59.102001}{\emph{Phys. Rev. D}
  {\bfseries 59} (1999) 102001}
  [\href{https://arxiv.org/abs/gr-qc/9710117}{{\ttfamily gr-qc/9710117}}].

\bibitem{Maggiore:1999vm}
M.~Maggiore, \emph{{Gravitational Wave Experiments and Early Universe
  Cosmology}}, \href{https://doi.org/10.1016/S0370-1573(99)00102-7}{\emph{Phys.
  Rept.} {\bfseries 331} (2000) 283}
  [\href{https://arxiv.org/abs/gr-qc/9909001}{{\ttfamily gr-qc/9909001}}].

\bibitem{Romano:2016dpx}
J.~D. Romano and N.~J. Cornish, \emph{{Detection Methods for Stochastic
  Gravitational-Wave Backgrounds: a Unified Treatment}},
  \href{https://doi.org/10.1007/s41114-017-0004-1}{\emph{Living Rev. Rel.}
  {\bfseries 20} (2017) 2} [\href{https://arxiv.org/abs/1608.06889}{{\ttfamily
  1608.06889}}].

\bibitem{Schmitz:2020syl}
K.~Schmitz, \emph{{New Sensitivity Curves for Gravitational-Wave Experiments}},
   \href{https://arxiv.org/abs/2002.04615}{{\ttfamily 2002.04615}}.

\bibitem{Cornish:2001bb}
N.~J. Cornish, \emph{{Detecting a Stochastic Gravitational Wave Background with
  the Laser Interferometer Space Antenna}},
  \href{https://doi.org/10.1103/PhysRevD.65.022004}{\emph{Phys. Rev. D}
  {\bfseries 65} (2002) 022004}
  [\href{https://arxiv.org/abs/gr-qc/0106058}{{\ttfamily gr-qc/0106058}}].

\bibitem{Thrane:2013oya}
E.~Thrane and J.~D. Romano, \emph{{Sensitivity Curves for Searches for
  Gravitational-Wave Backgrounds}},
  \href{https://doi.org/10.1103/PhysRevD.88.124032}{\emph{Phys. Rev. D}
  {\bfseries 88} (2013) 124032}
  [\href{https://arxiv.org/abs/1310.5300}{{\ttfamily 1310.5300}}].

\bibitem{Cornish:2018dyw}
T.~Robson, N.~Cornish and C.~Liu, \emph{{The Construction and Use of Lisa
  Sensitivity Curves}},
  \href{https://doi.org/10.1088/1361-6382/ab1101}{\emph{Class. Quant. Grav.}
  {\bfseries 36} (2019) 105011}
  [\href{https://arxiv.org/abs/1803.01944}{{\ttfamily 1803.01944}}].

\bibitem{Yagi:2013du}
K.~Yagi, \emph{{Scientific Potential of Decigo Pathfinder and Testing Gr with
  Space-Borne Gravitational Wave Interferometers}},
  \href{https://doi.org/10.1142/S0218271813410137}{\emph{Int. J. Mod. Phys. D}
  {\bfseries 22} (2013) 1341013}
  [\href{https://arxiv.org/abs/1302.2388}{{\ttfamily 1302.2388}}].

\bibitem{Kuroyanagi:2014qza}
S.~Kuroyanagi, K.~Nakayama and J.~Yokoyama, \emph{{Prospects of Determination
  of Reheating Temperature After Inflation by Decigo}},
  \href{https://doi.org/10.1093/ptep/ptu176}{\emph{PTEP} {\bfseries 2015}
  (2015) 013E02} [\href{https://arxiv.org/abs/1410.6618}{{\ttfamily
  1410.6618}}].

\bibitem{Yagi:2011yu}
K.~Yagi, N.~Tanahashi and T.~Tanaka, \emph{{Probing the Size of Extra Dimension
  with Gravitational Wave Astronomy}},
  \href{https://doi.org/10.1103/PhysRevD.83.084036}{\emph{Phys. Rev. D}
  {\bfseries 83} (2011) 084036}
  [\href{https://arxiv.org/abs/1101.4997}{{\ttfamily 1101.4997}}].

\bibitem{Aasi:2013wya}
{\scshape KAGRA, LIGO Scientific, VIRGO} collaboration, \emph{{Prospects for
  Observing and Localizing Gravitational-Wave Transients with Advanced Ligo,
  Advanced Virgo and Kagra}},
  \href{https://doi.org/10.1007/s41114-018-0012-9}{\emph{Living Rev. Rel.}
  {\bfseries 21} (2018) 3} [\href{https://arxiv.org/abs/1304.0670}{{\ttfamily
  1304.0670}}].

\bibitem{Nishizawa:2009bf}
A.~Nishizawa, A.~Taruya, K.~Hayama, S.~Kawamura and M.-a. Sakagami,
  \emph{{Probing Non-Tensorial Polarizations of Stochastic Gravitational-Wave
  Backgrounds with Ground-Based Laser Interferometers}},
  \href{https://doi.org/10.1103/PhysRevD.79.082002}{\emph{Phys. Rev. D}
  {\bfseries 79} (2009) 082002}
  [\href{https://arxiv.org/abs/0903.0528}{{\ttfamily 0903.0528}}].

\bibitem{Himemoto:2017gnw}
Y.~Himemoto and A.~Taruya, \emph{{Impact of Correlated Magnetic Noise on the
  Detection of Stochastic Gravitational Waves: Estimation Based on a Simple
  Analytical Model}},
  \href{https://doi.org/10.1103/PhysRevD.96.022004}{\emph{Phys. Rev. D}
  {\bfseries 96} (2017) 022004}
  [\href{https://arxiv.org/abs/1704.07084}{{\ttfamily 1704.07084}}].

\bibitem{Bertoldi:2019tck}
{\scshape AEDGE} collaboration, \emph{{Aedge: Atomic Experiment for Dark Matter
  and Gravity Exploration in Space}},
  \href{https://doi.org/10.1140/epjqt/s40507-020-0080-0}{\emph{EPJ Quant.
  Technol.} {\bfseries 7} (2020) 6}
  [\href{https://arxiv.org/abs/1908.00802}{{\ttfamily 1908.00802}}].

\bibitem{Badurina:2019hst}
L.~Badurina et~al., \emph{{Aion: an Atom Interferometer Observatory and
  Network}}, \href{https://doi.org/10.1088/1475-7516/2020/05/011}{\emph{JCAP}
  {\bfseries 05} (2020) 011}
  [\href{https://arxiv.org/abs/1911.11755}{{\ttfamily 1911.11755}}].

\bibitem{Seto:2001qf}
N.~Seto, S.~Kawamura and T.~Nakamura, \emph{{Possibility of Direct Measurement
  of the Acceleration of the Universe Using 0.1-Hz Band Laser Interferometer
  Gravitational Wave Antenna in Space}},
  \href{https://doi.org/10.1103/PhysRevLett.87.221103}{\emph{Phys. Rev. Lett.}
  {\bfseries 87} (2001) 221103}
  [\href{https://arxiv.org/abs/astro-ph/0108011}{{\ttfamily
  astro-ph/0108011}}].

\bibitem{Yagi:2011wg}
K.~Yagi and N.~Seto, \emph{{Detector Configuration of Decigo/Bbo and
  Identification of Cosmological Neutron-Star Binaries}},
  \href{https://doi.org/10.1103/PhysRevD.83.044011}{\emph{Phys. Rev. D}
  {\bfseries 83} (2011) 044011}
  [\href{https://arxiv.org/abs/1101.3940}{{\ttfamily 1101.3940}}].

\bibitem{Isoyama:2018rjb}
S.~Isoyama, H.~Nakano and T.~Nakamura, \emph{{Multiband Gravitational-Wave
  Astronomy: Observing Binary Inspirals with a Decihertz Detector, B-Decigo}},
  \href{https://doi.org/10.1093/ptep/pty078}{\emph{PTEP} {\bfseries 2018}
  (2018) 073E01} [\href{https://arxiv.org/abs/1802.06977}{{\ttfamily
  1802.06977}}].

\bibitem{Gao:2017rgh}
D.~Gao, J.~Wang and M.~Zhan, \emph{{Atomic Interferometric Gravitational-Wave
  Space Observatory (Aigso)}},
  \href{https://doi.org/10.1088/0253-6102/69/1/37}{\emph{Commun. Theor. Phys.}
  {\bfseries 69} (2018) 37} [\href{https://arxiv.org/abs/1711.03690}{{\ttfamily
  1711.03690}}].

\bibitem{Wang:2019oeu}
G.~Wang, D.~Gao, W.-T. Ni, J.~Wang and M.~Zhan, \emph{{Orbit Design for Space
  Atom-Interferometer Aigso}},
  \href{https://doi.org/10.1142/S0218271819400042}{\emph{Int. J. Mod. Phys. D}
  {\bfseries 29} (2020) 1940004}
  [\href{https://arxiv.org/abs/1905.00600}{{\ttfamily 1905.00600}}].

\bibitem{Ni:2019nau}
W.-T. Ni, G.~Wang and A.-M. Wu, \emph{{Astrodynamical Middle-Frequency
  Interferometric Gravitational Wave Observatory Amigo: Mission Concept and
  Orbit Design}}, \href{https://doi.org/10.1142/S0218271819400078}{\emph{Int.
  J. Mod. Phys. D} {\bfseries 29} (2020) 1940007}
  [\href{https://arxiv.org/abs/1909.04995}{{\ttfamily 1909.04995}}].

\bibitem{Hu:2017mde}
W.-R. Hu and Y.-L. Wu, \emph{{The Taiji Program in Space for Gravitational Wave
  Physics and the Nature of Gravity}},
  \href{https://doi.org/10.1093/nsr/nwx116}{\emph{Natl. Sci. Rev.} {\bfseries
  4} (2017) 685}.

\bibitem{Kuns:2019upi}
K.~A. Kuns, H.~Yu, Y.~Chen and R.~X. Adhikari, \emph{{Astrophysics and
  Cosmology with a Decihertz Gravitational-Wave Detector: Tiango}},
  \href{https://doi.org/10.1103/PhysRevD.102.043001}{\emph{Phys. Rev. D}
  {\bfseries 102} (2020) 043001}
  [\href{https://arxiv.org/abs/1908.06004}{{\ttfamily 1908.06004}}].

\bibitem{Luo:2015ght}
{\scshape TianQin} collaboration, \emph{{Tianqin: a Space-Borne Gravitational
  Wave Detector}},
  \href{https://doi.org/10.1088/0264-9381/33/3/035010}{\emph{Class. Quant.
  Grav.} {\bfseries 33} (2016) 035010}
  [\href{https://arxiv.org/abs/1512.02076}{{\ttfamily 1512.02076}}].

\bibitem{Hu:2018yqb}
X.-C. Hu, X.-H. Li, Y.~Wang, W.-F. Feng, M.-Y. Zhou, Y.-M. Hu et~al.,
  \emph{{Fundamentals of the Orbit and Response for Tianqin}},
  \href{https://doi.org/10.1088/1361-6382/aab52f}{\emph{Class. Quant. Grav.}
  {\bfseries 35} (2018) 095008}
  [\href{https://arxiv.org/abs/1803.03368}{{\ttfamily 1803.03368}}].

\end{thebibliography}\endgroup

\end{document}